\documentclass[a4paper, 12pt]{article}

\usepackage{latexsym,amsmath,amsfonts,amssymb}
\usepackage{mathrsfs}
\usepackage{tikz}
\usetikzlibrary{decorations.pathmorphing,cd,decorations.markings,calc}
\usepackage[utf8]{inputenc}
\usepackage[american]{babel}
\usepackage{graphicx}
\usepackage{bbm}
\usepackage{cite}
\usepackage[colorlinks=true, citecolor=blue, linkcolor=blue, linktocpage=true]{hyperref}
\usepackage{xcolor,colortbl}

\renewcommand{\baselinestretch}{1.2}
\newcommand{\ts}{\textstyle}

\newcommand{\wt}{\widetilde}
\newcommand{\wb}{\overline}
\newcommand{\wh}{\widehat}

\newcommand{\matht}[1]{\ensuremath{\boldsymbol{#1}}}
\newcommand{\tps}[2]{\texorpdfstring{#1}{#2}}

\newcommand{\fP}{\mathfrak{P}}

\newcommand{\eg}{\textit{e.g.}}

\newcommand{\ie}{\textit{i.e.}}

\numberwithin{equation}{section}

\newcommand{\nn}{\nonumber}
\newcommand{\mat}[1]{\begin{pmatrix} #1 \end{pmatrix}}
\newcommand{\smat}[1]{\big( \begin{smallmatrix} #1 \end{smallmatrix} \big)}
\newcommand{\be}{\begin{equation}} \newcommand{\ee}{\end{equation}}
\newcommand{\ben}{\begin{equation*}} \newcommand{\een}{\end{equation*}}
\newcommand{\bea}{\begin{equation} \begin{aligned}} \newcommand{\eea}{\end{aligned} \end{equation}}
\newcommand{\bean}{\begin{equation*} \begin{aligned}} \newcommand{\eean}{\end{aligned} \end{equation*}}

\newcommand{\cA}{\mathcal{A}}

\newcommand{\cC}{\mathcal{C}}

\newcommand{\cF}{\mathcal{F}}

\newcommand{\calL}{\mathscr{L}}

\newcommand{\cN}{\mathcal{N}}
\newcommand{\cO}{\mathcal{O}}

\newcommand{\cS}{\mathcal{S}}

\newcommand{\cW}{\mathcal{W}}

\newcommand{\bC}{\mathbb{C}}

\newcommand{\bR}{\mathbb{R}}

\newcommand{\bZ}{\mathbb{Z}}
\newcommand{\fD}{\mathfrak{D}}

\newcommand{\fg}{\mathfrak{g}}

\newcommand{\fs}{\mathfrak{s}}

\newcommand{\rme}{\mathrm{e}}
\newcommand{\rmm}{\mathrm{m}}
\newcommand{\sT}{\mathsf{T}}

\newcommand{\YM}{\mathrm{YM}}

\newcommand{\one}{^{[1]}}

\newcommand{\ed}{\;.}
\newcommand{\ec}{\;,}

\def\su{\mathfrak{su}}

\def\so{\mathfrak{so}}

\DeclareMathOperator{\Tr}{Tr}

\DeclareMathOperator{\re}{\mathbb{R}e}
\DeclareMathOperator{\im}{\mathbb{I}m}

\DeclareMathOperator{\PD}{PD}

\newcommand{\inv}{^{\raisebox{.2ex}{$\scriptscriptstyle-1$}}}

\definecolor{Gray}{gray}{0.85}
\definecolor{LightCyan}{rgb}{0.88,1,1}
\definecolor{YellowT}{rgb}{1,1,0.2}
\definecolor{GreenT}{rgb}{0.15,1,0.15}

\newcolumntype{a}{>{\columncolor{Gray}}c}
\newcolumntype{b}{>{\columncolor{LightCyan}}c}
\newcolumntype{y}{>{\columncolor{YellowT}}c}
\newcolumntype{g}{>{\columncolor{GreenT}}c}

\newcommand\scalemath[2]{\scalebox{#1}{\mbox{\ensuremath{\displaystyle #2}}}}

\begin{document}
	\thispagestyle{empty}
	\fontsize{12pt}{20pt}
	\begin{flushright}
		SISSA  07/2023/FISI
	\end{flushright}
	\vspace{13mm}  
	\begin{center}
		{\huge Non-invertible symmetries along 4d RG flows  
  }
		\\[13mm]
		{\large Jeremias Aguilera Damia$^a$, \, Riccardo Argurio$^a$, \, Francesco Benini$^{b , \, c , \, d }$, \\[6mm]\, Sergio Benvenuti$^{d}$, \, Christian Copetti$^{b , \, d}$ and Luigi Tizzano$^{a}$ }	
				\bigskip
				
				{\it
						$^a$ Physique Th\'eorique et Math\'ematique and International Solvay Institutes\\
Universit\'e Libre de Bruxelles, C.P. 231, 1050 Brussels, Belgium  \\[.2em]
						$^b$ SISSA, Via Bonomea 265, 34136 Trieste, Italy \\[.2em]
						$^c$ ICTP, Strada Costiera 11, 34151 Trieste, Italy \\[.2em]
						$^d$ INFN, Sezione di Trieste, Via Valerio 2, 34127 Trieste, Italy \\[.2em]
					}
		
	\end{center}

\begin{abstract}
 We explore novel examples of RG flows preserving a non-invertible self-duality symmetry. Our main focus is on $\cN=1$ quadratic superpotential deformations of 4d $\cN=4$ super-Yang-Mills theory with gauge algebra $\mathfrak{su}(N)$. A theory that can be obtained in this way is the so-called $\cN=1^*$ SYM where all adjoint chiral multiplets have a mass. Such IR theory exhibits a rich structure of vacua which we thoroughly examine. Our analysis elucidates the physics of spontaneous breaking of self-duality symmetry occurring in the degenerate gapped vacua. The construction can be generalized, taking as UV starting point a theory of class $\cS$, to demonstrate how non-invertible self-duality symmetries exist in a variety of $\cN=1$ SCFTs. We finally apply this understanding to prove that the conifold theory has a non-invertible self-duality symmetry.
\end{abstract}

\newpage
\pagenumbering{arabic}
\setcounter{page}{1}
\setcounter{footnote}{0}
\renewcommand{\thefootnote}{\arabic{footnote}}

{\renewcommand{\baselinestretch}{.88} \parskip=0pt
	\setcounter{tocdepth}{2}
	\tableofcontents}


\section{Introduction}
\label{sec: intro}

Global symmetries in quantum field theory (QFT) are an indispensable tool, especially at strong coupling. As such, symmetry considerations oftentimes have inspired far-reaching insights into non-perturbative physics.

The modern understanding of $p$-form global symmetries, pioneered in \cite{Gaiotto:2014kfa}, begins with the notion of topological operators $\fD_a(\gamma)$ supported on codimension-($p+1$) submanifolds that act on $p$-dimensional operators. From this topological perspective, it is natural to consider symmetries that do not obey the standard multiplication rules of a group, but that instead exhibit a more general categorical fusion algebra:
\be
\label{noninvfusionschematic}
\fD_a \times \fD_b = \sum\nolimits_{c} \cN_{ab}^c \, \fD_c \ed
\ee
The expression on the right-hand-side offers two important points of departure from the standard discussion of symmetry groups. First, it allows for the presence of multiple operators $\fD_c$. Second, the coefficients $\cN^c_{ab}$ are topological quantum field theories (TQFTs) instead of ordinary $c$-numbers. This type of symmetries are commonly referred to as \emph{non-invertible}. Their existence has been appreciated for a long time in the context of both $d=2$ conformal field theories (CFTs) \cite{Verlinde:1988sn, Petkova:2000ip} and $d=3$ TQFTs \cite{Witten:1988hf, Barkeshli:2014cna}, where many new insights have appeared recently \cite{Chang:2018iay, Thorngren:2019iar, Gaiotto:2020iye, Komargodski:2020mxz, Thorngren:2021yso, Huang:2021zvu, Burbano:2021loy,Lin:2023uvm}.
The generalization of these concepts to $d\geq 3$ QFTs has been the subject of intense research over the past few years, see for instance \cite{Choi:2021kmx, Kaidi:2021xfk, Apruzzi:2021nmk, Bhardwaj:2022yxj, Hayashi:2022fkw, Choi:2022zal, Kaidi:2022uux, Choi:2022jqy, Cordova:2022ieu, Antinucci:2022eat, Damia:2022rxw, Damia:2022bcd, Bhardwaj:2022lsg, Bartsch:2022mpm, Apruzzi:2022rei, Mekareeya:2022spm, Chen:2022cyw, Choi:2022fgx, Yokokura:2022alv, Bhardwaj:2022kot, Bhardwaj:2022maz, Bartsch:2022ytj, Heckman:2022xgu}. 

The starting point of this work is the recent discovery that $\cN=4$ super-Yang-Mills (SYM) theory exhibits a particular class of non-invertible symmetries implemented by the so called \emph{self-duality} defects \cite{Choi:2021kmx, Choi:2022zal, Kaidi:2021xfk, Kaidi:2022uux}. The basic idea behind the construction of these defects is as follows. An $\cN=4$ SYM theory $\mathsf{T}_\rho (\tau_{\YM})$ with gauge algebra $\mathfrak{g}$ is labelled by a choice $\rho$ of gauge group and by its complexified gauge coupling $\tau_{\YM}$. The theory enjoys $SL(2,\bZ)$ Montonen-Olive duality \cite{Montonen:1977sn}. The $S$ generator identifies theories with \mbox{$\tau_{\YM}' = -\frac{1}{\tau_{\YM}} $} and Langlands dual gauge group $\rho' = \rho^{L}$, while $T$ identifies $\tau_{\YM}'$ with $\tau_{\YM} + 1$ which means a shift of the theta angle for $\rho$ \cite{Aharony:2013hda}.
Besides, the theory admits topological manipulations $\varphi_g: \mathsf{T}_\rho(\tau_{\YM}) \to \mathsf{T}_{\rho_g}(\tau_{\YM})$ which correspond to a generalized gauging of the 1-form symmetry. We will discuss them in depth in Section~\ref{sec: SYM}. These act only on the global form $\rho$ and generically give rise to an interface between two inequivalent gauge theories. For special values $\tau_{\YM}^*$ of $\tau_{\YM}$ it might happen that $\mathsf{T}_{\rho_g}(\tau_{\YM}^*)$ is dual to $\mathsf{T}_{\rho}(\tau_{\YM}^*)$ by some element $g$ of $SL(2,\bZ)$. Denoting the duality interface by  $I_g: \mathsf{T}_\rho (\tau_\YM) \to \mathsf{T}_{\rho_{g}} (g\cdot\tau_\YM)$ we see that $\tau_{\YM}^*$ needs to be invariant under a discrete subgroup of $SL(2, \bZ)$. We can then define a defect $\fD_g  : \mathsf{T}_\rho(\tau_{\YM}^*) \to \mathsf{T}_\rho(\tau_{\YM}^*) $ by
\be
\fD_g = \varphi^\dagger \circ I_g \;,  \qquad\text{pictorially:}\qquad
\raisebox{-4em}{\begin{tikzpicture}
    \draw[color=white!80!blue, fill=white!80!blue] (0,0) node[below,color=blue] {$\varphi^\dagger_g$} -- (2.5,0) node[below,color=blue] {$I_g$} -- (2.5,2)  -- (0,2) ;
    \draw[line width=1] (0,0) -- (0,2);  \draw[line width=1, dashed] (2.5,0) -- (2.5,2); 
    \node at (-1.25,1) {$\mathsf{T}_\rho(\tau_{\YM}^*)$}; \node at (3.75,1) {$\mathsf{T}_\rho(\tau_{\YM}^*)$}; \node at (1.25,1) {$\mathsf{T}_{\rho_{g^{-1}}}(\tau_{\YM}^*)$};
\node at (5,1) {$=$}; 
\draw[color=blue, line width = 1] (6,0) node[below] {$\fD_g$} -- (6,2);
\end{tikzpicture}}
\ee
This defect is non-invertible since:
\be
\fD_g \times \overline{\fD}_g = \cC \, ,
\ee
with $\cC$ a 3d condensation defect of the 1-form symmetry \cite{Roumpedakis:2022aik}.
At $\tau_\YM^*= i$, $\fD_g$ implements a non-invertible $S$ duality symmetry while at $\tau_\YM^* = e^{\frac{2\pi i}{3}}$ it implements a non-invertible $ST$ triality symmetry. 

The main goal in this paper is to explore how the non-invertible symmetry implemented by $\fD_g$ behaves under RG flows triggered by mass deformations. Our primary focus is on $\cN=1$ preserving deformations: decomposing the $\cN=4$ vector multiplet into one vector and three chiral $\cN=1$ multiplets, we study what happens when some or all of the chiral multiplets get a mass. 
Since the $SL(2,\bZ)$ modular group has a non-trivial action on the supercharges of $\cN=4$ SYM \cite{Intriligator:1998ig, Kapustin:2006pk}, the only way to preserve the duality/triality symmetry is to combine $\fD_g$ with a suitable R-symmetry rotation inside the maximal torus of the $SU(4)_R$ R-symmetry group that is also explicitly broken by the massive deformation. In practice this amounts to a discrete rotation inside the superconformal $U(1)_R$. The fact that $\cN=1$ mass deformations can preserve a duality symmetry was already appreciated long ago \cite{Argyres:1999xu}. A similar idea was also exploited in the construction of 4d S-folds \cite{Garcia-Etxebarria:2015wns, Aharony:2016kai, Argyres:2016yzz}.

When all three chiral multiplets get a mass, one ends up with the so called $\cN=1^*$ SYM theory \cite{Donagi:1995cf, Dorey:1999sj, Dorey:2000fc, Dorey:2001qj}. The structure of vacua of this theory is quite rich, and comprises both gapped and gapless vacua. In this work we present a detailed description of how the non-invertible self-duality defects are realized in these vacua. An important point is that all gapped vacua are organized into orbits of the \emph{spontaneously broken duality/triality symmetry}. A spontaneously broken non-invertible symmetry can relate ground states featuring different physical properties. More precisely, we find that the domain walls associated with such a spontaneously broken symmetry can interpolate between different realizations of the $1$-form symmetry, {\it e.g.}, between a Higgs and a confining phase. Similar examples of this kind have already appeared in the literature \cite{Chang:2018iay, Thorngren:2019iar}, remarkably in the tricritical Ising model in two dimensions.%
\footnote{The tricritical Ising model admits a relevant duality-preserving deformation by the $\epsilon'$ operator. With a certain sign of the deformation, the theory flows to a gapped phase with three vacua. These should be interpreted as a direct sum $\bZ_2 \oplus \text{trivial}$, representing the spontaneous symmetry breaking (SSB) of the duality symmetry. We thank Yifan Wang for pointing this out to us.}
Our work provides an explicit example in four dimensions.%
\footnote{See also \cite{Kaidi:2021xfk} for a related analysis in the context of pure $\cN=1$ $SO(3)$ SYM.}

We exhaustively classify the TQFTs describing the gapped vacua and establish a bijective correspondence between gapped phases, global forms of the gauge group, and TQFTs with $\bZ_N$ 1-form symmetry. We provide detailed examples of these features for theories based on the Lie algebra $\mathfrak{su}(N)$ with $N=2$, 3, and 4. For gapless vacua in $\mathfrak{su}(N)$ $\cN=1^*$ SYM with $N\geq3$, we describe how to realize non-invertible symmetries in a Coulomb vacuum using a single $U(1)$ gauge field, while more general cases remain to be explored in future work. As a final consistency check, we compute the anomalies associated to duality and triality symmetries following \cite{Hsieh:2019iba, Hsieh:2020jpj}.%
\footnote{See also \cite{Debray:2023yrs} for similar computations.}
We conclude that, for all values of $N$ with a self-duality symmetric ground state, the anomaly vanishes. This follows from a nontrivial cancellation between the pure self-duality anomaly and the cubic anomaly associated with the superconformal $U(1)_R$.

Our analysis can be readily adapted to other interesting classes of RG flows that we explore. It is known that the self-duality defects are not limited to $\cN=4$ SYM and exist also in more general $\cN=2$ SCFTs \cite{Bashmakov:2022jtl, Bashmakov:2022uek, Antinucci:2022cdi, Carta:2023bqn}. As shown in \cite{Bashmakov:2022uek, Antinucci:2022cdi}, for those theories it is more convenient to consider the class $\cS$ formulation. We generalize previous analysis by considering theories whose UV curve is a torus with regular punctures associated to fundamental matter. As discussed above, giving mass to the adjoint scalars in the \mbox{$\cN=2$} vector multiplets allows one to define $\cN=1$ preserving RG flows which also preserve the non-invertible self-duality symmetry defects. In this context, we discuss the RG flow whose IR fixed point is the $\cN=1$ conifold theory (also known as Klebanov-Witten theory) \cite{Klebanov:1998hh}.%
\footnote{Self-duality symmetry defects in the conifold theory, as well as general $\cN=1$ SCFTs from D3-branes at conical singularities, have also been studied in \cite{Heckman:2022xgu}.}
Thanks to our construction, we can demonstrate that the conifold theory inherits a self-duality symmetry defect. To strengthen this perspective, we study the holographic realization of the model in type IIB string theory and show how it matches the expectations from field theory.
In the special case of the conifold theory with $\su(2) \times \su(2)$ gauge algebra, one can derive the action of S-duality on the global forms from Seiberg duality of $\so(4)$ super-QCD. Our construction can be generalized to class $\cS$ theories without a Lagrangian description since many details of their symmetry structure have already been studied \cite{DelZotto:2015isa, Bhardwaj:2021pfz, Bhardwaj:2021mzl}. This opens up the possibility to discover non-invertible duality defects in a wide variety of strongly coupled $\cN=1$ SCFTs.

We point out that our strategy to construct novel examples of non-invertible self-duality defects is in principle stable under the effect of mass deformations that break supersymmetry completely. An interesting example is the deformation by the bottom component of the $\cN=4$ Konishi supermultiplet. This point certainly deserves further study, in particular it would be interesting to understand its relevance for the IR physics of $N_f=4$ adjoint QCD.

The paper is organized as follows.
In Section~\ref{sec: SYM} we review the construction of non-invertible self-duality defects in $\cN=4$ SYM and their action on global variants. We then introduce the duality action on massive deformations and describe the construction of non-invertible symmetries preserved along the RG flow.
In Section~\ref{sec: onestar} we discuss the realization of self-duality symmetries on massive vacua of $\cN=1^*$ SYM and formulate a simple criterion to establish when these symmetries are spontaneously broken, including a description of the physical consequences of such spontaneous symmetry breaking (SSB) process. We then comment on some applications to gapless (Coulomb) vacua and the cubic anomalies of self-duality symmetries.
In Section~\ref{sec:gapless} we study generalizations of our construction in which the starting point is an $\cN=2$ SCFT. We give an explicit derivation of the self-duality defects for the $\bZ_r$ orbifolds of $\cN=4$ SYM using its class $\cS$ description, and discuss their fate under massive RG flows. Our main example is the flow to the $\cN=1$ conifold theory. We discuss the holographic description of such a duality and its relation to $\cN=1$ Seiberg duality.
Technical details, mainly involving formal manipulations of TQFTs with $\bZ\one_N$ 1-form symmetry, are included in appendices.

\section{Duality-preserving flows in 4d \tps{\matht{\cN=4}}{N=4} SYM}
\label{sec: SYM}

We begin with a review of several aspects of $\cN=4$ theories which are instrumental to our work. We start reviewing the classifications of global variants for a pure gauge theory based on the $\fg = \mathfrak{su}(N)$ gauge algebra, following \cite{Aharony:2013hda}. We then combine this with Montonen-Olive duality to describe in detail the non-invertible duality and triality defects of $\cN=4$ SYM. Finally, we outline the general strategy to introduce duality-preserving RG flows and give several key examples.

\subsection{Global variants and line operators}
\label{globalvariants}

Consider a four-dimensional gauge theory based on the simple Lie algebra $\mathfrak{g}$ and with connected gauge group $G$. We denote the universal cover of $G$ by $\wt{G}$ and the center of the latter by $\Gamma$. Then the gauge group is $G = \wt{G} / \Pi$ and its center is $\Gamma/\Pi$, where $\Pi \subset \Gamma$ is a subgroup. Different choices of $\Pi$ define different global variants of the gauge group based on the same Lie algebra. However, in order to completely characterize the gauge theory, in general one should also specify the set of allowed line operators \cite{Aharony:2013hda}. Each line falls in a conjugacy class specified by a pair $(a_\rme, b_\rmm) \in \wh\Gamma \times \Gamma$, where $\wh\Gamma = \{\text{homomorphisms}: \Gamma \to \bR/\bZ\}$ is the Pontryagin dual to $\Gamma$ (and is isomorphic to it). The set $\mathscr{L}$ of allowed conjugacy classes of lines is subject to a version of Dirac quantization (ensuring that correlation functions of line operators are local) which is expressed in terms of the natural pairing:
\be
0 = \bigl\langle (a_\rme, b_\rmm) , (c_\rme, d_\rmm) \big\rangle = a_\rme(d_\rmm) - c_\rme(b_\rmm) \ed
\ee
For example, if $\Gamma=\bZ_N$ then $(a_\rme, b_\rmm) = (n,m) \mod N$ and the quantization condition requires that for each pair of line operators $(n,m)$ and $(n',m')$: 
\be
nm'-mn'=0 \mod N \ed
\ee
The set of allowed classes should also be maximal. Satisfaction of the two conditions corresponds to the choice of a Lagrangian subgroup $\mathscr{L}$ of $\wh\Gamma\times\Gamma$. The set of Wilson lines, namely of elements in $\mathscr{L}$ of the form $(\hat\gamma, 0)$, is fixed by the global variant $G$: $\hat\gamma \in \wh\Gamma$ are such that $\Pi \subset \ker(\hat\gamma)$ (they form a subgroup isomorphic to the center $\Gamma/\Pi$ of $G$). Even after imposing this condition, for a given global variant there can be physically different choices of allowed line operators (\ie, different choices of $\mathscr{L}$). From the Lagrangian point of view, the different theories are distinguished by various types of theta terms \cite{Aharony:2013hda}.

In this section we will mostly focus on four-dimensional $\cN=4$ supersymmetric%
\footnote{Supersymmetry will play a role only when discussing dualities. Everything that pertains to the choice of global variant and line operators applies to generic gauge theories. However one should remember that, in the presence of matter fields, only those global variants such that the matter representation is invariant under $\Pi$ are allowed. Here we will have in mind the case that the matter representation is invariant under the full $\Gamma$.}
Yang-Mills theory with Lie algebra $\mathfrak{su}(N)$. In this case the maximal center is $\Gamma = \bZ_N$ and the gauge group is $SU(N)/\Pi$ with $\Pi = \bZ_k$ and $k$ is a divisor of $N$. When $\Pi$ is trivial, the gauge group is $SU(N)$ and the set of allowed line operators corresponds to the Lagrangian subgroup
\be
\mathscr{L}_{1,0} = \bigl\{ (a_\rme, b_\rmm) = (n,0) \text{ mod } N \bigm| n \in \bZ \bigr\} \ed
\ee
These are the standard Wilson line operators in representations of $N$-ality $n$, and in this paper we denote them by $D_{(n,0)}$. When $\Pi = \bZ_N$, the possible sets of allowed classes of line operators (\ie, the possible Lagrangian subgroups) are given by
\be
\label{linesets}
\mathscr{L}_{N,\ell} = \bigl\{ (a_\rme, b_\rmm) = (\ell m, m) \text{ mod } N \bigm| m \in \bZ \bigr\} \ec
\ee
where for each value of $\ell \in \bZ_N$ we have a different possible choice. Indeed there are distinct theories, that we dub $\bigl( PSU(N) = SU(N)/\bZ_N \bigr){}_\ell$, whose line operators have charges as in \eqref{linesets}. We denote such line operators by $D_{(\ell m, m)}$. Importantly, the $N$ distinct theories are related to each other by the Witten effect \cite{Witten:1979ey}. A shift $\theta \to \theta + 2\pi$ of the theta parameter induces a shift $(a_\rme, b_\rmm)\to (a_\rme + b_\rmm, b_\rmm)$ of the labels, implying the following identification between different $PSU(N)_\ell$ theories:
\be
PSU(N)^{\theta+2\pi}_\ell = PSU(N)^{\theta}_{\ell+1} \ed
\ee
In the general case, one has $\Pi = \bZ_k$ with $k$ a divisor of $N$, and we write $N= kk'$. The possible Lagrangian subgroups are
\be
\label{general Lagrangians subgs}
\mathscr{L}_{k,\ell} = \bigl\{ (a_\rme, b_\rmm) = n \, (k, 0) + m \, (\ell,k') \text{ mod } N \bigm| n,m \in \bZ \bigr \} \ec
\ee
with $\ell \in \bZ_k$. They correspond to the possible classes of line operators in the theories $\bigl( SU(N)/\bZ_k \bigr){}_\ell$. A shift $\theta \to \theta + 2\pi$ relates different theories according to:
\be
\bigl( SU(N)/\bZ_k \bigr)^{\theta+2\pi}_\ell = \bigl( SU(N)/\bZ_k \bigr)^{\theta}_{\ell + k'} \ed
\ee
Note that when $\gcd(k,k')= p > 1$, there exist $p$ sets of theories with the same gauge group that are not related by shifts of $\theta$. For instance, when $N=4$ the two theories $\bigl( SU(4)/\bZ_2 \bigr){}_0$ and $\bigl( SU(4)/\bZ_2 \bigr){}_1$ are not related by a shift of $\theta$. The total number of distinct theories (for fixed $\theta$) is given by the sum of the divisors of $N$:
\be
\label{number of theories}
\text{number of theories} = \sigma_1(N) = \sum_{d \,| N} \, d \ed
\ee
This number will come up again in later discussions.

For 4d $\cN=4$ SYM with gauge algebra $\su(N)$, theories with different global form are related by $SL(2,\bZ)$ S-duality. The duality group is generated by the operations $T = \smat{1 & 1 \\ 0 & 1}$ and $S = \smat{0 & -1 \\ 1 & 0}$. They act by fractional linear transformations on the complexified gauge coupling
\be
\label{gauge coupling}
\tau_\text{YM} = \frac{\theta}{2\pi} + \frac{4\pi i}{g_\YM^2}
\ee
as follows, $T:\tau_\text{YM} \mapsto \tau_\text{YM} +1$ and $S: \tau_\text{YM} \mapsto - \frac1{\tau_\text{YM}}$. They also act on the labels $\smat{ a_\rme \\ b_\rmm }$ of lines as described by their matrix form, and consequently on the Lagrangian subgroup $\calL$ as \cite{Aharony:2013hda, Bergman:2022otk}%
\footnote{Note that $S: \calL_{k,0} \mapsto \calL_{k',0}$ because $p=k$ and then $0^{-1} = 0$ in $\bZ_1$. \label{foo: inverse}}
\be
\label{TSactionL}
T: \calL_{k,\ell} \,\mapsto\, \calL_{k, \ell+k'} \;,\qquad S: \calL_{k,\ell} \,\mapsto\, \calL_{N/p,\; - k'(\ell/p)^{-1}_{\text{mod } k/p}} \quad \text{with } p = \gcd(k,\ell) \;.
\ee
The group that acts faithfully on the set of global forms is $PSL(2,\bZ_N)$.%
\footnote{The theories form orbits under $PSL(2,\bZ_N)$, one orbit for each square divisor $d^2$ of $N$. Each orbit contains a number of theories equal to $\psi\bigl(N/d^2\bigr)$, where $\psi(n) = n \prod_{p|n} \bigl(1 + 1/p \bigr)$ is the Dedekind psi function and the product is over all primes $p$ that divide $n$. The orbit of $\calL_{k,\ell}$ is identified by $d=\gcd(k, k', \ell)$.}
We will come back to it in Section~\ref{sec: self-duality}.

Another way to connect gauge theories with different global form, and that applies to more general 4d gauge theories than $\cN=4$ SYM because it does not rely on duality, is the following \cite{Gaiotto:2014kfa}. We note that the gauge theories are characterized by a 1-form symmetry. For simplicity we restrict to gauge algebra $\su(N)$ and to matter whose representation is invariant under the maximal center $\Gamma$. Then the 1-form symmetry is $\bZ_N$ for the global variants $SU(N)$ and $PSU(N)$, while more generally it is the Pontryagin dual to
\be
\label{eq: ellgroup}
\calL_{k,\ell} = \Bigl( \bZ_{k'} \times \bZ_{N/\gcd(k',\ell)} \Bigr) \text{\Large$/$} \bZ_{k' / \gcd(k',\ell)} \,\cong\, \bZ_{N/\gcd(k,k',\ell)} \times \bZ_{\gcd(k,k',\ell)}
\ee
and it has order $N$ in all cases. The first equality is apparent from (\ref{general Lagrangians subgs}) where the quotient comes from the relation $k(\ell, k') \sim (\ell k, 0)$ mod $N$, while the second isomorphism is shown in \cite{Bergman:2022otk}.%
\footnote{For $\cN=4$ SYM, as it should, the 1-form symmetry is invariant (up to isomorphisms) under S-duality.}
We can then associate to each theory its partition function in the presence of a background 2-form discrete gauge field $B$ for the 1-form symmetry. Different global variants are related by partially or completely gauging the 1-form symmetry. It also turns out that there is an anomaly in the space of couplings \cite{Cordova:2019uob} that involves the $\theta$ angle and the 1-form symmetry. For instance, for gauge group $SU(N)$ we have that under $\theta\to \theta+2\pi$ the partition function on $X$ changes as
\be
Z(B) \;\to\; \exp\biggl( \frac{2 \pi i(N-1)}{2 N} \int_X \fP(B) \biggr) \; Z(B) \ec
\ee
where $B \in H^2(X, \bZ_N)$, $\fP \!:\! H^2(X, \bZ_N) \to H^4 \bigl( X, \bZ_{N\gcd(N,2)} \bigr)$ is the Pontryagin square operation, and $\fP(B)$ is an even integer class on spin manifolds, which we will assume for the rest of the paper. Indeed, we can always add to the theory the following local counterterm in the background field $B$ (that we can think of as an invertible TQFT on $X$):
\be
\label{sptkb}
\text{SPT}_k(B) = \exp\biggl( \frac{2\pi i k}{2N}\int_X \fP(B) \biggr) \ed 
\ee
This produces a background-dependent phase.

We can thus refine the classification of global forms by keeping track of the coupling to the background field, in terms of the partition functions $Z(B)$. For simplicity%
\footnote{We either assume that $N$ does not have square divisors and thus the 1-form symmetry is $\bZ_N$ in all global forms of $\su(N)$, or we restrict to the orbit of $SU(N)$. When $N$ is not prime, one can also partially gauge the 1-form symmetry, \ie, restrict the sum in (\ref{eq: SL2 manip 1}) to $b \in H^2(X,\bZ_k)$ for a divisor $k$ of $N$. We will discuss some additional details of a non-prime $N$ in an example below.}
we restrict to the case that the 1-form symmetry is $\bZ_N$. We then introduce two topological manipulations $\sigma$ and $\tau$ \cite{Gaiotto:2014kfa, Ang:2019txy, Bhardwaj:2020ymp} that transform the partition function as follows:
\bea
\label{eq: SL2 manip 1}
\sigma &: \quad & [ \sigma Z](B) &= \frac{1}{\sqrt{\bigl\lvert H^2(X,\bZ_N) \bigr\rvert}} \sum_{b \, \in H^2(X,\bZ_N)} \exp\biggl( \frac{2 \pi i}{N} \int_X b \cup B \biggr) \; Z(b) \ec \\
\tau &: \quad & [ \tau Z](B) &= \exp\biggl( \frac{2 \pi i}{2 N} \int_X \fP(B) \biggr) \; Z(B) \ed
\eea
The operation $\sigma$ corresponds to gauging the 1-form symmetry, while $\tau$ acts by adding the local counterterm $\text{SPT}_1(B)$. Together, they generate the group $SL(2,\bZ_N)$.%
\footnote{Studying the operations $\sigma,\tau$ on generic (not necessarily spin) oriented 4-manifolds, one obtains a central extension of $SL(2,\bZ)$ in which the extension is by stacking with the invertible TQFTs $\bigl[ \bZ_N^{(2)} \bigr]_k$ \cite{Bhardwaj:2020ymp, Antinucci:2022cdi}. Here $\bigl[ \bZ_N^{(2)} \bigr]_k$ indicates the $\bZ_N$ 2-form Dijkgraaf-Witten theory whose partition function reads $Y_k = |H^2|^{-1/2} \sum_{b\in H^2} \exp \bigl( \frac{2\pi ik}{2N} \int_X \fP(b) \bigr)$ for $k \in \bZ_N^*$. For instance, one finds that $(\sigma\tau)^3 = Y_1$. On simply-connected spin manifolds all $Y_k$ reduce to 1 \cite{Choi:2022zal}, and the faithful action of $\sigma,\tau$ reduces to $SL(2,\bZ_N)$. In any case, one can consider the action of $\sigma,\tau$ on theories coupled to $B$ modulo stacking with invertible TQFTs, which is $SL(2,\bZ_N)$.}

The topological operations $\sigma$ and $\tau$ affect the spectrum of line operators, and this fact can be used to understand how they act on global variants. We use the notations $SU(N)_p$ and $PSU(N)_{k,p}$ where the extra label $p$ indicates the counterterm (\ref{sptkb}). We also define the following combined action:
\be
\Phi_k \equiv \sigma \, \tau^k \ed
\ee
The $SU(N)_0$ theory has genuine Wilson line operators $D_{(n,0)}$ supported on a curve $\gamma$ which we can always couple to the $\bZ_N$ background field as follows:
\be
\label{wilsoncoupled}
D_{(n,0)}(B) = D_{(n,0)}[\gamma] \; \exp\biggl( \frac{2\pi i n}{N}\int_{\Sigma}B\biggr) \ec \qquad\qquad \partial\Sigma=\gamma \ed
\ee
This gives $D_{(n,0)}$ the correct charge under $\bZ_N$ gauge transformations.
One can also consider non-genuine 't~Hooft operators $D^{U}_{(0,m)}[\gamma]$ which need to live at the boundary of the surface operators $U_\rme[\Sigma]$ that generate the electric 1-form symmetry \cite{Gaiotto:2014kfa}, in other words, the lines $D^U_{(0,m)}$ are in twisted sectors. If we act with $\Phi_0$ on $SU(N)_0$, the correlation functions of $U_\rme[\Sigma]$ become trivial and the roles of $D_{(n,0)}$ and $D^{U}_{(0,m)}$ get exchanged, namely the 't~Hooft operators $D_{(0,m)}$ become genuine line operators while the Wilson lines $D^U_{(n,0)}$ become non-genuine. To understand what is the effect of $\Phi_{k>0}$ on $SU(N)_0$ we first need to consider a $\bZ_N$ background gauge transformation $B\to B+\delta \lambda$. From \eqref{wilsoncoupled} we have
\be
D_{(n,0)}(B) \,\to\, \exp\biggl( \frac{2\pi i n}{N} \int_\gamma \lambda \biggr) \, D_{(n,0)}(B) \ec
\ee
expressing the $\bZ_N$ charge. On the other hand, the insertion of an 't~Hooft line operator $D^U_{(0,m)}$ supported on $\gamma$ corresponds to a variation of the background field by a $\Delta B$ such that $\delta (\Delta B) = m \PD[\gamma]$, where $\PD[\gamma]$ is a delta-3-form localized on $\gamma$ (its Poincar\'e dual). As a consequence, the counterterm $\text{SPT}_k(B)$ ceases to be gauge invariant and instead:
\be
\text{SPT}_k(B) \,\to\, \exp\biggl( - \frac{2\pi i k m}{N} \int_X \PD[\gamma] \cup \lambda \biggr) \, \text{SPT}_k(B) \ed
\ee
This shows that acting with $\Phi_k$ on $SU(N)_0$ gives a global form in which the line operators $D_{(km, m)} \equiv D_{(km,0)} D_{(0,m)}$ are gauge invariant and thus genuine. This fact characterizes the global variant $PSU(N)_k$ with Lagrangian subgroup $\calL_{N,k}$.

As an example, the set of actions for $\su(2)$ $\cN=4$ SYM is:
\be
\label{su2top}
\begin{tikzcd}
SU(2)_0 \arrow[d, leftrightarrow,"\tau"] \arrow[r, leftrightarrow, "\sigma"] & PSU(2)_{0,0} \arrow[d, shift left=1ex, leftrightarrow, "\tau" near end] & PSU(2)_{1,0} \arrow[d, leftrightarrow, "\tau"] \\
SU(2)_1 \arrow[rru, leftrightarrow, "\sigma" near start] & PSU(2)_{0,1} \arrow[r, leftrightarrow, "\sigma"]  & PSU(2)_{1,1}
\end{tikzcd}
\ee

\subsection{Non-invertible self-duality defects}
\label{sec: self-duality}

For $\cN=4$ SYM, one can discuss the interplay between the topological action described above and the action of $SL(2,\bZ)$ S-duality discussed around (\ref{gauge coupling}).
We will do that for Lie algebra $\su(N)$. As explained in \cite{Kapustin:2006pk} and reviewed above, the action of $SL(2,\bZ)$ generates duality orbits connecting theories with different global structure. The action on lines is
\be
\label{eq:STonlines}
S: (n,m) \,\mapsto\, (-m,n) \;,\qquad\qquad
T: (n,m) \,\mapsto\, (n+m,m) \ed
\ee
For instance, the duality orbits for $N=2,3,4$ are the following.

\noindent$\su(2)$: \vspace{-1.5em}
\be
\label{su2}
\begin{tikzcd}
\arrow[loop left, "T"] SU(2) \arrow[r, leftrightarrow, "S"] & PSU(2)_0 \arrow[r, leftrightarrow, "T"]  & PSU(2)_1 \arrow[loop right, "S"] 
\end{tikzcd}
\ee
The theories $PSU(2)_0$ and $PSU(2)_1$ are usually denoted as $SO(3)^+$ and $SO(3)^-$.

\noindent$\su(3)$: \vspace{-1.2em}
\be
\label{su3}
\begin{tikzcd}[row sep=tiny]
 & & PSU(3)_1 \arrow[dd, "T"] \arrow[dd, leftrightarrow, bend left = 50, "S",shift left=1.5ex] \\
SU(3) \arrow[loop left, "T"] \arrow[r, leftrightarrow, "S"] & PSU(3)_0 \arrow[ru, "T"]  \\
& & PSU(3)_2 \arrow[lu,"T"]
\end{tikzcd}
\ee

\noindent$\su(4)$: \vspace{-2.5em}
\bea
\begin{tikzcd}
\raisebox{0pt}[0.5em][0pt]{$(SU(4)/\bZ_2)_0 \quad\;\;$} \arrow[loop right, "{S,T}"]
\end{tikzcd} \\[-1.5em]
\begin{tikzcd}[row sep=tiny]
& & PSU(4)_1 \arrow[rd, "T"] \arrow[dd, leftrightarrow, "S"] \\
SU(4) \arrow[loop below, "T"] \arrow[r, leftrightarrow, "S"] & PSU(4)_0 \arrow[ru, "T"]  & & PSU(4)_2 \arrow[ld,"T"] \arrow[r, leftrightarrow, "S"] & {(SU(4)/\bZ_2)_1} \arrow[loop below, "T"] \\
& & PSU(4)_3 \arrow[lu,"T"]
\end{tikzcd}
\eea

As described in \cite{Kaidi:2022uux}, one can refine the description by taking into account the coupling to a $\bZ_N$ background field $B$ and combine the actions of $S,T$ with those of $\sigma,\tau$. For instance, in the case of $\su(2)$ $\cN=4$ SYM one obtains:
\be
\label{su2orbits}
\begin{tikzcd}
SU(2)_0 \arrow[d, leftrightarrow, "{T,\tau}"'] \arrow[r, leftrightarrow, "{S,\sigma}"] & PSU(2)_{0,0} \arrow[d, shift left=1ex, leftrightarrow, "\tau" near end] \arrow[r, leftrightarrow, "T"] & PSU(2)_{1,0} \arrow[d, leftrightarrow, "{S,\tau}"] \\
SU(2)_1 \arrow[rru, leftrightarrow, "\sigma" near start] \arrow[r, leftrightarrow, "S"'] & PSU(2)_{0,1} \arrow[r, leftrightarrow, "{T,\sigma}"]  & PSU(2)_{1,1}
\end{tikzcd}
\ee
Similar tables, of increasing complexity, can be drawn for groups of higher rank (see \cite{Kaidi:2022uux}). We describe a simple way to do that for $N$ prime in Section~\ref{ssec:globalvariantmatrix}.

The topological and duality operations can be used to construct self-duality defects. As an example, consider the theory $SU(2)_0$. From \eqref{su2orbits} we see that
\be
[\sigma S \, Z_{SU(2)_0}](\tau_\YM,B) = Z_{SU(2)_0} \Bigl( -\tfrac1{\tau_\YM}, B \Bigr) \ed
\ee
In particular, for $\tau_\YM=i$ the operation $\sigma S$ maps the theory to itself and therefore it becomes a symmetry, implemented by a topological defect in the theory.
The construction generalizes to any value of $N$. Let us denote by $\sT_\rho(\tau_\text{YM})$ the set of global forms of $\cN=4$ SYM with Lie algebra $\su(N)$, where $\rho$ labels the global form. The family of theories is acted upon by duality transformations implemented by topological interfaces $I_g: \mathsf{T}_\rho(\tau_\YM) \to \mathsf{T}_{\rho_g}(g\cdot\tau_\YM)$, where $g \in SL(2,\bZ)$ acts both on the global form and the coupling. Meanwhile, the global form $\rho_g$ can also be reached by a sequence of $\sigma$ and $\tau$ operations, implemented by another topological interface $\varphi_g: \mathsf{T}_\rho(\tau_\YM) \to \mathsf{T}_{\rho_g}(\tau_\YM)$. The action of the modular group on the conformal manifold of $\cN=4$ SYM has fixed points at orbifold singularities on the fundamental domain $\cF = H^+ / SL(2,\bZ)$, located at
\be
\tau^*_\YM = i ,\; e^{2\pi i /3} ,\; i \infty \ed
\ee
At these points the stabilizer group $H = \bigl\{ g \in SL(2,\bZ) \bigm| g \cdot \tau^*_\YM = \tau^*_\YM \bigr\}$ is%
\footnote{We have also included the action of charge conjugation $C$ which acts trivially on $\cF$.}
\be
H = \bZ_4 ,\; \bZ_6 ,\; \bZ \ec
\ee
and is generated by $S$, $ST$, $T$, respectively. At the special points, $H$ becomes a symmetry and for each of its elements we can construct a defect
$\fD_g = \varphi^\dag_g \circ I_g: \sT_\rho(\tau^*_\YM) \mapsto \sT_\rho(\tau^*_\YM)$. The defect $\mathfrak{D}_g$ satisfies the following relations \cite{Kaidi:2022uux}:
\bea
\fD_g \times \fD_h &= \cN_{g,h} \; \fD_{gh} \\
\fD_g \times \wb\fD_g &= \cC \ec 
\eea
where $\cN_{g,h}$ are some 3d TQFTs, and $\wb\fD_g$ is the orientation reversal of $\fD_g$. The topological operator $\cC$ is known as a 3d condensate, or higher gauging \cite{Roumpedakis:2022aik}, of the 1-form symmetry on a 3d submanifold.

\subsection{Duality-preserving RG flows}
\label{dualpreserve}

After reviewing the self-duality symmetries of $\cN=4$ SYM, we address the question of constructing relevant deformations that trigger an RG flow while preserving those symmetries.

In the language of $\cN=1$ supersymmetry, the field content of $\cN=4$ SYM is given by a vector multiplet $V = (A_\mu, \lambda_\alpha)$ and three chiral multiplets $\Phi_{i=1,2,3} = (\phi_i, \psi_{i\alpha})$ transforming in the adjoint representation of the gauge group. The theory has a superpotential
\be
W = \sqrt{2} \, \Tr\Phi_1[\Phi_2, \Phi_3] \ed
\ee
The R-symmetry group is $SU(4)_R$, although only $U(1) \times SU(3)$ is manifest in the $\cN=1$ description. The $SL(2,\bZ)$ duality group also acts on the supercharges $Q^A_\alpha$, where $A$ is an index in the fundamental representation of $SU(4)_R$ \cite{Intriligator:1998ig, Kapustin:2006pk}. Given $M = \smat{a & b \\ c & d}$ in $SL(2,\bZ)$ that acts as $\tau_\YM \mapsto \frac{a \tau_\YM + b}{c \tau_\YM + d}$ on the coupling, its action on the supercharges is
\be
\label{SU2Z action on Q}
M \cdot Q_\alpha^A = \biggl( \frac{\lvert c \, \tau_\YM + d \rvert }{c \, \tau_\YM + d} \biggr)^{1/2} Q^A_\alpha \ed
\ee
This action was chosen so as to commute with $SU(4)_R$. The square root signifies that the group that acts on the supersymmetries is a double cover of $SL(2,\bZ)$, centrally extended by $S^4 = C^2 = (-1)^F$ which changes the sign of all fermions. The extension is called the metaplectic group $Mp(2,\bZ)$. Notice that $(-1)^F$ is also a central element of $SU(4)_R$. The one in (\ref{SU2Z action on Q}) is a $U(1)$ action, and the one on $\wb{Q}$ is its conjugate. Indeed, an operator $\cO$ is said to have charge $q$ under modular transformations if it transforms as
\be
M \cdot \cO = \biggl( \frac{\lvert c \, \tau_\YM + d \rvert}{c \, \tau_\YM + d} \biggr)^{q/2} \cO \ec
\ee
and thus $Q^A_\alpha$ has charge $1$. For a $\bZ_k$ subgroup of $SL(2,\bZ)$ occurring at an invariant value of $\tau_\YM$, its action on the supercharges is
\be
\label{MQ}
M \cdot Q_\alpha^A = e^{-\frac{i\pi}{k}} \, Q_\alpha^A \ed
\ee
The action of the modular group on the field strength $F$ and its dual $\wt F$ is the standard one described by the matrix representation: $\smat{ \wt F \\ F} \mapsto M \smat{ \wt F \\ F}$. This implies that the operator $\frac1{\sqrt{\im\tau}} (\wt F - \tau F)$ has charge 2.
For the choice we made of the modular action, the real scalars $\phi_I$ (with $I$ an index in the rank-2 antisymmetric representation of $SU(4)_R$) are invariant, while the fermions transform as the $\wb{Q}$'s.

A simple class of superpotential deformations that we will analyze in this paper is:
\be
\label{supdef}
\delta W = \sum^3_{i=1} m_i \Tr \Phi^2_i \ed
\ee
These superpotentials partially break supersymmetry. If all the masses $m_i \neq 0$, the resulting theory only retains $\cN=1$ supersymmetry and it is typically referred to as the $\cN=1^*$ theory. When $m_1=m_2 \neq 0$ and $m_3=0$, the theory has $\cN=2$ supersymmetry. If instead $m_1=m_2=0$ but $m_3\neq0$, the theory flows to a $\cN=1$ superconformal fixed point with a conformal manifold \cite{Leigh:1995ep}.

We ask whether the superpotential deformations \eqref{supdef} are also invariant under the action of the duality defects $\fD_g$ introduced above. We notice that the superspace coordinates $\theta$ are also charged under S-duality, in particular they transform in the opposite way with respect to the supercharges \eqref{MQ}, and therefore
\be
\label{modular charge d2theta}
M \cdot d^2\theta = e^{-\frac{2\pi i}k} \, d^2\theta \ed
\ee
Here $d^2\theta$ is the (holomorphic) differential for the $\cN=1$ superspace coordinate.
It follows that the superpotential should transform as $M \cdot W = e^{\frac{2\pi i}k} W$ in order to lead to an invariant Lagrangian, however the chiral superfields $\Phi_i$ are neutral. To remedy, we combine $\fD_g$ with a suitable R-symmetry rotation $R_\varphi$ inside the maximal torus of $SU(4)_R$ such that 
\be
\label{newsym}
\fD_{\varphi} = R_\varphi \times \fD_g
\ee
is a symmetry of the deformed theory. The complex scalar fields $\phi_i$, the fermions $\psi_{i\alpha}$ and the gaugino $\lambda_\alpha$ have the following charges under the Cartan subalgebra $U(1)_{1,2,3}$ of $SU(4)_R$:
\be
\label{su4charges}
\begin{array}{c|c|c|c|c|c|c|c||c}
        & \phi_1 & \phi_2 & \phi_3 & \psi_1 & \psi_2 & \psi_3 & \lambda & d^2\theta \\
\hline
U(1)_1 & 1 & 0 & 0 & \frac12 & - \frac12 & - \frac12 & \frac12 & -1 \\
U(1)_2 & 0 & 1 & 0 & - \frac12 & \frac12 & - \frac12 & \frac12 & -1 \\
U(1)_3 & 0 & 0 & 1 & - \frac12 & - \frac12 & \frac12 & \frac12 & -1 \\[.1em]
\hline
U(1)_R & \frac23 & \frac23 & \frac23 & - \frac13 & -\frac13 & -\frac13 & 1 & -2
\end{array}
\ee
In the last column we indicated the charges of the differential $d^2\theta$, while in the last row we indicated the charges under the $\cN=1$ superconformal R-symmetry $U(1)_R$ which is a diagonal combination of the previews three.
Using these charges, we can discuss which non-invertible duality symmetries of the form \eqref{newsym} are preserved by the superpotential deformations  \eqref{supdef} and along the resulting RG flows. 
For the class of examples we consider, we can always choose $R_\varphi$ to lie within $U(1)_R$.

\paragraph{Example I.} Let us first consider the case $m_1=m_2=0$ and $m_3 \equiv m\neq 0$:
\be
\delta W = m \Tr \Phi^2_3 \ed
\ee
This example preserves $\cN=1$ supersymmetry. Under an S-duality transformation the integrated superpotential deformation $\int\! d^2\theta\, \delta W$ picks up a phase $e^{-\frac{2\pi i}{k}}$ because of (\ref{modular charge d2theta}).
From \eqref{su4charges}, its charge under $U(1)_R$ is
\be
\Bigl[ \, {\ts \int\! d^2\theta \, \delta W } \Bigr]_R = -\frac23 \;.
\ee
We can thus compensate the effect of the duality transformation by a superconformal $U(1)_R$ rotation $R_{-\frac{3 \pi}{k}}$, that applies a rotation by an angle $-\frac{3\pi}k$ and thus it gives a phase $e^{\frac{2\pi i}k}$ to the integrated superpotential. We conclude that
\be
\mathfrak{D}_{\frac{2\pi}{k}} = R_{-\frac{3 \pi}{k}} \times \fD_g
\ee
is a symmetry of the deformed theory.

\paragraph{Example II.} Let us now consider the case $m_1=m_2 \equiv m\neq 0$ and $m_3=0$:
\be
\delta W = m \, \bigl( \Tr \Phi^2_1 + \Tr \Phi^2_2 \bigr) \ec
\ee
which preserves $\cN=2$ supesymmetry. The argument is the same as before, because all scalars are neutral under S-duality and have the same charge under $U(1)_R$. It follows that the operator $\mathfrak{D}_{\frac{2\pi}{k}} = R_{-\frac{3 \pi}{k}} \times \fD_g$ is a symmetry of the deformed theory.

\paragraph{Example III.} In our last example we study the generic case $m_1, m_2, m_3 \neq 0$. This deformation leads to the $\cN=1^*$ theory. For special non-vanishing values of the masses one can have extra preserved symmetries, which however decouple in the IR. For instance, without loss of generality, one can consider the case $m_1 =m_2=m_3 \equiv m\neq 0$,
\be
\label{delta1*}
\delta W = m \, \bigl( \Tr \Phi^2_1 + \Tr \Phi^2_2+\Tr \Phi^2_3 \bigr) \ec
\ee
that preserves an $SO(3)$ flavor symmetry rotating the chiral multiplets $\Phi_i$ as a triplet.
As before, $\int\! d^2\theta\, \delta W$ has charge $-\frac23$ under $U(1)_R$ and so the operator
\be
\label{eq: dualityN1star}
\mathfrak{D}_{\frac{2\pi}{k}}= R_{-\frac{3 \pi}{k}} \times \fD_g
\ee
is a symmetry of the $\cN=1^*$ theory.

\paragraph{Example IV.} One could also consider a non-supersymmetric deformation triggered by the Konishi operator:%
\footnote{The possible interest of such a deformation was pointed out to us by Justin Kaidi and Kantaro Ohmori.}
\be
\label{Konishdef}
\delta S = \int\! d^4x\,  \mu \, \cO_{\text{Konishi}} \;,
\ee
which at weak coupling is given by $\cO_{\text{Konishi}} = \sum_{i=1}^3 \Tr \bigl( \phi_i \phi_i^\dag \bigr)$. This deformation at $\tau_{\YM} = i$ preserves both the non-invertible duality symmetry $\fD$, which satisfies $\fD^4 \sim (-1)^F \, \cC$ (here $\cC$ is the condensate operator), as well as the full $SU(4)_R$ R-symmetry. The Konishi operator is relevant at $\tau_\text{YM} = i$ for small enough values of $N$ (for instance for $N \leq 4$ \cite{Beem:2013hha}).
The flow is intrinsically strongly coupled and the theory in the IR cannot be gapped due to the cubic 't~Hooft anomaly of $SU(4)_R$. The correct description is either given by spontaneous symmetry breaking (SSB) of the chiral symmetry due to gaugino condensation, or by a strongly coupled CFT. If the deformation happens to interpolate smoothly between weak and strong coupling $\tau_{\YM}$, the CFT should be the same as the one conjectured for the pure $SU(N)$ gauge theory with $N_f = 4$ adjoint Weyl fermions.

Let us conclude this section with a brief remark about the case $k=4$, namely about the element $S\in SL(2,\bZ)$. Notice that, even though both $S$ and $R_{-\frac{3\pi}{4}}$ are of order 8 --- indeed $S^4 = R_{-\frac{3\pi}{4}}^4 = (-1)^F$ --- the preserved combined action $\fD_{\frac{2\pi}{k}}$ defines a $\bZ_4$ symmetry. In the IR of the deformed $\cN=4$ SYM theory at $\tau_\text{YM}=i$, the preserved symmetry is a non-invertible $\bZ_4$ times $\bZ_2^F$ fermion parity (as opposed to a $\bZ_8^F$). This will have consequences for the cancellation of cubic anomalies, which we examine in Section~\ref{ss: cubic}.

\section{Gapped flows: \tps{\matht{\cN=1^*}}{N=1*}}
\label{sec: onestar}

In this section we will focus on mass deformations of the form \eqref{delta1*} leading to the ${\cal N}=1^*$ SYM theory. When the three masses $m_{1,2,3}$ in (\ref{supdef}) are generic, the deformation completely breaks the (continuous part of the) $SU(4)_R$ symmetry of the UV theory but preserves both the non-invertible duality and triality symmetries $\fD$ for $\tau_{\YM}=i$ or $\tau_{\YM}=e^{\frac{2\pi i}{3}}$, respectively.
Recall that the preserved defects are combined with a discrete $R_{-\frac{3 \pi}{k}}$ R-symmetry rotation.

The $\cN=1^*$ SYM theory has been intensively studied in the past \cite{Donagi:1995cf, Dorey:1999sj, Dorey:2000fc, Polchinski:2000uf, Dorey:2001qj}. It has a discrete set of vacua, some of which are completely gapped while others include Coulomb phases. Let us briefly review them here. 
By a suitable rescaling of the chiral fields (or by choosing equal masses), the F-term equations read:
\be
[\Phi_i , \Phi_j] = \epsilon_{ijk} \; \Phi_k \;.
\ee 
Since the $\Phi_i$ are $N\times N$ traceless matrices, solutions to the above equations are given in terms of $N$-dimensional (possibly reducible) representations of the Lie algebra $\mathfrak{su}(2)$. For any positive integer $N$, the algebra $\mathfrak{su}(2)$ has a unique (up to isomorphisms) irreducible representation of that dimension. Therefore each classical vacuum corresponds to a partition of $N$ into positive integers. For a generic partition $N = \sum_{d=1}^N n_d \,d$ with $k$ non-vanishing terms, the gauge algebra is Higgsed to
\be
\label{eq:classicalHiggsing}
\bigoplus_{d} \mathfrak{su}(n_d) \; \oplus \, (k-1)\ \mathfrak{u}(1) \ed
\ee
The two extreme cases with $k=1$ are: $\{d\}=\{N\}$, $n_N = 1$ where the gauge group is fully Higgsed, and $\{d\}=\{1\}$, $n_1 = N$ where the $\mathfrak{su}(N)$ gauge symmetry is unbroken.  

Quantum mechanically, the dynamics is that of a pure ${\cal N}=1$ gauge theory with gauge group \eqref{eq:classicalHiggsing}. Each $\mathfrak{su}(n_d)$ factor flows to a confining phase with $n_d$ degenerate gapped vacua, while each Abelian factor hosts a Coulomb phase.  The latter are always present unless $k=1$, and thus the fully gapped vacua are in correspondence with the divisors $d$ of $N$. In this case, the gauge group is Higgsed to $\mathfrak{su}(N/d)$ leading to $N/d$ confining vacua. Then the total number of gapped vacua is
\be
\label{eq:Ngapped}
\text{number of gapped vacua}=\sum_{d \vert N} \; N/d = \sum_{d \vert N} \; d \;.
\ee
This is equal to the number $\sigma_1(N)$ of global variants of the gauge group, that we defined in \eqref{number of theories}. We will see that there is indeed a natural one-to-one correspondence between gapped vacua and global variants, as they are both labeled by a choice of line operators.

According to 't Hooft's classification of gapped phases in gauge theories, vacua are characterized by the condensation of dyons. For $N$ prime, there are exactly $N+1$ gapped vacua because of \eqref{eq:Ngapped}. These consist of a perturbative Higgs vacuum, where $D_{(1,0)}$ condenses, and $N$ confining vacua where $D_{(s,1)}$ condense, with $s=0$ corresponding to confinement of purely electric charges and $s=1,\ldots, N-1$ describing different patterns of oblique confinement. We will denote these vacua by $H$ and $C^{(s)}$ respectively. Using a modern terminology, whenever a genuine line operator condenses, the 1-form symmetry is spontaneously broken and the corresponding phase hosts a non-invertible TQFT. On the other hand, condensation of non-genuine line operators leads to vacua described by an SPT.

When $N$ has non trivial divisors, the structure is more involved due to the occurrence of combined Higgs/confining phases. More precisely, take a divisor $d$ of $N$ and consider vacua in which the gauge symmetry is Higgsed to $\mathfrak{su}(N/d)$. Then, quantum mechanically, there are $N/d$ confining vacua characterized by patterns of condensation of $\mathfrak{su}(N/d)$ dyons. These vacua will be denoted by $HC^{(s)}_{d,N/d}$, with $s=0,\ldots, N/d-1$. They exhibit a partially broken 1-form symmetry and host a non-trivial TQFT on top of an SPT. 

One of the main purposes of this work is to show that the vacua discussed above, which are characterized by different physical properties, are nevertheless related by non-invertible duality and triality symmetries. The latter are hence generically spontaneously broken along the RG flow triggered by the mass deformation.
For the sake of clarity, we will first describe the vacua of the $\cN=1^*$ theories based on the Lie algebra $\mathfrak{su}(2)$. Then we will present a general discussion, and carry out an explicit analysis of the $\mathfrak{su}(3)$ and $\mathfrak{su}(4)$ $\cN=1^*$ gauge theories.

\subsection[\tps{$\mathfrak{su}(2)$}{su(2)} gauge algebra]{\matht{\mathfrak{su}(2)} gauge algebra}
\label{ssec: su2}

The $\cN=1^*$ SYM theory with gauge algebra $\mathfrak{su}(2)$ possesses three degenerate gapped vacua, $\{H, \,C^{(0)}, \,C^{(1)}\}$ where the corresponding line operators $\{D_{(1,0)}, \,D_{(0,1)}, \,D_{(1,1)}\}$ condense. These vacua are in one-to-one correspondence with the global variants $SU(2)$, $PSU(2)_0$ and $PSU(2)_1$ respectively. In the $SU(2)$ global structure, the fundamental Wilson line $D_{(1,0)}$ is the unique genuine line.
In the Higgs vacuum $H$,  the $\mathbb{Z}_2$ 1-form symmetry is spontaneously broken and it is realized by a $\mathbb{Z}_2$ gauge theory. In the remaining two vacua, the $\mathbb{Z}_2$ 1-form symmetry is unbroken and the low energy dynamics is described by a $\mathbb{Z}_2$ SPT. This analysis can be easily repeated for other choices of global structure.
We will now focus on two special values of the UV modular parameter $\tau_{\YM}$, namely $\tau_{\YM}=i$ and $\tau_{\YM}=e^{\frac{2\pi i}{3}}$ where the duality element $S$ and the triality element $ST$ become non-invertible global symmetries, respectively.

The action of $S$ and $ST$ on the spectrum of line operators is summarized by the following diagrams: 
\be
\begin{tikzcd}[column sep=small]
	D_{(1,0)} \arrow[rr, leftrightarrow, "\text{\small$S$}"] & & D_{(0,1)} \\
	& D_{(1,1)} \arrow[loop above, "\text{\small$S$}"]&
\end{tikzcd}  \qquad\qquad
\begin{tikzcd}[column sep=small]
D_{(1,0)} \arrow[rr, "\text{\small$ST$}"] & & D_{(0,1)} \arrow[dl, "\text{\small$ST$}"]\\
	& D_{(1,1)} \arrow[ul, "\text{\small$ST$}"]&
\end{tikzcd}
\ee
We thus find that, at $\tau_{\YM}=i$ and under the action of the non-invertible global symmetry defect associated to $S$ duality, the three degenerate vacua form a doublet $\{H,C^{(0)}\}$ and a singlet $\{C^{(1)}\}$. On the other hand, at $\tau_{\YM}=e^{\frac{2\pi i}{3}}$, these vacua form a triplet, with its elements permuted under the action of $ST$ triality. 

In this example we found two or more degenerate vacua with different physical properties that are permuted under the action of a non-invertible symmetry. We interpret this phenomenon as a spontaneous breaking of non-invertible symmetry. On the other hand,  if a vacuum happens to be a singlet (such as $\{C^{(1)}\}$ for $\tau_{YM}=i$), the corresponding non-invertible symmetry is unbroken in such a vacuum.

Whenever a symmetry, even if non-invertible, is spontaneously broken, it is important to identify a corresponding local order parameter. In fact, these symmetries have an invertible action on local operators and we expect to characterize a spontaneously-broken phase in a standard way.

Let us recall that, in order for the symmetry to be preserved by a massive deformation, the topological defect is constructed by composing a modular transformation (either $S$ or $ST$) with a discrete R-symmetry rotation acting on the matter fields. From the discussion in Section~\ref{dualpreserve}, we find this rotation to be by a $k$-th root of unity, with $k=4$ for $S$ duality at $\tau_{\YM}=i$ and $k=6$ for $ST$ triality at $\tau_{\YM}=e^{\frac{2\pi i}{3}}$. The simplest chiral operator which is non-trivially charged under this action is 
\be
\label{eq:Worderparameter}
\cO =  \Tr \Phi_i^2 \ec
\ee
for some value of $i=1,2,3$. Using the charges in (\ref{su4charges}), one finds
\begin{align}
\mathfrak{D}_{-\frac{2\pi}{4}} \; \cO &= e^{i\pi} \; \cO \; \mathfrak{D}_{-\frac{2\pi}{4}} \quad , &&\quad (\tau_{\YM}=i) \label{eq:OPaction}\\
\mathfrak{D}_{-\frac{2\pi}{6}} \; \cO &= e^{\frac{2\pi i}{3}}\;\cO \; \mathfrak{D}_{-\frac{2\pi}{6}} \quad , &&\quad (\tau_{\YM}=e^{\frac{2\pi i}{3}})\nonumber
\end{align}
Expectation values of the order parameter \eqref{eq:Worderparameter} need to comply with the above selection rules when evaluated on the different vacua. Furthermore, $\langle \cO\rangle=\partial_m \, W$, with $W$ the superpotential for which an exact expression in any of the gapped vacua has been derived in \cite{Dorey:1999sj, Dorey:2001qj}:
\be
\label{eq:onestarsuperpot}
W^{p, q, s}(\tau_{\YM})\equiv \bigl\langle HC^{(s)}_{p,q} \big| W \big| HC^{(s)}_{p,q} \bigr\rangle = \frac{N^3}{24 } m^3 \im(\tau_\YM) \biggl[ E_2(\tau_\YM) - \frac{p}{q}E_2\biggl( \frac{p \tau_\YM + s}{q} \biggr) \biggr] \ed
\ee
Here the relevant values of $\{p,q,s\}$ are $H=\{2,1,0\}$, $C^{(0)}=\{1,2,0\}$ and $C^{(1)}=\{1,2,1\}$ and $E_2(x)$ denotes the quasi-modular second Eisenstein series. By exploiting the modular properties of these functions, one can easily show (for details see Appendix \ref{app: orderpar}) that
\bea
\label{eq: chiralvev}
 &\cO^{1,2,0}(\tau_\YM = i ) = e^{i\pi} \, \cO^{2,1,0}(\tau_\YM = i) \ec \quad \cO^{1,2,1}(\tau_\YM = i) = 0 \ec \\
&\cO^{1,2,1} \bigl( \tau_\YM = e^{\frac{2\pi i}{3}} \bigr) = e^{\frac{2\pi i}{3}} \, \cO^{1,2,0} \bigl( \tau_\YM = e^{\frac{2\pi i}{3}} \bigr) = e^{\frac{4\pi i}{3}} \,\cO^{2,1,0} \bigl( \tau_\YM = e^{\frac{2\pi i}{3}} \bigr) \ec
\eea
hence in perfect agreement with \eqref{eq:OPaction}. The remarkable fact that these functions satisfy the required properties only at $\tau_\YM=i$ and $\tau_\YM=e^{\frac{2\pi i}{3}}$ should be regarded as a direct consequence of the presence of non-invertible symmetry defects at these specific points in the UV conformal manifold.

\paragraph{Action on lines and TQFTs.}
We now give a detailed description of gapped phases in $\mathfrak{su}(2)$ $\cN=1^*$ SYM theory. Vacua with different physical properties are mapped into each other by broken duality/triality symmetries. This is due to the non-invertibility of such global symmetries which act non-trivially on the line operators. In the following we focus on the case of $S$ duality. The analysis can be repeated for the $ST$ triality symmetry.

If we choose the global variant $SU(2)$, the symmetry defect acts implementing both $S$ and $\sigma$. It follows that a genuine line is necessarily mapped to a non-genuine line, attached to a topological surface in order to be gauge invariant, as depicted in the following figure:%
\footnote{Indeed, $S$ takes the Wilson line (the genuine line of $SU(2)$) to the 't~Hooft line (the genuine line of $PSU(2)_0$), while $\sigma$ brings back to the global variant $SU(2)$, thus making the 't~Hooft line no longer genuine.}
\bea
\label{eq:lineaction}
\begin{tikzpicture}
\draw[color=black, line width=1] (-0.75,0.25) node[below] {$D_{(1,0)}$} to [bend left=30] (-0.75,1.25) to [bend right=30] (-0.75,2.25);
\node at (-2,2.5) {$\bZ_2$};
\node at (3,2.5) {$\text{SPT}$};
\filldraw[fill=white!70!cyan,opacity=0.5] (0,0) -- (2,0.5) -- (2,2.5) -- (0,2) -- cycle;
\node at (5,1) {\huge $\leadsto$};	
\node at (2.25,0.25) {$\fD$}; 
\begin{scope}[shift={(8,0)}]
\node at (-1,2.5) {$\bZ_2$};
\node at (4,2.5) {$\text{SPT}$};
\filldraw[fill=white!70!cyan,opacity=0.5] (0,0) -- (2,0.5) -- (2,2.5) -- (0,2) -- cycle;
\filldraw[color=red, fill=white!50!red,opacity=0.5] (1,0.25) to [bend left=30] (1,1.25) to [bend right=30] (1,2.25) to (3,2.25) to [bend left=30] (3,1.25) to [bend right=30] (3,0.25) --cycle;
\draw[color=black, line width=1] (3,0.25) node[below] {$D_{(0,1)}$} to [bend left=30] (3,1.25) to [bend right=30] (3,2.25);
\node at (-0.25,-0.25) {$\fD$}; 
\end{scope}
\end{tikzpicture}
\eea
This implies that vacua related by the action of the non-invertible duality symmetry $\fD$ can be physically inequivalent. For instance, in the Higgs vacuum $|H\rangle$ the (genuine) fundamental Wilson line condenses, $\langle H| D_{(1,0)}|H\rangle\neq0$, so as to play the role of an order parameter for the spontaneously broken 1-form symmetry $\mathbb{Z}_2^{[1]}$. This topological order is described by a $\mathbb{Z}_2$ gauge theory. 
The duality defect maps the genuine line $D_{(1,0)}$ to the non-genuine line $D_{(0,1)}$ corresponding to a vacuum $S \, |H\rangle=|C^{(0)}\rangle$. Of course, the fact that $\langle H| D_{(1,0)} |H\rangle\neq 0$ implies that $\langle C^{(0)}|D_{(0,1)}|C^{(0)}\rangle\neq 0$. However, since the 't~Hooft line is not genuine in the $SU(2)$ global variant, $D_{(0,1)}$ cannot serve as an order parameter for the 1-form symmetry. Hence $\mathbb{Z}_2^{[1]}$ is preserved in $|C^{(0)}\rangle$ and confinement takes place. There is no topological order and the effective topological theory is accounted by a $\mathbb{Z}_2^{[1]}$ SPT phase.
The fact that the duality symmetry defect $\fD$ maps a genuine order parameter to a non-genuine (twisted) order parameter gives strong evidence for the spontaneous breaking of a non-invertible symmetry. 

We now explicitly construct the topological theories describing each gapped vacuum. As explained above, the Higgs vacuum hosts a $\mathbb{Z}_2$ gauge theory, as it should be for a phase with spontaneously broken $\mathbb{Z}_2^{[1]}$ 1-form symmetry. Including the conventional normalization, the partition function for this theory is 
\be
Z[\mathbb{Z}_2](B) = \frac{1}{\sqrt{ \lvert H^2(X,\mathbb{Z}_2) \rvert}} \sum_{b\in H^2(X,\mathbb{Z}_2)} e^{\frac{2\pi i}{2} \int\! b \,\cup B} = \sqrt{ \lvert H^2(X,\mathbb{Z}_2) \rvert} \;\; \delta_{H^2(X,\bZ_2)}(B) \ed
\ee
Because of the delta function in cohomology, this theory does not suffer from any ambiguity in the form of an SPT factor.
On the other hand, the vacua $C^{(0)}$ and $C^{(1)}$ are characterized by a $\mathbb{Z}_2$ SPT phase. There are two possibilities for such a topological phase, namely
\be
\label{eq:Z2SPT}
\text{SPT}(B)_{-k} \; : \;\; -\frac{2\pi i k}{4}\int \fP(B)\ec \qquad k=0,1 \ed
\ee
The question of which value of $k$ in \eqref{eq:Z2SPT} corresponds to $C^{(0)}$ depends on a choice of local counterterm in the UV and, as such, is not universal. We choose this counterterm in such a way that to $C^{(0)}$ is assigned $\text{SPT}_0$, {\it i.e.}, the trivially gapped vacuum.

In fact, since $S$ duality maps $|H\rangle$ to $|C^{(0)}\rangle$, and the symmetry defect involves the action of $\sigma$ (given by gauging the $\mathbb{Z}_2^{[1]}$ 1-form symmetry), then the choice is fixed by the following computation:
\be
[\sigma Z](B) = \frac{1}{\sqrt{ \lvert H^2 \rvert}} \sum_{b\in H^2(X,\mathbb{Z}_2)} Z[\mathbb{Z}_2](b) \; e^{\frac{2\pi i}{2} \!\int\! b \, \cup B}
= 1 =\text{SPT}_0 \ed
\ee
We can similarly determine which topological phase corresponds to $C^{(1)}$. This state is a singlet  under $S$ duality and it turns out that this condition is enough to fix its low energy description. The only $\mathbb{Z}_2^{[1]}$ SPT phase that is left invariant by $\sigma$ is precisely $\text{SPT}_1$: $\bigl[ \sigma Z[\text{SPT}_1] \bigr](B) = Z[\text{SPT}_1](B)$. Alternatively, we can use the property that $C^{(0)}$ and $C^{(1)}$ are mapped to each other by the action of the modular transformation $T$. In the UV, this is accounted for by a shift $\theta\to\theta+2\pi$, under which the partition function gets a background dependent phase $\exp\left(\frac{2\pi i}{4} \!\int\! \fP(B) \right)$. We then conclude that, having assigned $\text{SPT}_0$ to $C^{(0)}$, there is no other choice but assigning $\text{SPT}_1$ to $C^{(1)}$. We can finally summarize the analysis with the following formal expression for the topological field theory assigned to each vacuum:
\be
\label{eq:SU(2)TQFT}
\text{TQFT}= \begin{cases} \mathbb{Z}_2 \bigoplus \text{SPT}_0 \qquad\qquad &\text{for } \{ H , \, C^{(0)} \} \ec \\
\text{SPT}_1 \qquad &\text{for } C^{(1)}\ed
\end{cases}
\ee
The duality symmetry $S$ is spontanously broken in the vacua $\{ H, C^{(0)}\}$, while it is preserved in the vacuum $C^{(1)}$. Note that the ones in (\ref{eq:SU(2)TQFT}) are the only topological field theories with $\mathbb{Z}_2$ 1-form symmetry. Each gapped vacuum of $SU(2)$ $\cN=1^*$ SYM theory realizes one of them. We will see that this is a general feature also for theories with gauge group $SU(N)$.   

The story is similar for the triality symmetry at $\tau_{\YM}= e^{\frac{2 \pi i}{3}}$. It is clear that the TQFT assignment to the three vacua must be the same as the one for $\tau_{\YM}=i$. The triality symmetry in the $SU(2)$ theory is implemented by composing the UV duality $ST$ with a topological manipulation $\sigma \tau$. A quick computation shows that the three $\bZ_2^{[1]}$ TQFTs form a unique orbit under the action of triality:
\be
\begin{tikzcd}[column sep = small]
\bZ_2 \arrow[rr, "\text{\small$\sigma \tau$}"] & & \text{SPT}_0 \arrow[dl, "\text{\small$\sigma \tau$}"]\\
	& \text{SPT}_1 \arrow[ul, "\text{\small$\sigma \tau$}"]&    
\end{tikzcd}
\ee
signalling the spontaneous breaking of the non-invertible triality symmetry.

\paragraph{Comments on the other global variants.}
Let us discuss the other global variants associated with the $\mathfrak{su}(2)$ algebra, namely $PSU(2)_0$ and $PSU(2)_1$ (also commonly referred to as $SO(3)^+$ and $SO(3)^-$, respectively). Notice that the condensation of dyons is a dynamical process, not affected by the choice of global variant. On the other hand, the realization of the 1-form global symmetry in each phase, either being preserved or broken,  does depend on the global aspects of the theory. Consequently, the effective low-energy description of each vacuum changes when discussing $PSU(2)_0$ or $PSU(2)_1$. The descriptions will actually be reshuffled, as we still expect a one-to-one correspondence between the three vacua and the three possible topological field theories with $\mathbb{Z}^{[1]}_2$ 1-form symmetry. For instance, in each global variant there will always be a vacuum where the condensing dyon corresponds to the genuine line, leading to a $\mathbb{Z}_2$ topological order. 

Global variants are simply related by the topological manipulations $\sigma$ and $\tau$: starting from $SU(2)$, $PSU(2)_0$ is obtained by acting with $\sigma$, while $PSU(2)_1$ by acting with $\sigma\tau$. Note that here we are {\em not} combining these operations with an $S$ duality, so that we do not change the behavior of local operators in each vacuum, only the TQFT describing them. In practice, we perform the corresponding topological manipulations on the effective topological field theories. 

Let us begin with  $G=PSU(2)_0$. We act with $\sigma$ on the topological theories describing the three vacua of $SU(2)$. We get $\text{SPT}_0$ for $H$, the $\mathbb{Z}_2$ gauge theory for $C^{(0)}$, and $\text{SPT}_1$ for $C^{(1)}$. The 't~Hooft line $D_{(0,1)}$ becomes a genuine operator, and in the vacuum $C^{(0)}$ where it condenses, the $\mathbb{Z}_2$ 1-form symmetry is spontaneously broken and described by a $\mathbb{Z}_2$ gauge theory. Since $H= S\cdot C^{(0)}$, we conclude that this phase is now realized by a trivial SPT phase. Finally, since $D_{(1,1)}$ is invariant under $S$ duality, the vacuum $C^{(1)}$ is accounted for by $\text{SPT}_1$. 

Consider finally $G=PSU(2)_1$. This variant is interesting because it is invariant under $S$ duality, and thus the $S$ duality defect becomes invertible. The effective description in each of the three vacua is obtained by acting with $\sigma\tau$ on the phases of $G=SU(2)$, leading to $\text{SPT}_0$ for $H$, $\text{SPT}_1$ for $C^{(0)}$, and the $\mathbb{Z}_2$ gauge theory for $C^{(1)}$. Alternatively, one can assign the $\mathbb{Z}_2$ gauge theory to $C^{(1)}$ by the argument based on the condensation of the genuine line $D_{(1,1)}$, and the SPT phases to $H$ and $C^{(0)}$ by the argument below.

In $PSU(2)_1$ there is a mixed anomaly between the invertible duality symmetry and the $\mathbb{Z}_2$ 1-form symmetry. From the perspective of the UV theory, this stems from the fact that $PSU(2)_1$ is not exactly invariant in the presence of a 1-form symmetry background. Indeed the action on the Lagrangian subgroup is $S \, \smat{1\\1} = \smat{-1 \\ 1 } \cong \smat{1 \\ 1}$. The equivalence $\smat{-1 \\ 1} \cong \smat{1\\1}$ is implemented by a shift of the $\theta$ angle, and this corresponds to an 't~Hooft anomaly
\be
\label{eq:SO3-mixedanomaly}
I_\text{anom} = \frac{2 \pi i }{4} \int A_S \cup \fP(B) \ec
\ee
where $A_S \in H^1(X, \bZ_4)$ is the $S$ duality gauge field. Indeed, this reproduces the phase factor arising when acting with $S$ on $\text{SPT}_0$ (vacuum $H$) to get $\text{SPT}_1$ (vacuum $C^{(0)}$).%
\footnote{As reviewed in Appendix~\ref{app: N1TQFTs}, vacua related by a broken invertible symmetry display different SPT phases if the broken symmetry participates in a mixed anomaly.}
On the other hand, this is also the anomaly of the $\mathbb{Z}_2$ gauge theory in the vacuum $C^{(1)}$.

\begin{table}{t}
\centering
\begin{gather*}
\begin{array}{c|b|b|c}
\multicolumn{4}{c}{\boldsymbol{SU(2)}} \\[0.25em]
\text{Vacuum} & H & C^{(0)} & C^{(1)} \\ \hline \hline
\text{Cond. line} & D_{(1,0)}^{\phantom{U}} & D_{(0,1)}^U & D_{(1,1)}^U  \\ \hline
\text{TQFT} & \bZ_2 & \text{SPT}_0 & \text{SPT}_1 \\ \hline
 \text{SSB} \; \fD &  \color{red} \checkmark \color{black} & \color{red} \checkmark \color{black} & \text{\texttimes} \\ \hline \hline
\end{array}
\hspace{4em}
\begin{array}{c|b|b|c} 
\multicolumn{4}{c}{\boldsymbol{PSU(2)_0}} \\[0.25em]
\text{Vacuum} & H & C^{(0)} & C^{(1)} \\ \hline \hline
\text{Cond. line} & D_{(1,0)}^U & D_{(0,1)}^{\phantom{U}} & D_{(1,1)}^U  \\ \hline
\text{TQFT} & \text{SPT}_0 & \bZ_2 & \text{SPT}_1 \\ \hline
\text{SSB} \; \fD &  \color{red} \checkmark \color{black} & \color{red} \checkmark \color{black} & \text{\texttimes} \\ \hline \hline
\end{array}
\\[1.5em]
\begin{array}{c|b|b|c} 
\multicolumn{4}{c}{\boldsymbol{PSU(2)_1}} \\[0.25em]
\text{Vacuum} & H & C^{(0)} & C^{(1)} \\ \hline \hline
\text{Cond. line} & D_{(1,0)}^U & D_{(0,1)}^U & D_{(1,1)}^{\phantom{U}}  \\ \hline
\text{TQFT} & \text{SPT}_0 & \text{SPT}_1 & \bZ_2 \\ \hline
\text{SSB} \; \fD &  \color{red} \checkmark \color{black} & \color{red} \checkmark \color{black} & \text{\texttimes} \\ \hline \hline
\end{array}
\hspace{4em}
\begin{array}{c|b|b|b}
\multicolumn{4}{c}{\boldsymbol{SU(2)} \textbf{ -- triality}} \\[0.25em]
\text{Vacuum} & H & C^{(0)} & C^{(1)} \\ \hline \hline
\text{Cond. line} & D_{(1,0)}^{\phantom{U}} & D_{(0, 1)}^U & D_{(1,1)}^U  \\ \hline
\text{TQFT} & \bZ_2 & \text{SPT}_0 & \text{SPT}_1 \\ \hline
\text{SSB} \; \fD &  \color{red} \checkmark \color{black} & \color{red} \checkmark \color{black} & \color{red} \checkmark \color{black} \\ \hline \hline
\end{array}
\end{gather*}
\caption{The first three tables summarize the gapped vacua and the patterns of non-invertible $S$ duality symmetry breaking in the three global variants of $\su(2)$. The last table --- with the same vacua as in the first one --- shows the pattern of $ST$ triality symmetry breaking.
\label{tab: su(2) phases}}
\end{table}

We can summarize the content of this section in Table~\ref{tab: su(2) phases}.%
\footnote{Condensed lines which are non-genuine are indicated by a superscript $U$, reflecting the fact that they are attached to a $U$ surface for the $\bZ_N$ 1-form symmetry. Vacua which are connected by spontaneous symmetry breaking (SSB) of the duality symmetry are indicated by the same background color.}
We also include a table for the triality symmetry at $\tau_{\YM}=e^{\frac{2\pi i}{3}}$ in the $SU(2)$ variant. Note that all three vacua form a single orbit under the symmetry action, though the effective topological field theories remain the same as in the previous case.

\subsection[\tps{$\mathfrak{su}(N)$}{su(N)} gauge algebra]{\matht{\mathfrak{su}(N)} gauge algebra}
\label{subsec: gappedTQFT}

Here we generalize the previous analysis to higher-rank gauge algebras $\mathfrak{su}(N)$. We exploit some interesting patterns that were already encountered in the simple case of $\mathfrak{su}(2)$, in particular the correspondence between gapped vacua and topological field theories with a given 1-form symmetry, and, more importantly, the realization of the duality action in terms of simple topological manipulations performed over such topological field theories.

\paragraph{TQFTs with \matht{\bZ_N} 1-form symmetry.}
For global variants with $\Gamma=\bZ_N$ 1-form symmetry, such as $SU(N)$, the TQFTs describing the gapped vacua are completely characterized by the symmetry-breaking pattern $\bZ_N \rightarrow H = \bZ_m$. Not all global variants fall in this category: the 1-form symmetry may take the form $\Gamma = \bZ_{m_1} \times \bZ_{m_2}$ as it may happen in $SU(N)/\bZ_k$, see (\ref{eq: ellgroup}). In the following we will focus on the case $\Gamma = \bZ_N$, and then make some comments about more complicated examples at the end.

A TQFT with $\bZ_N$ 1-form symmetry can be specified by the symmetry-breaking pattern $\bZ_N \rightarrow \bZ_m$. The broken symmetry is described by a $\bZ_N / \bZ_m$ gauge theory, while an SPT phase $\alpha_m$ is assigned to the preserved subgroup $\bZ_m$.%
\footnote{More precisely, one should also specify the coupling to the $\bZ_N$ background field. The choice of coupling can be absorbed by a rescaling $B \to r B$, with $\gcd(r,N)=1$.}
The latter is determined by a class
\be
\alpha_m \,\in\, H^4 \bigl( B^2 \bZ_m , \, U(1) \bigr) = \bZ_{\gcd(2,m) \, m} \ed
\ee 
As generator we take the Pontryagin square operation $\fP \!:\! H^2(X,\bZ_m) \to H^4 \bigl( X, \bZ_{\gcd(2,m) \, m} \bigr)$:
\be
\alpha_m(b) = \frac{2 \pi i}{2 m} \, \fP(b) \qquad\text{with}\qquad b \in H^2(X, \bZ_m) \ed
\ee
On spin manifolds, $\fP(b)$ is an even class and thus there are only $m$ SPTs given by $\ell \, \alpha_m$, regardless of whether $m$ is even or odd.%
\footnote{More precisely, for spin theories these classes should be read from $\Omega_4^{\text{spin}}(B^2 \bZ_N) = \bZ \oplus \bZ_N$, where the first generator is the signature $\sigma(X)$. See, \eg, \cite{Hayashi:2022fkw} for a derivation.}
For $H=\bZ_N$ we have the SPTs
\be
\text{SPT}_\ell(B) = \exp \biggl( \frac{2 \pi i \ell}{2 N} \int \fP(B) \biggr) \qquad\text{with } \ell \,\in\, \bZ_N \ed
\ee
On the other hand, for $H = \{0\}$ the TQFT is a $\bZ_N$ gauge theory with partition function%
\footnote{In order to fix the normalization of the various TQFTs, we work under the assumption that the spacetime manifold is closed, orientable, connected and simply connected.}
\be
Z[\bZ_N](B) = \frac{1}{\sqrt{ \lvert H^2(X, \bZ_N) \rvert}} \, \sum_{b \in H^2(X, \bZ_N)} \!\! \exp \biggl( \frac{2 \pi i}{N} \!\int\! b \, \cup B \biggr) = \sqrt{ \lvert H^2(X, \bZ_N) \rvert} \;\; \delta_{H^2}( B) \ed
\ee
For $N$ prime, the cases above exhaust all realizations of the 1-form symmetry, and are in correspondence with the $N+1$ gapped vacua \eqref{eq:Ngapped}.

More generaly, for $N=kk'$ but still considering a global variant with $\Gamma=\bZ_N$, partially broken phases with $H = \bZ_k$ and SPT given by $\ell \alpha_k$ are described by theories that we dub $\bZ_{N \vert k}^{\ell}$ with $\ell \in \bZ_k$, whose partition function is
\bea
\label{part func of ZNkl}
Z \bigl[ \bZ_{N \vert k}^\ell \bigr](B) &= \exp \biggl( \frac{2 \pi i \ell}{2N k'} \!\int\! \fP(B) \biggr) \, Z[\bZ_{k'}](B) \\
&= \sqrt{ \lvert H^2(X, \bZ_{k'}) \rvert} \; \exp\biggl( \frac{2 \pi i \ell}{2N k'} \!\int\! \fP(B) \biggr) \; \delta_{H^2(X,\bZ_{k'})}(B) \ed
\eea
The partition function of the $\bZ_{k'}$ gauge theory vanishes unless $B = k' \tilde B$ with $\tilde B \in H^2(X, \bZ_k)$, and in that case the SPT phase takes the well-defined form $\exp\bigl( \frac{2\pi i \ell}{2k} \!\int\! \fP(\tilde B) \bigr)$. In particular
\be
\bZ_{N|1}^0 \,\equiv\, \bZ_N \qquad\text{and}\qquad \bZ_{N|N}^\ell \,\equiv\, \text{SPT}_\ell \;.
\ee
Some more details concerning these topological field theories are presented in Appendix~\ref{app: topmanip}.

There is a correspondence between the TQFTs we described and global variants of the $\su(N)$ gauge theory with 1-form symmetry $\bZ_N$.
Both are classified by doublets $(H, \alpha)$ where $H \subset \bZ_N$ and $\alpha \in H^4 \bigl( B^2 H, U(1) \bigr)$. Indeed the data $(H, \alpha)$ determines how to reach a given global variant starting from the electric $SU(N)$ via discrete gauging, but also specifies the possible patterns of spontaneous symmetry breaking. As the set of global variants is also isomorphic to the gapped vacua of the $\cN=1^*$ theory, one might expect that each $\bZ_N$ gauge theory is realized exactly once in a certain massive vacuum of $\cN=1^*$ with gauge group $G=SU(N)$. This expectation is indeed correct, as we will see.

As mentioned at the beginning, there are some global variants of the kind $\bigl( SU(N)/\bZ_k \bigr){}_s$ in which the 1-form symmetry takes a factorized form, see \eqref{eq: ellgroup}. The correspondence between TQFTs and gapped vacua breaks down in this case, and it turns out to be simpler to fix this data through discrete gauging starting from the $SU(N)$ variant.

\paragraph{Assigning TQFTs to vacua.} Let us list the TQFTs that describe the massive vacua of the $SU(N)$ $\cN=1^*$ theory.
As already explained in Section~\ref{ssec: su2}, in the Higgs vacuum $H$ the Wilson line $D_{(1,0)}$ --- which is a genuine line operator --- condenses and the 1-form symmetry is completely broken. This is matched in the IR by a $\bZ_N$ gauge theory. We summarize the properties of this vacuum by the triplet
\be
\bigl( H, \; \langle D_{(1,0)} \rangle \neq 0 , \;  \bZ_N \bigr) \ed
\ee
In each of the confining vacua $C^{(p)}$ ($p=0, \dots, N-1$) the non-genuine dyon $D_{(p,1)}$ condenses. Since this is not a genuine line, the 1-form symmetry is preserved. The IR physics is reproduced by a $\bZ_N$ SPT phase. The standard choice is to assign to $C^{(p)}$ the theory $\text{SPT}_{-p}(B)$. This is in accordance with  pure $\cN=1$ SYM, which can be approached from $\cN=1^*$ by sending $m\to\infty$ and $g_{\YM}\to 0$ while keeping the dynamical scale $\Lambda = m^3 e^{2 \pi i \tau}$ fixed.
We stress that this choice is affected by UV counterterms of the form $\fP(B)$, which do not affect the physics in the Higgs vacuum because they become trivial in the $\bZ_N$ gauge theory. We summarize these vacua by
\be
\bigl( C^{(p)} , \; \langle D_{(p ,1)} \rangle  \neq 0 , \; \text{SPT}_{-p} \bigr) \ed
\ee
In the mixed vacua --- that we indicate as $HC^{\ell}_{k',k}$ --- the condensation pattern is more involved. The $\bZ_k$ subgroup of the 1-form symmetry is confined, and only the genuine lines $D_{(r k ,  0)}$ which are uncharged under it may get an expectation value. 
As reviewed in Section~\ref{sec: SYM}, the generic lattice describing such a process is generated by
\be
D_{(\ell, k')} \oplus D_{(k, 0)}
\ee 
with $\ell \in \bZ_k$. To this lattice we associate the partially broken TQFT $\bZ_{N|k}^{-\ell}$, and thus the vacua are summarized by
\be
\Bigl( HC_{k' , k}^{\ell} \;, \; \langle D_{(\ell, k')} \rangle \oplus \langle D_{(k, 0)} \rangle \neq 0 \;,\; \bZ_{N \vert k}^{-\ell} \Bigr) \ed
\ee
For instance, for $\ell =0$ the lattice of condensed lines is generated by $D_{(0, k')} \oplus D_{(k, 0)}$ and the TQFT in the IR is a pure $\bZ_{k'}$ gauge theory.

Having understood the case $G=SU(N)$, we can study other global variants by discrete gauging. We can reach the theory $\bigl( SU(N)/\bZ_q \bigr){}_s$ from $SU(N)$ by gauging a $\bZ_q$ subgroup of the electric 1-form symmetry after stacking with an $\text{SPT}_s^{\bZ_q} = \exp \bigl( \frac{2 \pi i s}{2 q} \int \fP(B_{\bZ_q}) \bigr)$ phase:
\begin{multline}
\label{eq: pfothervariant}
Z_{\left( \rule{0pt}{0.6em} SU(N)/\bZ_q \right){}_s} (B_{\bZ_p} , B_{\bZ_q}) = \frac{1}{\sqrt{\lvert H^2(X, \, \bZ_q) \rvert}} \; \sum_{b_{\bZ_q} \in H^2(X, \, \bZ_q)} \; Z_{SU(N)} \bigl( B_{\bZ_p} + p \, b_{\bZ_q} \bigr) \times {} \\[.2em]
{} \times \exp\biggl( \frac{2 \pi i s}{2 q} \int \fP\bigl( b_{\bZ_q} \bigr) + \frac{2 \pi i}{q} \int b_{\bZ_q} \cup B_{\bZ_q} \biggr) \ec
\end{multline}
where $N = pq$.

\paragraph{SSB of duality symmetries.} As in Section~\ref{ssec: su2}, the gapped vacua form orbits under the action of the duality/triality symmetries. Here we will mostly focus on the action of $S$ duality, while similar statements for $ST$ triality can also be obtained.

The condensation patterns of dyons --- hence gapped vacua --- are associated to Lagragian sublattices $\mathscr{L}_{k,\ell}$ of line operators spanned by $D_{(\ell,k')}\oplus D_{(k,0)}$ ($N = kk'$). In addition, there is a natural action \eqref{TSactionL} of S-duality on these lattices.
The relation \eqref{TSactionL} determines the duality orbits comprising several subsets of vacua, together with the occurrence of singlets labelled by duality invariant sublattices, which chracterize the spontaneous breaking of $S$ duality symmetry at $\tau_{\YM}=i$ (or of $ST$ triality at $\tau_\text{YM} = e^{\frac{2\pi i}3}$) in the $SU(N)$ $\cN=1^*$ theory.
The local order parameter $\cO=\Tr \Phi_i^2$ was described in (\ref{eq:Worderparameter}), and is charged under the action of $\fD$. The non-trivial relations between expectation values taken by this operator in different vacua can be  obtained using the modular properties of the IR superpotential \eqref{eq:onestarsuperpot} listed in Appendix~\ref{app: orderpar}, as we did in (\ref{eq: chiralvev}) for $\su(2)$.

We can also study the realization of the duality symmetry in the IR by inspecting its action on the low-energy TQFTs. For instance, in the $SU(N)$ Higgs vacuum the low energy TQFT is a $\bZ_N$ gauge theory that describes the SSB of the 1-form symmetry by the condensation of the Wilson line $D_{(1,0)}$. This is mapped by the duality defect to a non-genuine 't~Hooft line and the dual vacuum is in a $\bZ_N$ SPT phase. Below we characterize such action on vacua completely and find that it matches with the UV duality action \eqref{TSactionL} on the lines.

\paragraph{Symmetry action on TQFTs.}
The non-invertible duality symmetry acts in the IR on the $SU(N)$ variant by a discrete gauging $\sigma$ of the $\bZ_N$ 1-form symmetry. For consistency, this should mimic the UV action of $S$ duality on the dyons:
\be
\begin{tikzcd}
	\mathscr{L} \arrow[rr, leftrightarrow, "S"] \arrow[d, "\text{IR}"] & & \mathscr{L}'  \arrow[d, "\text{IR}"] \\
	\text{TQFT}(B) \arrow[rr, leftrightarrow, "\sigma"] & & \text{TQFT}'(B)
\end{tikzcd}
\ee
It is simple to show the following actions of $\sigma$:
\bea
\label{eq:gaugingTQFTs}
\bigl[ \sigma Z[\bZ_N] \bigr] (B) &= \text{SPT}_0(B) \\
[\sigma \, \text{SPT}_\ell] (B) &= \begin{cases} Z[\bZ_N] (B) \,, \qquad &\text{if } \ell=0 \text{ mod } N \\
              \text{SPT}_{- \ell\inv}(B) \,, \qquad &\text{if } \ell \neq 0 \text{ mod }N \;\; \text{and} \;\; \gcd(\ell,N) = 1 \\
              Z \bigl[ \bZ_{N \vert N/p}^{\tilde\ell} \bigr](B) \,, \qquad &\text{if } \gcd(\ell,N) = p \,,\;\text{with } \tilde\ell = - (\ell/p)^{-1}_{\text{mod } N/p}
\end{cases}
\eea
where $\bZ_{N|k}^\ell$ was defined in (\ref{part func of ZNkl}).
When $N$ is prime, the action of $\sigma$ on the TQFTs is compatible with the action of $S$ on the condensed dyons, since $S D_{(\ell  , 1)} = D_{(-1, \ell)}$ and the latter generates the same lattice as $D_{(- \ell\inv, 1)}$ (where $\ell^{-1}$ is taken in $\bZ_N$).

The action (\ref{eq:gaugingTQFTs}) implies that, in a vacuum described by a $\bZ_N$ SPT, the non-invertible $S$ duality symmetry is spontaneously broken unless the equation $\ell^2 = -1 \text{ mod } N$ has a solution, and in that case $\text{SPT}_\ell$ is invariant.%
\footnote{One can compare with the discussion in Section~\ref{ssec: su2}.}
This is related to the existence of a global variant $G = PSU(N)_\ell$ in which the duality symmetry becomes invertible. Indeed, symmetry defects of this kind have been dubbed ``non-intrinsic'' in \cite{Kaidi:2022cpf} (see also \cite{Antinucci:2022vyk}).

A similar story applies to mixed vacua in which the $\bZ_N$ 1-form symmetry is spontaneously broken to $\bZ_k$. As detailed in Appendix~\ref{app: topmanip}, one finds 
\be
\label{sigmaZNkell}
\bigl[ \sigma \, Z[\bZ_{N \vert k}^\ell] \big](B)  = Z[\bZ_{N \vert N/p }^{\tilde{\ell}}](B) \quad\text{with}\quad
k' = \frac{N}{k} \;,\quad
p = \gcd(\ell,k) \;,\quad
\tilde{\ell}=-k' \bigl( \tfrac \ell p \bigr)\inv_{\text{mod } k/p}
\ee
with the proviso of footnote~\ref{foo: inverse}. This formula includes (\ref{eq:gaugingTQFTs}), and is consistent with the duality action \eqref{TSactionL} on dyonic lattices. As a byproduct, for $N=k^2$ one finds
\be
\label{N=k2}
\bigr[ \sigma \, Z[\bZ_{k^2 \vert k}^0] \bigl] (B) = Z[\bZ_{k^2 \vert k}^0] (B) \ed
\ee
This shows that $\bZ_{k^2|k}^0$, which is a $\bZ_k$ gauge theory coupled to a $\bZ_{k^2}$ 1-form symmetry, is invariant under the full modular group.%
\footnote{Indeed using (\ref{eq: SL2 manip 1}) and (\ref{part func of ZNkl}), in general $\tau$ maps $\bZ_{N|k}^\ell$ to $\bZ_{N|k}^{\ell + k'}$, and thus $\bZ_{k^2|k}^0$ is also $\tau$-invariant.}
In turn, this implies that the vacuum $HC_{k,k}^{0}$ --- described by the $\bZ_{k^2|k}^0$ gauge theory --- is invariant under the full modular group and in particular it preserves the $S$ duality symmetry.

More generally, the presence of $S$ duality-invariant vacua is in correspondence with the existence of TQFTs which are self-dual under discrete gauging of the $\bZ_N$ 1-form symmetry. A thorough study of this problem was initiated in \cite{Choi:2021kmx} and completed in \cite{Apte:2022xtu}, finding the following condition:
\be
\label{self-duality condition Cordova}
\exists \; k', s, \tilde\ell \qquad \text{such that} \qquad N= (k')^2 s \qquad\text{and}\qquad \tilde\ell^2 + 1 = 0 \text{ mod } s \ed
\ee
Indeed, identifying $k = k's$ and $\ell = k'\tilde\ell$, one can prove that (\ref{self-duality condition Cordova}) is equivalent to
\be
\exists \; k, \ell \quad\text{such that}\quad k = \frac Np \;,\quad \ell = - p \, (\ell/p)\inv_{\text{mod } k/p} \mod k \;,\quad \text{with} \quad p = \gcd(\ell, k) \;,
\ee
\ie, to the condition that there exists a phase $\bZ_{N|k}^\ell$ invariant under $\sigma$ according to \eqref{sigmaZNkell}. Note that for $N$ prime one necessarily has to set $k'=1$ and $s=N$, thus reproducing the analysis above. On the other hand, for $N=k^2$ there is the solution $s=1$ that leads to \eqref{N=k2}. Another instance where these TQFTs appear is in $\mathfrak{su}(8)$, for which $k'=2$, $s=2$ and $\ell= \pm 2$ provide solutions. Indeed, the global variant $G= \bigl( SU(8)/\bZ_4 \bigr){}_2$ has a duality-invariant lattice of genuine line operators generated by the dyons  $D_{(4  ,  0)} \oplus D_{(2  ,  2)}$.%
\footnote{The same problem can be studied for a generic 1-form symmetry group $\Gamma$ and one is led to similar conclusions. These results will appear in \cite{Antinucci:2023ezl}.}

\subsection{Duality symmetries on generic global variants}
\label{ssec:globalvariantmatrix}

In Section~\ref{subsec: gappedTQFT} we showed that the action of  $S$ duality on dyons is reproduced by a discrete gauging in the low-energy TQFTs assigned to the vacua of $G= SU(N)$ $\cN=1^*$ SYM. 
For different choices of global variants, such as $G= PSU(N)_k$, the technique becomes slightly more involved. In this section we refine our approach from Section~\ref{sec: SYM} in order to construct non-invertible defects associated with generic modular transformations over different global variants. This procedure allows one to detect the occurrence of global variants for which the defects become invertible, and to diagnose the presence of mixed anomalies between the latter and the 1-form symmetry. Unfortunately, a systematic approach comprising all possible gauge-group ranks is out of the scope of this paper, so we will restrict to $N$ prime. 

First, recall that in theories with a $\bZ_N$ 1-form symmetry there is a natural topological action of $SL( 2, \bZ_N)$ \cite{Choi:2021kmx}.%
\footnote{More precisely, the action is $SL\bigl( 2, \bZ_{\gcd(2,N) N} \bigr)$. The $\gcd$ factor is only important on non-spin manifolds and it corresponds to $\tau^{2N} =1$, while if $\omega_2(TM)=0$ then $\tau^N=1$. We will thus ignore it in our analysis.}
For $N$ prime, this group has two generators $\sigma$, $\tau$ that we already defined in (\ref{eq: SL2 manip 1}). Notice that we can also define elements $\nu(u)$ as
\be
\nu(u) : \  \left[ \nu(u) Z\right](B) = Z(u B) \qquad\text{for}\qquad u \in \bZ_N^\times \;,
\ee
where $\bZ_N^\times$ is the multiplicative group of integers modulo $N$ coprime with $N$. The global variants $G$ of $\mathfrak{su}(N)$ ($N$ prime) are classified by the quotient $SL(2, \, \bZ_N)/\Gamma_\infty$ \cite{Bashmakov:2022uek, Antinucci:2022cdi}, where
\be
\Gamma_\infty = \biggl\{ M \in SL(2, \, \bZ_N) \biggm| M = \biggl( \begin{matrix} u & k \\ 0 & u^{-1} \end{matrix} \biggr) \biggr\}
\ee
is the parabolic subgroup of $SL(2, \bZ_N)$. In this matrix representation, the first column is a generator of the sublattice of genuine line operators, while the second column is a generator of the sublattice of symmetry defects. We can choose the following representatives:
\be
\label{reps of global forms}
SU(N) \;\to\; \biggl( \begin{matrix} 1 & 0 \\ 0 & 1 \end{matrix} \biggr) \;,\qquad\qquad PSU(N)_r \;\to\; \biggl( \begin{matrix} r & -1 \\ 1 & 0 \end{matrix} \biggr) \;.
\ee
As we discussed in Section~\ref{globalvariants} after (\ref{eq: SL2 manip 1}), one can refine the classification of global forms by keeping track of the coupling to the background field $B$ and of possible contact terms. This is classified (for $N$ prime) by $SL(2,\bZ_N)/\bZ_N^\times$ where we quotient by the matrices $\smat{ u & 0 \\ 0 & u^{-1}}$ with $u \in \bZ_N^\times$. We can choose as representatives:
\be
\label{refined reps of global forms}
SU(N)_p \;\to\; \biggl( \begin{matrix} 1 & p \\ 0 & 1 \end{matrix} \biggr) \;,\qquad\qquad PSU(N)_{r,p} \;\to\; \biggl( \begin{matrix} r & pr-1 \\ 1 & p \end{matrix} \biggr) \;.
\ee
In this representation, the topological $SL(2,\, \bZ_N)$ acts by right multiplication, where $\sigma$ and $\tau$ are represented by the same matrices as for $S$ and $T$ respectively, whilst the duality group acts by left multiplication. We used this fact to construct the representatives (\ref{reps of global forms}) and (\ref{refined reps of global forms}), following the definitions after (\ref{eq: SL2 manip 1}). The condition that the matrix for $G$ is modular implies that the lattice of genuine lines is Lagrangian, while the quotient by $\Gamma_\infty$ takes care of the choice of lattice generator. With this recipe one easily reproduces the diagrams in (\ref{su2top}), (\ref{su2}), (\ref{su3}), (\ref{su2orbits}), and constructs the ones for higher rank.

In order to define a duality symmetry operator $\fD$ for a variant $G$ and a duality element $M \in SL(2, \bZ)$, we must solve the matrix equation
\be
M \cdot G \cdot \Sigma = G \label{eq: generalD}
\ee
which determines the topological manipulation $\Sigma$ that compensates for the action of $M$ on the global form. The global variant $G$ could be invariant under $M$: this implies that \eqref{eq: generalD} is solved by $\Sigma \in \Gamma_\infty$. Hence the $M$ duality defect is invertible in $G$, but the presence of a nontrivial parabolic element $\Sigma = \tau^n \nu(u) $ implies that the partition function transforms as 
\be
[M \, Z](B) = \exp\biggl(\frac{2\pi i n}{2 N} \int \fP(B) \biggr) \; Z(u B) \;.
\ee
When $M=S$, $N=2$, $G = PSU(2)_1$, and $\Sigma = \tau$ (namely $u, n=1$) this just becomes the familiar anomaly between 1-form symmetry and 0-form symmetry
\be
I = \frac{2 \pi i}{4} \int S \cup \fP(B) \;,
\ee
as was first noticed in \cite{Kaidi:2022uux}.%
\footnote{For $N>2$ a similar interpretation is possible, however now the 0-form symmetry comes with an action $\rho \in \text{Aut}(\bZ_N): B \to u B$ making the symmetry a split 2-group \cite{Benini:2018reh}. Likewise, the anomaly must now be classified by an appropriate class $I \in H^5\bigl( B^2\bZ_N \rtimes_\rho B\bZ_4 \,,\, U(1) \bigr)$.}

Whenever $u\neq1$, the rescaling implemented by $\nu(u)$ is important because it allows the stacking operation to have an order compatible with $M$. Consider for instance the case of $S$ duality and gauge group $PSU(N)_u$ for $N$ prime. Then, in order for the duality defect to be invertible, we need for the topological manipulation $\Sigma$ to square to charge conjugation: $\Sigma^2=C$. Furthermore, it imposes a condition on the discrete torsion $u$, namely $u^2+1=0$ mod $N$. In these cases the $PSU(N)_u$ variant has an invertible self-duality symmetry which acts via the element%
\footnote{Using the matrix formulation, $M = \smat{0 & -1 \\ 1 & 0}$ and $G = \smat{ u & -1 \\ 1 & 0}$ give $\Sigma = \smat{ -u & 1 \\ -u^2 -1 & u}$. The condition $\Sigma \in \Gamma_\infty$ for the defect to be invertible gives $u^2 + 1 = 0 \text{ mod } N$, and then $\Sigma = \smat{ u^{-1} & 1 \\ 0 & u}$. This matrix can be decomposed as $\Sigma = \smat{u^{-1} & 0 \\ 0 & u} \smat{1 & u \\ 0 & 1}$ corresponding to the element $\tau^u \, \nu(u)$ with right action. It is also possible to check the result directly by using the explicit formula for the $PSU(N)_u$ partition function \eqref{eq: pfothervariant}.}
\be
\label{eq: parabolic}
\Sigma_u = \tau^u \; \nu(u)  \quad\in \Gamma_\infty \;.
\ee

As emphasized before, this simple representation of global forms does not apply when $N$ is not prime (\eg, because some Lagrangian lattices have two generators). In those cases the correct transformation $\Sigma$ must be computed by brute force.

\subsection[Examples: \tps{$\mathfrak{su}(3)$}{su(3)} and \tps{$\mathfrak{su}(4)$}{su(4)}]{Examples: \matht{\su(3)} and \matht{\su(4)}}

With these tools at hand, we can provide explicit examples of SSB in the $\cN=1^*$ vacua. A detailed exposition of the $\mathfrak{su}(2)$ case was provided in Section~\ref{ssec: su2}. Here we describe the examples of $\mathfrak{su}(3)$ and $\mathfrak{su}(4)$, which are slightly more complicated and lead to some additional important insights. As before, we mostly focus on the $S$ duality symmetry at $\tau_{\YM}=i$ and make some comments regarding $ST$ triality on the way.

\paragraph{\matht{\su(3)}.} There are four global variants $SU(3)$, $PSU(3)_0$, $PSU(3)_1$, $PSU(3)_2$ which form two orbits under $S$ duality. The non-invertible symmetry is always spontaneously broken. The map \eqref{TSactionL} among lattices yields
\be
\langle D_{(1,0)} \rangle \;\overset{S}{\longleftrightarrow}\; \langle D_{(0,1)} \rangle \;,\qquad\qquad \langle D_{(1,1)} \rangle \;\overset{S}{\longleftrightarrow}\; \langle D_{(2,1)} \rangle
\ee
and thus
\ben
\begin{subarray}{c} \scalemath{1}{\boldsymbol{SU(3)}} \\[0.25em]
\begin{array}{c|b|b|y|y} 
 \text{Vacuum} & H & C^{(0)} & C^{(1)} & C^{(2)} \\ \hline\hline 
		 \text{Cond. line} & D_{(1,  0)}^{\phantom{U}} & D_{(0  ,  1)}^U & D_{(1  ,  1)}^U & D_{(2  ,  1)}^U \\[.2em] \hline
		 \text{TQFT} & \bZ_3 & \text{SPT}_0 & \text{SPT}_2 & \text{SPT}_1 \\ \hline
		 \text{SSB} \; \fD &  \color{red} \checkmark \color{black} & \color{red} \checkmark \color{black} &  \color{blue} \checkmark \color{black} & \color{blue} \checkmark \color{black} \\ \hline \hline
\end{array} \end{subarray}
\ \ \ \ \
\begin{subarray}{c} \scalemath{1}{\boldsymbol{PSU(3)_1}} \\[0.25em]
\begin{array}{c|b|b|y|y} 
\text{Vacuum} & H & C^{(0)} & C^{(1)} & C^{(2)} \\ \hline\hline 
		 \text{Cond. line} & D_{(1,0)}^U & D_{(0,1)}^U & D_{(1,1)}^{\phantom{U}} & D_{(2  , 1)}^U \\[.2em] \hline
		 \text{TQFT} & SPT_0 & \text{SPT}_2 & \bZ_3 & \text{SPT}_1 \\ \hline
		 \text{SSB} \; \fD &  \color{red} \checkmark \color{black} & \color{red} \checkmark \color{black} &  \color{blue} \checkmark \color{black} & \color{blue} \checkmark \color{black} \\ \hline \hline
\end{array} \end{subarray}
\een
This case is similar to $\mathfrak{su}(2)$, since $N=3$ is a prime number. Notice that the $S$ duality defect on $PSU(3)_1$ is implemented by including the topological manipulation $\tau\sigma\tau^2$ (instead of just $\sigma$). This follows either from direct computation or from the matrix formalism we introduced in Section~\ref{ssec:globalvariantmatrix}.

The $\mathfrak{su}(3)$ theories also have a Coulomb vacuum, corresponding to the partition $3 = 1 + 2$. We will comment on the physics of such vacua in Section~\ref{subsec: Coulomb}. The case of generic $N$ prime is also very similar. In the $SU(N)$ variant we have a spontaneously broken $S$ duality symmetry unless we are in a confining vacuum $C^{(r)}$ with $r^2 + 1 = 0 \text{ mod } N$, in which case both the duality symmetry and the 1-form symmetry are preserved. Other non-invariant variants work similarly, but with some of the TQFTs permuted, while in the invariant variant(s) the invertible action in the IR is implemented by \eqref{eq: parabolic}.

For the $ST$ triality symmetry the orbits instead are:
\be
\begin{tikzcd}[row sep=tiny]
\langle D_{(1,0)} \rangle \arrow[rr, "ST"] & & \langle D_{(0,1)} \rangle \arrow[dl, "ST"] \\
    & \langle D_{(2,1)} \rangle \arrow[ul,"ST"]
\end{tikzcd}
\qquad\qquad
\raisebox{-1.1em}{\begin{tikzcd}
    \langle D_{(1,1)} \rangle \arrow[loop above, "ST"]
\end{tikzcd}}
\ee
In the $SU(3)$ variant the symmetry is implemented by $\sigma \tau^{-1}$ and we find:
\ben \begin{subarray}{c} \scalemath{1}{\boldsymbol{SU(3)} \textbf{ -- triality}} \\[0.25em]
\begin{array}{c|b|b|c|b} 
 \text{Vacuum} & H & C^{(0)} & C^{(1)} & C^{(2)} \\ \hline\hline 
		 \text{Cond. line} & D_{(1,0)}^{\phantom{U}} & D_{(0  ,  1)}^U & D_{(1  ,  1)}^U & D_{(2  ,  1)}^U \\[.2em] \hline
		 \text{TQFT} & \bZ_3 & \text{SPT}_0 & \text{SPT}_2 & \text{SPT}_1 \\ \hline
		 \text{SSB} \; \fD &  \color{red} \checkmark \color{black} & \color{red} \checkmark \color{black} &  \text{\texttimes}    & \color{red} \checkmark \color{black} \\ \hline \hline
\end{array} \end{subarray}
\een
This is consistent with the action on dyons.

\paragraph{\matht{\su(4)}.} Since $N=4$ is not prime, the $\su(4)$ theory has two new gapped vacua: $HC_{2,2}^{(0)}$ and $HC_{2,2}^{(1)}$ which correspond to the condensation of the lattices of dyons $\langle D_{(2,0)} \oplus D_{(0,2)} \rangle$ and $\langle D_{(1,2)} \rangle$, respectively. According to \eqref{TSactionL}, their orbits under $S$ are:
\be
\begin{tikzcd}
\langle D_{(1,0)} \rangle \arrow[d, leftrightarrow, "S"] & \langle D_{(1,1)} \rangle \arrow[d, leftrightarrow, "S"] & \langle D_{(2,1)} \rangle \arrow[d, leftrightarrow, "S"] & \\
\langle D_{(0,1)} \rangle & \langle D_{(3,1)} \rangle & \langle D_{(1,2)} \rangle & \langle D_{(2,0)} \oplus D_{(0,2)} \rangle \arrow[loop above, "S"]
\end{tikzcd}
\ee
By assigning the corresponding TQFTs to these vacua, we find the following table:
\ben
\begin{subarray}{c} \scalemath{1}{\boldsymbol{SU(4)}} \\[0.25em]
	\begin{array}{c|b|b|y|g|y|g|c}
\text{Vacuum} & H & C^{(0)} & C^{(1)} & C^{(2)} & C^{(3)} & HC_{2,2}^{(1)} &  HC_{2,2}^{(0)} \\[.1em] \hline\hline 
\text{Cond. line} & D_{(1,0)}^{\phantom{U}} & D_{(0,1)}^U & D_{(1,1)}^U & D_{(2,1)}^U & D_{(3,1)}^U & D_{(1,2)}^U & D_{(2,0)}^{\phantom{U}} \oplus D_{(0,2)}^U \\[.2em] \hline
		 \text{TQFT} & \bZ_4 & \text{SPT}_0 & \text{SPT}_3 & \text{SPT}_2 & \text{SPT}_1 & \bZ_{ 4 \vert 2}^1 & \bZ_{4|2}^0 \simeq \bZ_2 \\ \hline
		 \text{SSB} \; \fD &  \color{red} \checkmark \color{black} & \color{red} \checkmark \color{black} &  \color{blue} \checkmark \color{black} & \color{brown} \checkmark \color{black} &  \color{blue} \checkmark \color{black} &  \color{brown} \checkmark \color{black} & \text{\texttimes} \\ \hline \hline
\end{array}\end{subarray}
\een
The theory $\bZ_{N|k}^\ell$ was defined in (\ref{part func of ZNkl}). Notice the presence of phases in which $\bZ_4$ is spontaneously broken to $\bZ_2$. To show that the vacua transform in the correct way under gauging the 1-form symmetry, we use the results of Section~\ref{subsec: gappedTQFT}. For the doublets $\{H,C^{(0)}\}$ and $\{C^{(1)},C^{(3)}\}$ it follows directly from the first three lines in \eqref{eq:gaugingTQFTs}. Regarding the doublet $\{C^{(2)},HC_{2,2}^{(1)}\}$ and the singlet $HC_{2,2}^{(0)}$, their relation follows from the last line of \eqref{eq:gaugingTQFTs} and from \eqref{sigmaZNkell}. In particular, notice that
\be
[ \sigma \; \text{SPT}_2] (B) = Z\bigl[ \bZ_{4 \vert 2}^1 \bigr] (B) \qquad\text{and}\qquad \bigl[ \sigma \, Z[\bZ_{4|2}^0] \bigr] (B) = Z[\bZ_{4|2}^0] (B) \;.
\ee

There are three other variants to consider, of which two are obtained from the table above by gauging $\bZ_4$ with a discrete torsion, and the last one by gauging only the $\bZ_2$ subgroup. The former two give the following configurations:
\ben
\begin{subarray}{c} \scalemath{1}{\boldsymbol{PSU(4)_1}} \\[0.25em]
	\begin{array}{c|b|b|y|g|y|g|c}
\text{Vacuum} & H & C^{(0)} & C^{(1)} & C^{(2)} & C^{(3)} & HC_{2,2}^{(1)} &  HC_{2,2}^{(0)} \\[.1em] \hline\hline 
\text{Cond. line} & D_{(1,0)}^U & D_{(0,1)}^U & D_{(1,1)}^{\phantom{U}} & D_{(2,1)}^U & D_{(3,1)}^U & D_{(1,2)}^U & D_{(2,2)}^{\phantom{U}} \oplus D_{(0,2)}^U \\[.2em] \hline
\text{TQFT} & \text{SPT}_0 & \text{SPT}_3 & \bZ_4 & \text{SPT}_1 &  \bZ_{ 4 \vert 2}^1 & \text{SPT}_2 & \bZ_{4|2}^0 \simeq \bZ_2 \\ \hline
\text{SSB} \; \fD &  \color{red} \checkmark \color{black} & \color{red} \checkmark \color{black} &  \color{blue} \checkmark \color{black} & \color{brown} \checkmark \color{black} &  \color{blue} \checkmark \color{black} &  \color{brown} \checkmark \color{black} & \text{\texttimes} \\ \hline \hline
\end{array}\end{subarray}
\een
and
\ben
\begin{subarray}{c} \scalemath{1}{\boldsymbol{PSU(4)_2}} \\[0.25em]
	\begin{array}{c|b|b|y|g|y|g|c}
\text{Vacuum} & H & C^{(0)} & C^{(1)} & C^{(2)} & C^{(3)} & HC_{2,2}^{(1)} &  HC_{2,2}^{(0)} \\[.1em] \hline\hline 
\text{Cond. line} & D_{(1,0)}^U & D_{(0,1)}^U & D_{(1,1)}^U & D_{(2,1)}^{\phantom{U}} & D_{(3,1)}^U & D_{(1,2)}^U & D_{(2,0)}^U \oplus D_{(0,2)}^{\phantom{U}} \\[.2em] \hline
\text{TQFT} & \text{SPT}_0 & \bZ_{4 \vert 2}^1 & \text{SPT}_3 & \bZ_4 & \text{SPT}_1 & \text{SPT}_2 & \bZ_{4|2}^0 \simeq \bZ_2 \\ \hline
\text{SSB} \; \fD &  \color{red} \checkmark \color{black} & \color{red} \checkmark \color{black} &  \color{blue} \checkmark \color{black} & \color{brown} \checkmark \color{black} &  \color{blue} \checkmark \color{black} &  \color{brown} \checkmark \color{black} & \text{\texttimes} \\ \hline \hline
\end{array}\end{subarray}
\een
On the first one, the IR $S$ duality symmetry acts through $\Sigma \bigl( PSU(4)_1 \bigr) =\sigma \tau^2 \sigma \tau$ while on the second one it acts through $\Sigma \bigl( PSU(4)_2 \bigr) = \sigma \tau^2 \sigma \tau^2 \sigma$.

Not surprisingly, the most involved case is the one in which the duality symmetry is invertible corresponding to the global variant $[SU(4)/\bZ_2]_0$. To find the table of vacua we must gauge the $\bZ_2$ subgroup of $\bZ_4$. Since $\bZ_4$ is a non-trivial extension
\be
1 \longrightarrow \bZ_2 \longrightarrow \bZ_4 \longrightarrow \bZ_2 \longrightarrow 1 \;,
\ee		
then the backgrounds $C_\rme$ for the $\bZ_2$ subgroup and $B_\rme$ for the $\bZ_2$ quotient are related by
\be
\label{eq: extbundle}
\delta C_\rme = \beta(B_\rme) \;,
\ee
with $\beta$ the Bockstein map associated to the above sequence (explicitly $\beta(B_\rme) = \frac{1}{2} \delta B_\rme$). According to \cite{Tachikawa:2017gyf}, gauging the $\bZ_2$ subgroup results in a theory with $\bZ_2 \times \bZ_2$ 1-form symmetry and a mixed anomaly
\be
\label{eq: mixedz2z2}
I = \pi i \int B_\rmm \; \beta(B_\rme) = \pi i \int B_\rme \; \beta(B_\rmm) \;,
\ee
where $B_\rmm \in H^2(X, \bZ_2)$ is the gauge field for the dual 1-form symmetry. The mixed anomaly \eqref{eq: mixedz2z2} cannot be matched by a $\bZ_2 \times \bZ_2$ SPT phase, so we expect the $[SU(4)/\bZ_2]_0$ vacua in $\cN=1^*$ to always display (partial) SSB of the 1-form symmetry. The theory also has a $\bZ_2$ 0-form symmetry $\fs$ that acts as a $\bZ_2$ automorphism exchanging $B_\rme$ with $B_\rmm$.	
		
Let us first give the table of the results and then describe in detail some subtleties of the gauging procedure. We find
\ben
\begin{subarray}{c} \scalemath{1}{\boldsymbol{[SU(4)/\bZ_2]_0}} \\[0.25em]
\begin{array}{c|b|b|y|g|y|g|c}
\text{Vacuum} & H & C^{(0)} & C^{(1)} & C^{(2)} & C^{(3)} & HC_{2,2}^{(1)} &  HC_{2,2}^{(0)} \\[.1em] \hline\hline 
\text{Cond. line} & D_{(1  ,  0)}^U & D_{(0  ,  1)}^U & D_{(1  ,  1)}^U & D_{(2 ,  1)}^U & D_{(3  ,  1)}^U & D_{(1,2)}^U & D_{(2,0)}^{\phantom{U}} \oplus D_{(0,2)}^{\phantom{U}} \\[.2em] \hline
\text{TQFT} & \bZ_2 & \bZ_2 & \bZ_2 & \bZ_2 &  \bZ_2 & \bZ_2 & \bZ_2 \times \bZ_2 \\ \hline
\text{SSB} \; \fD &  \color{red} \checkmark \color{black} & \color{red} \checkmark \color{black} &  \color{blue} \checkmark \color{black} & \color{brown} \checkmark \color{black} &  \color{blue} \checkmark \color{black} &  \color{brown} \checkmark \color{black} & \text{\texttimes} \\ \hline \hline
\end{array}\end{subarray}
\een
The lattices generated by the condensed dyons in each vacuum always contains at least one local line (not counting the identity). This is consistent with the fact that at least a $\bZ_2$ must be spontaneously broken. In the last vacuum, in which we have a $\bZ_2 \times \bZ_2$ theory, the duality symmetry is unbroken, but it acts nontrivially as an outer automorphisms exchanging the two $\bZ_2$ factors.
To understand the matching of the mixed anomaly we need to carefully follow the discrete gauging procedure and understand the coupling of the $\bZ_2 \times \bZ_2$ symmetry with the $\bZ_2$ gauge theory. 
We describe $C_e$ in \eqref{eq: extbundle} as a torsor over $H^2(X, \, \bZ_2)$, by choosing a splitting $C_\rme = C_\rme^\beta + C_\rme'$, with $C_\rme' \in H^2(X; \bZ_2)$ and $\delta C_\rme^\beta = \beta(B_\rme)$. We can gauge the 1-form symmetry as usual by summing over $C_\rme'$. This means that
\be
\left[\sigma_{\bZ_2} \cdot Z \right](B_\rme , B_\rmm) = \frac{1}{\sqrt{H^2(X, \bZ_2)}} \sum_{c' \, \in \, H^2(X, \, \bZ_2)} e^{\pi i \int c \, \cup  B_\rmm} \; Z( B_\rme + 2 c) \;,\quad c = C^\beta_\rme + c' \,.
\ee
The fact that the above has a mixed $\bZ_2 \times \bZ_2$ anomaly follows from considering the gauge transformation $B_\rmm \to B_\rmm + \delta \lambda$. This kind of mixed anomaly allows for a Bardeen-like counterterm:
\be
2 \pi i \, K(B_\rme, B_\rmm) = \frac{2 \pi i}{4} \int B_\rme \, \cup \, B_\rmm \;.
\ee
With this notation we find:
\be
S \cdot Z_{[SU(4)/\bZ_2]_0}(B_\rme, B_\rmm) = Z_{[SU(4)/\bZ_2]_0}(B_\rmm, B_\rme)   \; e^{2 \pi i \; K(B_\rme, B_\rmm) } \equiv \fs \cdot Z_{[SU(4)/\bZ_2]_0}(B_\rme, B_\rmm)
\ee
where we have chosen to include the  counterterm in the definition of $\fs$. Thus the invertible $S$ duality is implemented by the discrete $\bZ_2$ symmetry. This is the action that we must match in the IR vacua.

To consistently assign vacua, we must compute the action of the $\bZ_2$ gauging on the $SU(4)$ TQFTs. We find
\be
\left[\sigma_{\bZ_2} \cdot \text{SPT}_k \right](B_\rme, B_\rmm) = \sqrt{H^2(X, \, \bZ_2)} \; \delta_{\bZ_2}(B_\rmm + k B_\rme) \; e^{\frac{\pi i k}{4} \int \fP(B_\rme) + \pi i \int C^\beta_\rme \cup (B_\rmm + k B_\rme)} \;.
\ee
This is a $\bZ_2$ gauge theory coupled to $B_\rme + k B_\rmm$. The Pontryagin square term is in principle ill-defined, however this is resolved by the term $k C_\rme^\beta \cup B_\rme$. The mixed anomaly is instead encoded in the coupling $C_\rme^\beta \cup B_\rmm$.
On the other hand
\be
\left[\sigma_{\bZ_2} \cdot \bZ_4 \right] (B_\rme , B_\rmm) = \sqrt{H^2(X, \, \bZ_2)} \; \delta_{\bZ_2}(B_\rme) \; e^{\pi i \int C^\beta_\rme \cup B_\rmm + 2 \pi i \, K(B_\rme, \, B_\rmm)}
\ee
is the image under $\fs$ of $\sigma_{\bZ_2} \!\cdot \text{SPT}_0$. To see it explicitly, we use the counterterm to write $\pi i \int C^\beta_\rme \cup B_\rmm + 2 \pi i \, K(B_\rme, B_\rmm) = \pi i \int C_\rmm^\beta \cup B_\rme$, with $\delta C_\rmm^\beta = \beta(B_\rmm)$. We also have that
\be
\left[\sigma_{\bZ_2} \cdot \bZ_{4| 2}^1 \right] (B_\rme , B_\rmm) = \sqrt{H^2(X, \, \bZ_2)} \; \delta_{\bZ_2}(B_\rme) e^{\frac{\pi i}{2} \int \fP(B_\rmm) + \pi i \int C_\rmm^\beta \cup B_\rme  } \;,
\ee
describing the action of $\fs$ on $\sigma_{\bZ_2} \!\cdot \text{SPT}_2$. Finally,
\be
\left[\sigma_{\bZ_2} \cdot \bZ_2 \right](B_\rme, B_\rmm) = \bigl\lvert H^2(X; \bZ_2) \bigr\rvert \; \delta_{\bZ_2}(B_\rme) \; \delta_{\bZ_2}(B_\rmm) \; e^{\pi i \int C_\rme^\beta \cup B_\rmm}
\ee
is the $\bZ_2 \times \bZ_2$ gauge theory which is invariant under $\fs$.

\subsection{Comments on Coulomb vacua}
\label{subsec: Coulomb}

All $\mathfrak{su}(N)$ theories with $N\geq 3$ also have gapless Coulomb vacua. Such vacua often appear dressed by SPT phases or discrete gauge theories that follow from the precise patterns of Higgsing or confinement. Focusing on the pure Coulomb vacuum --- which is labeled by the partition $N= 1 + (N-1)$ and in which the $SU(N)$ symmetry is Higgsed to $U(1)$ --- since at $\tau_{\YM}=i$ there is a non-invertible self-duality symmetry in the UV, this must also be true in the Coulomb vacuum.

It is known that pure Maxwell theory hosts a wealth of non-invertible defects coming from the composition of $SL(2, \bZ)$ modular transformations with the gauging of a discrete subgroup $\Gamma \subset U(1)\one_\rme \times U(1)\one_\rmm$ of the 1-form symmetry \cite{Choi:2021kmx, Niro:2022ctq}.
Once the subgroup $\Gamma$ and the modular transformation $M$ are identified, this fixes the Coulomb coupling $\tau_\text{C}$ of the Maxwell theory. In a pure Coulomb vacuum the 1-form symmetry is spontaneously broken, while the duality symmetry is preserved.
Here $M=S$, while the choice of $\Gamma$ is dictated by the UV global variant $G$, with the convention that the 1-form symmetry of the $SU(N)$ variant is associated to the subgroup $\bZ_{N,\rme}\one \subset U(1)\one_\rme$ generated by
\be
U_l = e^{\frac{i l}{N} \!\int\! \tilde{F}} \;,\qquad\qquad \tilde{F} = \frac{1}{e^2} \, {\star F} + \frac{i\theta}{2 \pi} \, F \;.
\ee
Gauging the electric 1-form symmetry projects onto Wilson lines $W_n$ with $n = N m$ but also allows for fractionalized 't~Hooft lines $T_{m'/N}$. We can restore the standard quantization by defining $A = \frac{1}{N} A'$ and $\tilde{A} = N \tilde{A}'$ for the dual gauge field. The action in the $A'$ variables has a coupling $\tau'_\text{C} = \tau_\text{C}/N^2$. By composing with $S$ duality, one finds that a duality symmetry defect exists if
\be
\tau_\text{C} = - \frac{N^2}{\tau_\text{C}} \qquad\Rightarrow\qquad \tau_\text{C}^{(\rme)} = i N \;.
\ee
The magnetic gauging can be treated similarly, finding $\tau_\text{C}^{(\rmm)}= i/N$.
The above argument fixes the Maxwell coupling $\tau_\text{C} = i N$ in the pure Coulomb vacuum of the $SU(N)$ $\cN=1^*$ theory. 

Other global variants can be treated using the appropriate discrete gauging. We can do it directly at the level of the action by writing it as follows:
\be
S = \int \biggl[ \frac{1}{2 e^2} (dA- b) \wedge \star (dA-b) + \frac{i}{2 \pi} G  \wedge (N b - d l) + \frac{i N p}{4 \pi} \, b \wedge b + \frac{i N}{2 \pi} b \wedge C \biggr] \;.
\ee
Here $b$ is the dynamical $\bZ\one_{N,\rme}$ gauge field, $C$ is the background field for the dual 1-form symmetry ($dC=0$), whilst $G$ and $l$ are dynamical fields acting as a Lagrange multiplier restricting $b$ to be in $\bZ_N$ by the equations of motion. We set the theta angle in the electric frame to zero for simplicity.%
\footnote{This is sufficient to deal with duality transformations, since the self-dual coupling is purely imaginary.}
We define shifted variables $b' = b - dA$ and $G' = G + p \, dA$, so that the action reads:
\be
\label{eq: gaugedMax}
S = \int \biggl[ \frac{1}{2 e^2} b' \wedge \star \, b' + \frac{i N p}{4 \pi} b' \wedge b' + \frac{i N}{2 \pi} b' \wedge (C + G') + 
 \frac{i N}{2 \pi} dA \wedge G' - \frac{i}{2\pi} G' \wedge d l \biggr] \;.
\ee
Integrating out $l$ sets $G'=dV$, and then the term $d A \wedge d V$ can be dropped.
The field $b'$ can be integrated out by its equation of motion:
\be
 \frac{1}{e^2} \!\star b' + \frac{iN}{2 \pi} \bigl( p \, b' + d V + C \bigr) = 0 \;,
\ee
which sets $b' = - \frac{iN}{2\pi} \bigl( \frac{1}{e^2} \!\star + \frac{i N p}{2 \pi} \bigr)^{-1} (d V + C)$. The action becomes
\be
\label{eq:dualMaxwell}
S = \int \biggl[ \frac{1}{2e^2 \beta} (dV + C) \wedge \star (dV + C) - \frac{i N p}{4 \pi \beta} (dV + C) \wedge (d V + C) \biggr]
\ee
with $\beta = p^2 + \frac{4\pi^2}{e^4 N^2}$.
If we start from the self-dual coupling $\frac{1}{e^2} = \frac{N}{2\pi}$ then the complexified coupling $\tau_\text{C} = \frac{\theta}{2\pi} + \frac{2 \pi i}{e^2}$ is given by
\be
\tau_\text{C}^{(\rme)} = \frac{N(i-p)}{(1 + p^2)} \;.
\ee
This is in an electric duality frame (since $dV$ couples electrically to $C$). In the magnetic frame the answer instead reads
\be
\tau_\text{C}^{(\rmm)} = \frac{(i + p)}{N}
\ee
and it reproduces the naive answer for $PSU(N)_0$ (magnetic gauging) $\tau_\text{C} = i/N$. This reasoning allows us to fix the coupling $\tau_\text{C}$ in pure Coulomb vacua with different global forms.

\subsection{Cubic anomalies and invariant vacua}
\label{ss: cubic}

Lastly we perform some further consistency checks based on 't~Hooft anomalies. It is known \cite{Hsieh:2019iba, Hsieh:2020jpj} (see also \cite{Kaidi:2023maf, Debray:2023yrs} for a recent discussion) that the $SL(2,\bZ)$ group --- or more appropriately its spin cover $Mp(2,\bZ)$ --- can have nontrivial 't~Hooft anomalies. 
The 't~Hooft anomaly for a single Maxwell field, $I_{\text{Maxwell}}$,  for a given cyclic subgroup $\bZ_k$ of the metaplectic group is related to the chiral anomaly of a Weyl fermion $I_{\text{Weyl}}$ by:
\be
I_{\text{Maxwell}} = 56 \, I_{\text{Weyl}} \ed
\ee
The latter must be carefully defined. Since $S^4 = (-1)^F$, the correct structure to consider is 
\be
\text{Spin}-Mp(2,\bZ) = \frac{\text{Spin}(d) \times Mp(2, \bZ)}{\bZ_2^F}
\ee
where the quotient is by the diagonal $\bZ_2^F$ generated by $(-1)^F$, and the possible anomalies are captured by a bordism group
\be
\Omega^{\text{Spin}-Mp(2,\bZ)}_5(\text{pt}) = \bZ_{32} \oplus \bZ_2 \oplus \bZ_9  \ed
\ee
The first two summands stem from the cubic anomaly of $S$, while the third one from the cubic anomaly of $ST$. On the other hand, if the symmetry group were only $SL(2,\bZ) \times \bZ_2^F$, its anomalies would be classified by \cite{Debray:2023yrs}
\be
\Omega^{\text{Spin}}_5\bigl( BSL(2,\bZ) \bigr) = \bZ_4 \oplus \bZ_9 \;,
\ee
the first factor stemming from the subgroup generated by $S$ while the second from $ST$.
The anomaly for the $\su(N)$ $\cN=4$ theory can be computed on the Coulomb branch (since the scalars are not charged under S-duality and thus the symmetry is unbroken on the Coulomb branch), where we have $\text{rank}(G)=(N-1)$ $\cN=4$ vector multiplets. Each vector multiplet contributes $56 I_\text{Weyl}$ from the vector plus $4 I_\text{Weyl}$ from the 4 Weyl fermions (which transform as if they had charge 1 under a chiral rotation, see eqn.~(\ref{MQ})) \cite{Hsieh:2020jpj}. Therefore
\be
I_{\cN=4} = 60 \, (N-1) \, I_{\text{Weyl}} \ed
\ee
On the other hand, the chiral anomaly for $U(1)_R$ is given by%
\footnote{The rotation $R_{-\frac{3\pi}k}$ gives phases $e^{- \frac{3\pi i}k}, e^{\frac{\pi i}k}, e^{\frac{\pi i}k}, e^{\frac{\pi i}k}$ to the four fermions in the $\cN=4$ vector multiplet, see Table~(\ref{su4charges}). Equivalently, we can use a normalization in which the charges of all fermions are integer and the periodicity of $U(1)_R$ is $2\pi$. This amounts to assigning charges $(3,-1,-1,-1)$.}
\be
I_{\text{chiral}} = - \sum_{i=1}^4 q_i^3 \, (N^2 -1) \, I_{\text{Weyl}} = -24 \, (N^2-1) \, I_{\text{Weyl}} \ec
\ee
the minus sign stemming from the opposite action of $SL(2,\bZ)$ and $U(1)_R$ on the fermions. Given a cyclic subgroup $\bZ_k$ of $Mp(2,\bZ)$, the presence of such cubic anomalies would prevent the existence of a gapped vacuum invariant under the symmetry.%
\footnote{One might be concerned that such cubic anomalies could have a different description when the symmetry is non-invertible. On way to think about it is to work on a manifold with trivial $H^2(X)$, so that we are blind to the 1-form symmetry and the duality defect looks invertible. In such a case we can turn on a standard background gauge field in $ H^1(X, \bZ_k)$.}
For this to be consistent with our results, the fact that the preserved subgroup is diagonal in both $Mp(2,\bZ)$ and $U(1)_R$ is essential.

Consider first the duality generated by $S$. A crucial point is that the duality symmetry preserved in the IR starting from $\tau_{\YM}=i$ is not $\bZ_8^F$~%
\footnote{We use the superscript $F$ to indicate that $(-1)^F$ is inside the symmetry group.}
but rather $\bZ_4 \times \bZ_2^F$, see the comments at the end of Section~\ref{dualpreserve}. Thus the anomaly is valued in $\bZ_4$, and since $60 = 24 = 0$ modulo 4, the anomaly must vanish and it does not give any constraint on the $\cN=1^*$ physics.%
\footnote{Notice however that, were it not for the $U(1)_R$ compensating transformation, the $\cN=4$ anomaly would be $60 (N-1) = -4 (N-1) \text{ mod }32$ which is generically nontrivial. This happens for example in the Konishi deformation \eqref{Konishdef}.}

A stronger constraint comes from considering the non-invertible symmetry generated by $ST$, which is a symmetry of the theory at $\tau_{\YM}= e^{\frac{2\pi i}{3}}$ and can be preserved by the $\cN=1^*$ deformation using our methods. 
In this case, the preserved symmetry is $\bZ_3 \times \bZ_2^F$ and we must cancel the anomaly mod 9. Thus it must be that
\be
A_{ST} = 6 (N-1) + 3 (N^2 - 1) = 0 \mod 9
\ee
whenever the theory admits a massive vacuum which preserves the triality symmetry --- \ie, when there is an $ST$-invariant Lagrangian lattice of dyons. We can write
\be
A_{ST} = 3 N (N-1) \text{ mod }9 = \begin{cases}
    6 \text{ mod } 9 \;, \qquad &\text{for } N=3m +2 \;, \\
    0 \text{ mod } 9 \;, \qquad &\text{otherwise} \;.
\end{cases}
\ee
The $ST$-invariant Lagrangian lattice of dyons is present if and only if~%
\footnote{This criterium for the existence of a TQFT invariant under $ST$ is the analog of (\ref{self-duality condition Cordova}) for $S$. Its derivation is a simple generalization of the case for $S$, see \cite{Apte:2022xtu} for a proof at the level of invariant TQFTs.} 
\be
\label{eq: STinv}
\exists \; k', s, \tilde\ell \qquad\text{such that}\qquad N = (k')^2 s \qquad\text{and}\qquad \tilde\ell( \tilde\ell - 1) + 1 = 0 \text{ mod }s \;.
\ee
The following table contains information about the anomaly and the presence of (possibly multiple) invariant Lagrangian lattices of dyons for small values of $N$:
\bea
\begin{array}{c |  c | c}
N & A_{ST} & D_{\text{invariant}} \\ \hline \hline
2 & 6 &  \text{\texttimes} \\
3 & 0 & \langle D_{(1,1)} \rangle \\
4 & 0 & \langle D_{(2  , 0)} \oplus D_{(0,  2)} \rangle \\
5 & 6 & \text{\texttimes} \\
6 & 0 & \text{\texttimes}\\
7 & 0 & \langle D_{(2,1)} \rangle \, , \, \langle D_{(4,1)} \rangle
\end{array}
\qquad
\begin{array}{c | c | c}
N & A_{ST} & D_{\text{invariant}} \\ \hline \hline
8  & 6 & \text{\texttimes} \\
9  & 0 & \langle D_{(3,0)} \oplus D_{(0,3)} \rangle \\
10 & 0 & \text{\texttimes} \\
11 & 6 & \text{\texttimes} \\
12 & 0 & \langle D_{(6,0)} \oplus D_{(2,2)} \rangle \\
13 & 0 & \langle D_{(3,1)} \rangle \, , \, \langle D_{(9,1)} \rangle 
\end{array}
\eea
We indicated the generators of the lattice, when it exists.
Notice that the anomaly cancellation condition is necessary but \emph{not sufficient} to conclude that there is an invariant vacuum, as the cases $N=6$ and $N=10$ show.%
\footnote{The values of $s$ such that $\bigl( \ell^2 - \ell + 1 = 0 \text{ mod }s \big)$ has solutions, are those that can be represented as $s = x^2 + xy + y^2$ for coprime $x,y$ and are $s=1,3,7,13,19,21,31,37,39,43,49,\ldots$ From the representation, one sees that $s=0,1 \text{ mod 3}$. Since $q^2 = 0,1 \text{ mod }3$, if $N= 2 \text{ mod }3$ then it should be $s = 2 \text{ mod }3$, which is not possible. This shows that anomaly cancelation is necessary to have an invariant vacuum.}
Here too it is important to include the R-symmetry anomaly in order to have the required cancellation. For example, in the case $N=3$ we have $24 + 12 = 0 \text{ mod } 9$, but neither term is vanishing on its own.

\section{Gapless flows: \tps{\matht{\cN=2}}{N=2} orbifolds and the conifold}
\label{sec:gapless}

The existence of non-invertible duality defects is not limited to the example of $\cN=4$ SYM theory. In this section we first construct new examples of self-duality defects in a class of $\cN=2$ Lagrangian SCFTs and then, similarly to Section~\ref{sec: SYM}, we study how the defects can be preserved under deformations and what effects they have on the low-energy description.

\subsection[Dualities from class \tps{$\cS$}{S}]{Dualities from class \matht{\cS}}
\label{classS}

We consider Lagrangian $\cN=2$ SCFTs described by a necklace quiver with $r$ $\mathfrak{su}(N)$ nodes and bifundamental hypermultiplets. For our purposes, it is convenient to recall their class $\cS$ description which is given in terms of the 6d $\cN=(2,0)$ SCFT of type $A_{N-1}$ compactified on a torus $T^2_r$ with $r$ distinct regular punctures. The generalized S-duality group acting on the marginal couplings is given by a quotient of the mapping class group $\mathrm{MCG}(T^2_r)$ of the Riemann surface \cite{Witten:1997sc}. For $r=1$, this is the $SL(2,\bZ)$ S-duality group of $\cN=4$ SYM. 

The group $\mathrm{MCG}(T^2_r)$ can be described following \cite{Halmagyi:2004ju}. We denote by $\tau$ the complex structure of the torus and by $p_i$ (with $i=1, \dots , r$) the positions of the punctures. We decompose $p_i = x_i + \tau \, y_i$ with $0 < x_i, y_i \leq 1$. The complexified gauge couplings of the necklace quiver are related to these parameters in the following way:
\be
\label{cpxcouplings}
\tau_i = p_{i+1} - p_i \quad\text{for } 1 \leq i \leq r-1 \;,\qquad \tau_r = \tau + p_1 - p_r \;,\qquad \sum_{i=1}^r \tau_i = \tau \ed
\ee
In order to obtain positive gauge couplings, the punctures must be ordered such that $\im(p_{i+1}) > \im(p_i)$. In this parametrization, a set of generators of $\mathrm{MCG}(T^2_r)$ are:
\bea
\label{transmodular}
&S: &&(\tau,\, p_1 ,\, ... ,\, p_r) \; \mapsto\; \Bigl( - \frac{1}{\tau} ,\, \frac{p_1}{\tau} , \, ... , \, \frac{p_r}{\tau} \Bigr) \,, \\
&T: &&(\tau, \, p_1 , \, ... , \, p_r) \; \mapsto\; \left(\tau + 1, \, p_1 , \, ... , \, p_r \right) \, , \\
&\ell_i: &&(\tau, \, p_1 , \, ... , \, p_r) \; \mapsto\; \left( \tau , \, p_1 , \, ... , \, p_i + 1 , \, ... , \, p_r \right)\ec\\
&\ell^{(\tau)}_i : \, &&(\tau, \, p_1 , \, ... , \, p_r) \; \mapsto\; \left( \tau , \, p_1 , \, ... , \, p_i + \tau , \, ... , \, p_r \right)\ec \\
&s_{i<r} : &&(\tau, \, p_1 , \, ... , \, p_r) \; \mapsto\; \left( \tau , \, p_1 , \, ... , \, p_{i+1} , \, p_i , \,  ... , \, p_r \right)\ec \\
&s_r : &&(\tau, \, p_1 , \, ... , \, p_r) \; \mapsto\; \left( \tau , \, p_r - \tau , \, p_2 , \, ... , \, p_{r-1} , \,  p_1 + \tau \right) \ed
\eea 
In the gauge-coupling parametrization $\tau_1, \dots , \tau_r$ these transformations act as:
\bea
\label{gaugecouplingbase}
&S: &&(\tau_1, \, ... , \, \tau_r) \; \mapsto\; \Bigl(\frac{\tau_1}{\tau} , \, ... , \, \frac{\tau_{r-1}}{\tau} , \,    \frac{\tau_r - \tau - 1}{\tau} \Bigr) \;, \\
&T: &&(\tau_1, \,  ... ,\, \tau_{r}) \; \mapsto\; \left(\tau_1, \, ...   , \, \tau_{r-1} \, , \tau_r + 1 \right) \;,
\eea
\bea
&\ell_i: &&(\tau_1, \, ... , \, \tau_r) \; \mapsto\;  \left( \tau_1 , \, ...  , \, \tau_{i-1} + 1 , \,  \tau_i -1 , \,  ... , \, \tau_r \right) \ec \\
&\ell^{(\tau)}_i : \, &&(\tau_1, \, ... , \, \tau_r) \; \mapsto\;  \left( \tau_1 , \, ...  , \, \tau_{i-1} + \tau , \,  \tau_i -\tau , \,  ... , \, \tau_r \right) \ec \\
&s_i : &&(\tau_1, \, ... , \, \tau_r) \; \mapsto\;  \left( \tau_1 , \, ...  , \, \tau_{i-1} + \tau_i , \,  -\tau_i , \, \tau_{i+1} + \tau_i , \,  ... , \, \tau_r \right) \;.
\eea
These formulas have to be interpreted with the cyclic convention $\tau_0 \equiv \tau_r$ and $\tau_{r+1} \equiv \tau_1$. Note that these elements satisfy relations, and one could restrict to a smaller set of generators \cite{Halmagyi:2004ju}. We can also introduce the element $\omega = \ell_j^{(\tau)} s_{j-1} s_{j-2} \dots s_{j+1}$ (for any $j$) that acts as a cyclic permutation $\tau_i \to \tau_{i+1}$.
The case $r=2$ is somewhat special. The transformations $\ell_1$ and $\ell_2$ are inverses of each other, as are $\ell_1^{(\tau)}$ and $\ell_2^{(\tau)}$. Thus we can consider only $\ell \equiv \ell_1$ and $\ell^{(\tau)} \equiv \ell_1^{(\tau)}$, which act as:
\bea
\label{eq: r2ells}
&\ell : &&(\tau_1, \, \tau_2) \; \mapsto\; (\tau_1 - 1 ,\, \tau_2 + 1) \,, \\
&\ell^{(\tau)} : \, &&(\tau_1, \, \tau_2) \; \mapsto\; (\tau_1 - \tau, \, \tau_2 + \tau) \ed
\eea
In the same way, $s_1$ and $s_2$ are not independent and we can keep $s \equiv s_1$ which acts by
\be
\label{eq: srtwo}
s : \ \ (\tau_1, \, \tau_2) \; \mapsto\; (\tau_2 - \tau, \, \tau_1 + \tau) \ec
\ee
so that $\omega = s \, \ell^{(\tau)} $ simply acts as a permutation of the two gauge couplings.

From the structure of the quiver it follows that for gauge group $SU(N)^r$ there is a single $\bZ_N$ 1-form symmetry, which is the diagonal of the would-be center symmetry of each $SU(N)$ factor. Accordingly, there is only a set of $N$ genuine Wilson lines. Using the topological manipulation $\sigma$ we can gauge the 1-form symmetry, exactly as we did for $\cN=4$ SYM based on a single node, and we end up with a theory with global form $SU(N)^r/\bZ_N$ and a dual magnetic $\bZ_N$ 1-form symmetry. The genuine magnetic line operators end up also carrying electric charge if one stacks with an SPT (acting with the $\tau$ operation) before performing $\sigma$. Our discussion here is very brief since all details work exactly as in the single gauge group case.
The action of $S$ must map a fundamental Wilson line to a fundamental 't~Hooft line. Hence $S$ takes us from the global form $SU(N)^r$ to $\bigl( SU(N)^r/\bZ_N \bigr)_0$ (where the suffix specifies that the 't~Hooft line does not carry electric charge), and vice versa. Similarly, $T$ takes us cyclically along all the $\bigl( SU(N)^r/\bZ_N \bigr)_k$ global forms for $k=0, \dots, N-1$.

One should be careful with the physical interpretation of $\mathrm{MCG}(T^2_r)$. Its elements in general do not preserve the ordering of the punctures, \ie, they do not preserve the fact that $\im\tau_i > 0$. Such elements do not map a physical point on the conformal manifold to another one, and thus do not qualify as elements of the S-duality group. However, it is always possible to combine those elements with a suitable string of permutations $s_i$ and shifts $\ell_i^{(\tau)}$ such that all $\im\tau_i > 0$ after the transformation.%
\footnote{In the gauge coupling parametrization, the permutations $s_i$ correspond to a change in the sign of $\tau_i$. In the type IIA construction of \cite{Witten:1997sc} they are interpreted as an exchange of two adjacent NS5 branes on $S^1$ (modulo cyclic permutations).}
These, and only these, are the elements of the S-duality group.%
\footnote{The particular string to use depends on the point $\{\tau_i\}$ we start with, not just on the element we consider. Thus the S-duality group is a quotient of $\mathrm{MCG}(T^2_r)$.}
We then see that only the generators $S$ and $T$ act on the global form, in the very same way as in $SL(2,\bZ)$.

We can now identify the fixed points of S-duality.

\paragraph{Example: \matht{r=2}.} This is the $\cN=2$ SCFT consisting of two $\mathfrak{su}(N)$ gauge nodes. It corresponds to a torus $T^2_2$ with two punctures. Since $S$ and $T$ act in the standard way on $\tau$, we fix it at the $S$ duality invariant point $\tau=i$. The action of $S$ duality can be geometrically interpreted as a 90-degree clockwise rotation around the bottom-left corner of the torus. The invariant position of the punctures, up to a shift, is
\be
p_1 = \frac{i}{2} \;,\qquad p_2 = \frac12 + i \;,\qquad\Rightarrow\qquad \tau_1 = \frac{i+1}2 \;,\qquad \tau_2 = \frac{i-1}2 \ed
\ee
Indeed, the transformation that leaves them invariant is as shown in the picture below:
\bea
\begin{tikzpicture}
    \draw[color=green, fill=white!90!green, line width=1] (0,0) -- (2,0) -- (2,2) -- (0,2) --cycle;
    \draw[fill=black] (0,1) circle (0.05); \node[left] at (0,1) {$p_1$};
     \draw[fill=black] (1,2) circle (0.05); \node[above] at (1,2) {$p_2$};
     \node at (2.8,1) {\large$\xrightarrow{\ell_2^{(\tau)} S}$};
     \begin{scope}[shift={(3.6,0)}]
          \draw[color=green, fill=white!90!green, line width=1] (0,0) -- (2,0) -- (2,2) -- (0,2) --cycle;
    \draw[fill=black] (2,1) circle (0.05); \node[right] at (2,1) {$p_2$};
     \draw[fill=black] (1,0) circle (0.05); \node[below] at (1,0) {$p_1$};
     \node at (3.2,1) {\large$\xrightarrow{\ell_2^{-1}}$};
     \end{scope}
      \begin{scope}[shift={(8,0)}]
          \draw[color=green, fill=white!90!green, line width=1] (0,0) -- (2,0) -- (2,2) -- (0,2) --cycle;
    \draw[fill=black] (0,1) circle (0.05); \node[left] at (0,1) {$p_2$};
     \draw[fill=black] (1,0) circle (0.05); \node[below] at (1,0) {$p_1$};
     \node at (3,1) {\large$\xrightarrow{\ell_2^{(\tau)} s_1}$};
     \end{scope}
      \begin{scope}[shift={(12.5,0)}]
          \draw[color=green, fill=white!90!green, line width=1] (0,0) -- (2,0) -- (2,2) -- (0,2) --cycle;
    \draw[fill=black] (1,2) circle (0.05); \node[above] at (1,2) {$p_2$};
     \draw[fill=black] (0,1) circle (0.05); \node[left] at (0,1) {$p_1$};
     \end{scope}
\end{tikzpicture}
\eea
At each step we have dressed the transformations in such a way that the gauge couplings are positive.
Following the discussion in Section~\ref{sec: SYM}, we construct a topological interface $I_\Sigma$ that acts on the marginal couplings of the $\cN=2$ SCFT as follows:
\be
\Sigma \,\equiv\,  \ell_2^{(\tau)} s_1 \ell_2^{-1} \ell_2^{(\tau)} S = \omega \, \ell \, \ell^{(\tau)-1} S = s \, \ell \, S \ed
\ee
Composing such topological interface with another topological interface $\varphi_\sigma$ that implements the topological manipulation $\sigma$, we define a topological operator $\mathfrak{D}_\Sigma = \varphi^\dagger_\sigma\circ I_\Sigma$ that implements the non-invertible self-duality symmetry of the theory. A similar reasoning applies to the triality defects at $\tau = e^{\frac{2 \pi i}{3}}$.
As we will see, the defect $\mathfrak{D}_\Sigma$ can be preserved under a massive $\cN=1$ RG flow.

\paragraph{Example: \matht{r=4}.}
The previous discussion generalizes to the case of $T_4^2$, \ie, an $\cN=2$ SCFT with a necklace quiver with four $\mathfrak{su}(N)$ gauge nodes. In this case there is a one-dimensional complex line of points within the conformal manifold which are fixed under $S$ duality, given by $\tau = i$ and
\be
(p_1, \, p_2, \, p_3, \, p_4) = \bigl( \alpha ,\; 1 + i\alpha ,\  i - i \alpha ,\; 1 + i - \alpha  \bigr) \quad\text{with}\quad 0 < \im(\alpha) < \re(\alpha) < \frac{1}{2} \ed
\ee
Using \eqref{cpxcouplings}, this configuration corresponds to the following complexified gauge couplings:
\be
(\tau_1, \, \tau_2, \, \tau_3, \, \tau_4) = \bigl( 1 + (i-1)\alpha ,\;  i -1 -2i\alpha ,\; 1 + (i-1)\alpha ,\;  -1 + 2 \alpha \bigr) \ed
\ee
The gauge couplings are all positive. We act on the punctures with \eqref{transmodular} in such a way to leave them invariant:
\begin{multline}
\bigl( \alpha ,\; 1 + i\alpha ,\; i - i \alpha ,\; i + 1 -\alpha  \bigr) \; \xrightarrow{S} \; \bigl( -i\alpha ,\; -i + \alpha ,\; 1-\alpha ,\; 1 -i +i\alpha  \bigr) \;\to \\
\xrightarrow{\prod_{i=1}^4 \ell_i^{(\tau)}} \; \bigl( i - i\alpha ,\; \alpha ,\; 1 + i -\alpha ,\; 1 + i \alpha \bigr) \; \xrightarrow{\pi} \; \bigl( \alpha ,\; 1 + i\alpha ,\; i - i \alpha ,\; i + 1 -\alpha  \bigr) \ec
\end{multline}
where $\pi = s_2 s_3 s_1$. We visualize these transformations in the picture below:
\bea
\begin{tikzpicture}
\coordinate (A) at (0.4,0.25); \coordinate (IA) at (-0.25, 0.4);
\coordinate (p1) at (A); \coordinate (p2) at ($(2,0) + (IA)$); \coordinate (p3) at ($(0,2) - (IA)$); \coordinate (p4) at ($(2,2) - (A)$);
     \draw[color=green, fill=white!90!green, line width=1] (0,0) -- (2,0) -- (2,2) -- (0,2) --cycle;
      \draw[fill=black] (p1) circle (0.05); \node[left] at (p1) {$p_1$};
       \draw[fill=black] (p2) circle (0.05); \node[right] at (p2) {$p_2$};
      \draw[fill=black] (p3) circle (0.05); \node[left] at (p3) {$p_3$};
       \draw[fill=black] (p4) circle (0.05); \node[right] at (p4) {$p_4$};
     \node at (3.5,1) {\large $\xrightarrow{\prod_i \ell_{i}^{(\tau)} S}$}; 
     \begin{scope}[shift={(5,0)}]
     \coordinate (A) at (0.4,0.25); \coordinate (IA) at (-0.25, 0.4);
\coordinate (p1) at (A); \coordinate (p2) at ($(2,0) + (IA)$); \coordinate (p3) at ($(0,2) - (IA)$); \coordinate (p4) at ($(2,2) - (A)$);
           \draw[color=green, fill=white!90!green, line width=1] (0,0) -- (2,0) -- (2,2) -- (0,2) --cycle;
      \draw[fill=black] (p1) circle (0.05); \node[left] at (p1) {$p_2$};
       \draw[fill=black] (p2) circle (0.05); \node[left] at (p2) {$p_4$};
      \draw[fill=black] (p3) circle (0.05); \node[left] at (p3) {$p_1$};
       \draw[fill=black] (p4) circle (0.05); \node[left] at (p4) {$p_3$};
        \node at (3.0,1) {\large $\xrightarrow{\pi}$}; 
     \end{scope}
       \begin{scope}[shift={(9,0)}]
     \coordinate (A) at (0.4,0.25); \coordinate (IA) at (-0.25, 0.4);
\coordinate (p1) at (A); \coordinate (p2) at ($(2,0) + (IA)$); \coordinate (p3) at ($(0,2) - (IA)$); \coordinate (p4) at ($(2,2) - (A)$);
           \draw[color=green, fill=white!90!green, line width=1] (0,0) -- (2,0) -- (2,2) -- (0,2) --cycle;
      \draw[fill=black] (p1) circle (0.05); \node[left] at (p1) {$p_1$};
       \draw[fill=black] (p2) circle (0.05); \node[left] at (p2) {$p_2$};
      \draw[fill=black] (p3) circle (0.05); \node[left] at (p3) {$p_3$};
       \draw[fill=black] (p4) circle (0.05); \node[left] at (p4) {$p_4$};
     \end{scope}
\end{tikzpicture}
\eea
Once again we construct a topological interface $I_\Sigma$, where $\Sigma = \pi \prod_i \ell_i^{(\tau)}S$, which we compose with a topological interface that implements the action of $\sigma$ in order to construct a non-invertible self-duality symmetry defect $\mathfrak{D}_\Sigma$.

We conclude this section with some extra comments on higher values of $r$. For $r = 4 k$ and $r=4k+2$ the story is similar, as we can always organize the punctures into $k$ squares as in $r=4$, possibly plus two more punctures as in $r=2$. Note that all punctures must be distinct. For odd values of $r = 2n+1$ we can place $2n$ punctures as in the previous case of $r$ even, plus one puncture in the middle of the torus, \eg:
\bea
\begin{tikzpicture}
    \coordinate (A) at (0.6,0.4); \coordinate (IA) at (-0.4, 0.6);
\coordinate (p1) at (A); \coordinate (p2) at ($(3,0) + (IA)$); \coordinate (p3) at ($(0,3) - (IA)$); \coordinate (p4) at ($(3,3) - (A)$);
     \draw[color=green, fill=white!90!green, line width=1] (0,0) -- (3,0) -- (3,3) -- (0,3) --cycle;
       \draw[fill=black] (1.5,1.5) circle (0.05) node[above] {$p_3$};
      \draw[fill=black] (p1) circle (0.05); \node[left] at (p1) {$p_1$};
       \draw[fill=black] (p2) circle (0.05); \node[right] at (p2) {$p_2$};
      \draw[fill=black] (p3) circle (0.05); \node[left] at (p3) {$p_4$};
       \draw[fill=black] (p4) circle (0.05); \node[right] at (p4) {$p_5$};
       
       \draw[dashed] (p1) -- ($0.5*(p1) + 0.5*(p2)$) node[below = 0.1] {$s_1$}; \draw[dashed] ($0.5*(p1) + 0.5*(p2)$) -- (p2);
             \draw[dashed] (p2) -- ($0.75*(p2) + 0.25*(p3)$) node[right=0.1] {$s_2$} ; 
               \draw[dashed] (p2) -- ($0.5*(p2) + 0.5*(p3)$) ;
               
             \draw[dashed] ($0.25*(p2) + 0.75*(p3)$) node[left] {$s_3$} -- (p3) ;
 \draw[dashed] ($0.5*(p2) + 0.5*(p3)$) --  ($0.25*(p2) + 0.75*(p3)$);
             
                   \draw[dashed] (p3) -- ($0.5*(p3) + 0.5*(p4)$) node[above=0.1] {$s_4$}; \draw[dashed] ($0.5*(p3) + 0.5*(p4)$) -- (p4);
                         \draw[dashed] (p4) -- ($0.75*(p4) + 0.25*(p1)$) node[right=0.1] {$s_5$}; \draw[dashed] ($0.75*(p4) + 0.25*(p1)$) -- (p1);
\end{tikzpicture}
\eea
Using the same techniques we find, for example if $r=5$: $\Sigma = s_5\, s_4 \, s_1 \prod_{i=1}^5 \ell_i^{(\tau)} \, S$.

\subsection{Non-invertible duality defects on the conifold}
\label{sec: conifold}

Starting from the $\cN=2$ quiver SCFTs and their self-duality defects introduced above, we can adopt the same strategy as in Section~\ref{sec: SYM} and introduce relevant deformations that partially break supersymmetry to $\cN=1$, while preserving the non-invertible symmetry. A particularly interesting example is the $r=2$ theory considered above, \ie, the $\cN=2$ SCFT consisting of two $\mathfrak{su}(N)$ gauge nodes and two bifundamental hypermultiplets. We decompose the hypermultiplets in terms of $\cN=1$ chiral superfields $A_\alpha$, $B_\beta$ with $\alpha,\beta=1,2$ transforming respectively in the $(N, \wb{N})$ and $(\wb{N},N)$ representation, while we call $\Phi$ and $\tilde\Phi$ the adjoint chiral multiplets in the $\cN=2$ vector multiplets. The superpotential is
\be
W = \epsilon^{\alpha\beta} \Tr \bigl( \Phi  A_\alpha B_\beta + \tilde\Phi  B_\alpha A_\beta \bigr)
=\Tr \Phi(A_1 B_2 - A_2B_1) + \Tr \tilde{\Phi}(B_1A_2 - B_2A_1) \ed
\ee
With two gauge nodes there is a non-Abelian  $SU(2)$ flavor symmetry under which $A_\alpha$ and $B_\beta$ transform as doublets.

We can show that there exist massive deformations of the $\cN=2$ SCFT that preserve the non-invertible symmetry defect $\mathfrak{D}_\Sigma$. Consider the superpotential deformation
\be
\label{KWmass}
\delta W = \frac{m}{2} \, \bigl( \Tr\Phi^2 - \Tr \tilde{\Phi}^2 \bigr)
\ee
that partially breaks supersymmetry to $\cN=1$. Any $\cN=2$ SCFT has $SU(2)_R\times U(1)_\mathsf{r}$ R\nobreakdash-symmetry group. The supercharges $Q_\alpha^i$ ($i=1,2$) transform as doublets of $SU(2)_R$ and with charge $[Q^i_\alpha]_\mathsf{r}=-1$ under $U(1)_\mathsf{r}$. In our conventions the superspace coordinate measure $d^2\theta$ transforms with charge $[d^2\theta]_\mathsf{r}=-1$. Both $\Tr \Phi^2$ and $\Tr\tilde{\Phi}^2$ are neutral under $SU(2)_R$ and have $U(1)_\mathsf{r}$ charge $[\Tr \Phi^2]_\mathsf{r}=[\Tr \tilde{\Phi}^2]_\mathsf{r}=2$. As a result, $ \int\! d^2\theta \, \delta W$ has charge $1$.
Under a $\Sigma$-duality transformation, the integrated superpotential deformation $\int\! d^2\theta \, \delta W$ picks up a phase $-e^{-\frac{i\pi}{2}}$ (notice the sign difference with respect to our discussion in Section~\ref{dualpreserve}). This is due to the fact that $\Sigma$ involves the action of $\omega$ that sends $\Phi \to \tilde{\Phi}$ and $m \to - m$.
Since the mass deformation \eqref{KWmass} breaks the superconformal $U(1)_\mathsf{r}$ explicitly, we can use a $U(1)_\mathsf{r}$ transformation to compensate for the action of $\Sigma$:
\be
\mathfrak{D}_{-\frac{\pi}{2}}= \mathsf{r}_{-\frac{\pi}{2}}\times \mathfrak{D}_{\Sigma}
\ee
is a symmetry of the deformed theory.
The effect of \eqref{KWmass} can be understood more clearly upon integrating out the adjoint chiral fields $\Phi$ and $\tilde \Phi$ \cite{Klebanov:1998hh}. This leads to an $\cN=1$ theory with $\mathfrak{su}(N)\times \mathfrak{su}(N)$ gauge group, bifundamental chiral fields $A_\alpha$, $B_{\dot\alpha}$ ($\alpha,\dot\alpha=1,2$) and a quartic superpotential
\be
\label{WKW}
W_{\mathrm{KW}} = \frac\lambda2 \, \epsilon^{\alpha\beta}\epsilon^{\dot\alpha\dot\beta} \, \Tr A_\alpha B_{\dot\alpha} A_\beta B_{\dot\beta} 
= \lambda \Tr \bigl( A_1B_1A_2B_2 - A_1B_2A_2B_1 \bigr) \ec
\ee
where $\lambda = -1/m$, exhibiting an enhanced $SU(2)\times SU(2)$ global symmetry. The choice of relative sign in \eqref{KWmass} is crucial to obtain such enhancement. There is another choice of mass deformation that preserves the non-invertible symmetry: $\delta W = \frac{m}{2} \bigl( \Tr\Phi^2 + \Tr\tilde{\Phi}^2 \bigr)$.

Using the methods of \cite{Leigh:1995ep} one can show that the $\cN=1$ theory flows to a non-trivial IR fixed point, where the dimension of the chiral superfields is $3/4$ (as determined from their $R$-charge $1/2$). The operator \eqref{WKW} has $R$-charge $2$ and is an exaclty marginal deformation of the $\cN=1$ SCFT which, together with the gauge couplings, gives rise to manifold of fixed points of complex dimension 2. We refer to this theory, introduced in \cite{Klebanov:1998hh} to describe a stack of $N$ D3-branes probing a conifold singularity, as the Klebanov-Witten (KW) theory or as the conifold theory. 

As opposed to the case of the $\cN=1^*$ theory discussed in Section~\ref{sec: onestar}, the adjoint mass deformation \eqref{KWmass} leads to an interacting gapless IR theory with no spontaneous breaking of the non-invertible symmetry. We should then understand how S-duality acts on the conformal manifold of the conifold theory. To address this, we will first resort to AdS/CFT, and then perform a purely field theoretic analysis.

For general $\bZ_r$ orbifold SCFTs, we can also turn on adjoint mass deformations
\be
\delta W = \sum\nolimits_{i=1}^r \, m_i \Tr(\Phi_i^2)
\ee
that are preserved by the duality $\Sigma$ up to a phase (then reabsorbed by an R-symmetry rotation). For small masses, the $\cN=2$ duality group acts as \cite{Halmagyi:2004ju}:
\bea
&S: &&(m, \, m_1, \, ..., \, m_{r}) \;\mapsto\; \Bigl( \frac{m}{\tau} ,\, m_1 - m \frac{\tau_1}{\tau} ,\, ... ,\, m_{r-1} - m \frac{\tau_{r-1}}\tau ,\, m_r - m \frac{\tau_r - 1}{\tau} \Bigr) \, , \\
&\ell_i^{(\tau)}: \, &&(m, \, m_1, \, ..., \, m_{r}) \;\mapsto\; (m, m_1, \, ... , \, m_{i-1} + m, \, m_i - m , \, ... , \, m_r) , \, \\
&s_i : && (m, \, m_1, \, ..., \, m_{r}) \;\mapsto\; (m, m_1, \, ... , \, m_{i-1} + m_i, \, -m_i, \, m_{i+1} + m_i , \, ... , \, m_r) \ed
\eea
We introduced $m= \sum_{i=1}^r m_i$, and in the last two formulas one has to use the cyclic convention $m_0 \equiv m_r$ and $m_{r+1} \equiv m_1$. We then look for mass configurations that at $\tau=i$ are invariant under $\Sigma$ up to a phase. For $r=2$ (at $\tau_1 = \frac{i+1}2$) one finds two solutions for $m_1, m_2$:
\be
m_2 = - m_1 : \; (m_1, m_2) \,\stackrel{\Sigma}\mapsto\, - (m_1, m_2) \qquad\text{and}\qquad  m_2 = m_1 : \; (m_1, m_2) \,\stackrel{\Sigma}{\mapsto}\, -i \, (m_1, m_2) \ed
\ee
This reproduces what we discussed above. For $r=4$ we find that $m_1=-m_3 = m$ and $m_2 = m_4 = 0$ gives a consistent duality-preserving deformation (but there might be more). It would be interesting to study this problem further.

\subsection{Holographic aspects}
\label{holosec}

Both the $\cN=2$ quiver SCFT with $r=2$ nodes and the $\cN=1$ KW SCFT are paradigmatic examples of AdS/CFT. They arise from D3-branes probing toric Calabi-Yau singularities, here \mbox{$\bC^2/\bZ_2\times \bC$} and the conifold, respectively. These two singularities have a single vanishing 2\nobreakdash-cycle, where D5-branes can be wrapped yielding fractional branes. A regular D3-brane, when placed at the singularity, breaks into two types of mutually BPS states, each corresponding to one node of the quiver. As long as the NSNS 2-form gauge potential $B_2$ is in a certain range --- (\ref{range for B2}) below --- the two states are a D5-brane wrapped on the 2-cycle, and an anti-D5-brane on the 2-cycle with $-1$ units of worldvolume flux \cite{Diaconescu:1997br, Bertolini:2000dk, Polchinski:2000mx}. If the number of D5 and $\overline{\text{D5}}$ is the same, the two nodes have the same rank and the quiver gauge theory is a SCFT, while if it is not the same, then an interesting RG flow occurs, where the ranks of the gauge groups gradually diminish towards the IR \cite{Klebanov:1999rd, Klebanov:2000nc, Klebanov:2000hb}. Here we will be interested in the conformal case.

Both theories above have a (symmetry preserving) conformal manifold that can be parameterized by two complex couplings. In the $\cN=2$ case these are simply the two gauge couplings, while in the $\cN=1$ KW theory the conformal manifold is embedded in the space of the two gauge couplings and the quartic superpotential coupling. In the latter case the two complex parameters are not exactly identified with the $\cN=1$ gauge couplings, although, following \cite{Strassler:2005qs}, we still assume that they have a positive imaginary part. In both SCFTs there is a simple way to determine the holographic description of the exactly marginal couplings via probe branes.

Let us first consider the $\cN=2$ SCFT. By looking at the DBI + WZ action of a D3-brane at the $\bC^2/\bZ_2\times \bC$ singularity, and at a D5-brane wrapped on the vanishing 2-cycle, one obtains the following map between field theory and supergravity parameters:
\be
\label{formulas for gauge couplings}
\tau_1+\tau_2 = \frac{\tau_\text{SG}}{g_s} \equiv \frac{1}{g_s} \, \bigl( C_0 + i \, e^{-\phi} \bigr) \;,\qquad\quad \tau_1 = - \frac{1}{4\pi^2\alpha' g_s} \int_{S^2} \bigl( C_2 - \tau_\text{SG} \, B_2 \bigr) \ec
\ee
where $\tau_j = \theta_j/2\pi+  4\pi i / g_j^2$ are the $\cN=2$ complexified gauge couplings, $C_0$ and $C_2$ are the RR 0-form and 2-form fields of type IIB supergravity, $\phi$ is the dilaton, and $B_2$ the NSNS 2-form field.  Given that $\im \tau_j$ should be positive, these formulas are correct only if
\be
\label{range for B2}
0 < \frac{1}{4\pi^2\alpha'} \int_{S^2} B_2 < 1 \ed
\ee
Indeed, only in this range a wrapped D5 and a wrapped anti-D5 with $-1$ units of worldvolume flux have positive D3 charge and can be mutually BPS. Outside of the range the decomposition is different: for $n < \frac{1}{4\pi^2\alpha'} \!\int\! B_2 < n+1$ the mutually BPS states are a wrapped D5 with $-n$ units of flux, and a wrapped anti-D5 with $-n-1$ units of flux. The formulas (\ref{formulas for gauge couplings}) for the gauge couplings should be accordingly modified. We will take the alternative point of view that one should first perform a large gauge transformation of $B_2$, corresponding to the integer shift $\delta \, \frac{1}{4\pi^2\alpha'} \int_{S^2} B_2 = -n$ in order to map it inside the range (\ref{range for B2}), and then apply (\ref{formulas for gauge couplings}).

The relevant type IIB $SL(2,\bZ)$ transformations are the following:
\be
\tau_\text{SG} \;\mapsto\; \frac{a \, \tau_\text{SG} + b}{c \, \tau_\text{SG} + d} \;,\qquad\qquad \biggl( \begin{matrix} C_2 \\ B_2 \end{matrix} \biggr) \;\mapsto\;  \biggl( \begin{matrix} a & b \\ c & d \end{matrix} \biggr) \biggl( \begin{matrix} C_2 \\ B_2 \end{matrix} \biggr)  \ec
\ee
under which we have 
\be
C_2 - \tau_\text{SG} \, B_2 \;\mapsto\; \frac1{c \, \tau_\text{SG} + d} \, \bigl( C_2 - \tau_\text{SG} \, B_2 \bigr) \ed
\ee
We can then deduce the $SL(2,\bZ)$ transformations of the complexified gauge couplings. We set $g_s=1$ and identify $\tau = \tau_\text{SG} = \tau_1 + \tau_2$. Naively, we obtain the transformations:
\be
\label{S-duality in the conifold}
\tau \;\mapsto\; \frac{a\tau + b}{c\tau + d} \;,\qquad\qquad \tau_1 \;\mapsto\; \frac{\tau_1}{c\tau + d}\ec
\ee
and $\tau_2 = \tau - \tau_1$. They agree with the transformations in \eqref{transmodular}.%
\footnote{To see this more clearly, one should use the gauge coupling basis $(\tau_1,\tau_2)$ introduced in \eqref{gaugecouplingbase} which is more natural for the comparison with supergravity.}
As discussed in Section~\ref{classS}, one should be careful since these transformations do not always respect $\im\tau_j>0$. In fact, from the point of view of supergravity we also have large gauge transformations of $B_2$ and $C_2$ which shift their respective integrals by integer amounts.
These give rise to transformations $\ell^{(\tau)}$ and $\ell$ acting as in \eqref{eq: r2ells}.
The transformation $\ell$ has a clear field-theoretical interpretation as a shift of the theta angles. The transformation $\ell^{(\tau)}$ is more subtle: since $\im\tau > 0$, $\ell^{(\tau)}$ is such that in its orbit there is one and only one representative with $\im\tau_{1,2}>0$. Therefore the transformation in (\ref{S-duality in the conifold}) has to be combined with a suitable repetition of $\ell^{(\tau)}$ (whose order depends on $\tau_1, \tau_2$) in \eqref{eq: r2ells} in order to respect $\im\tau_j > 0$. We find:
\begin{align}
C &= -\smat{1 & 0 \\ 0 & 1} \;,\quad& \ell^{(\tau)} C: \quad \tau &\to \tau \;,\quad& \tau_1 &\to \tau_2 \;,\quad& \tau_2 &\to \tau_1 \\
T &= \smat{1 & 1 \\ 0 & 1} \;,\quad& T: \quad \tau &\to \tau+1 \;,\quad& \tau_1 &\to \tau_1 \;,\quad& \tau_2 &\to \tau_2 + 1 \nn \\
&& \ell T: \quad \tau &\to \tau+1 \;,\quad& \tau_1 &\to \tau_1 + 1 \;,\quad& \tau_2 &\to \tau_2 \nn \\
S &= \smat{0 & -1 \\ 1 & 0} \;,\quad&  \left[\ell^{(\tau)}\right]^n S: \quad \tau &\to - \frac1\tau \;,\quad& \tau_1 &\to \frac{\tau_1}\tau + n\tau \;,\quad& \tau_2 &\to - \frac{\tau_1 + 1}\tau - n \tau \;. \nn
\end{align}
In the last line the integer power $n$ is a function of $\tau$ and $\tau_1$. Notice that $C$ is the holographic avatar of $s$ from \eqref{eq: srtwo}.

One can repeat the holographic analysis verbatim for the $\cN=1$ KW SCFT by studying a D3-brane probing the conifold singularity and a D5-brane wrapped on the vanishing 2\nobreakdash-cycle. In this context we continue using the notation $\tau_1$ and $\tau_2$ to parametrize the conformal manifold although, as emphasized before, in the Klebanov-Witten model $\tau_1$ and $\tau_2$ are not simply related to the gauge couplings. The holographic description indicates that the $\cN=2$ generalized $S$-duality transformations are inherited by the Klebanov-Witten theory. This is what we expected from the analysis of Section~\ref{sec: conifold}.

\subsection[\tps{$\cN=1$}{N=1} non-invertible symmetry from Seiberg duality]{\matht{\cN=1} non-invertible symmetry from Seiberg duality}

As shown in Section~\ref{sec: conifold}, the Klebanov-Witten theory has a non-invertible self-duality symmetry inherited from its parent $\cN=2$ SCFT. Further evidence of this symmetry comes from the holographic realization of the theory, as discussed in Section~\ref{holosec}. It is natural to ask whether it is possible to describe the S-duality symmetry in purely $\cN=1$ field theoretic terms. It turns out that, at least for $N=2$, one can understand S-duality of the KW theory in terms of Seiberg duality. The argument does not apply to larger values of $N$, and it would be worth exploring this question further.

In the specific case of $N=2$, the $\mathfrak{su}(2) \times \mathfrak{su}(2)$ conifold theory can be described as an $\mathfrak{so}(4)$ theory with $N_f =4$ flavors in the fundamental representation. The global variant $Spin(4)$ corresponds to $SU(2) \times SU(2)$, while $SO(4)_{\pm}$ are identified with $P \bigl( SU(2) \times SU(2) \bigr){}_{0,1}$.%
\footnote{By $P \bigl( SU(2) \times SU(2) \bigr)$ we mean $\bigl( SU(2) \times SU(2) \bigr)/\bZ_2$ where the quotient is by the diagonal center. See \cite{Aharony:2013hda} for an in-depth treatment of $\mathfrak{so}$ theories with matter in the vector representation.}
The $Spin(4)$ theory with fundamental matter has a $\bZ\one_2$ electric 1\nobreakdash-form symmetry acting on the spinor Wilson line that we denote by $s$, while the Wilson line in the vector representation $V$ is screened. In $SU(2)^2$ language, the line $s$ corresponds to a fundamental Wilson line $W_1$ for the first gauge group, or $W_2$ for the second gauge group, while the Wilson line $V = W_1 W_2$ can be screened by the bifundamental matter.
The $SO(4)_+$ variant has an 't~Hooft line $T$ which is charged under the dual magnetic $1$-form symmetry, while the $SO(4)_-$ variant has a genuine dyonic line denoted by $s T$.
We denote the chiral multiplets by $v^i_\alpha$ where $\alpha$ is an index in the fundamental representation of the flavor symmetry group $SU(4)$, while $i$ is in the vector representation of the gauge algebra $\mathfrak{so}(4)$. The Klebanov-Witten superpotential can be written as
\be
W_\text{KW} = \frac{1}{2} \, \lambda \, \epsilon_{ijkl} \, \epsilon^{\alpha \beta \gamma \delta} \, v^i_\alpha \, v^j_\beta \, v^k_\gamma \, v^l_\delta = \lambda \, B \ec
\ee
where $B$ is the baryon operator. The action of Seiberg duality on $\mathfrak{so}(N)$ gauge theories with $N_f \geq N \geq 4$ and vanishing superpotential has been studied in \cite{Aharony:2013hda} and it displays a nontrivial mapping between global variants:
\bea
& Spin(N) + N_f \ &&\leftrightarrow \quad\; SO(N_f - N + 4)_-  + N_f \\ 
& SO(N)_+ + N_f \  &&\leftrightarrow \quad\; SO(N_f - N + 4)_+ + N_f \, .
\eea
Notice that this duality action is \emph{different} from the one of $S$ duality, which maps $Spin$ to $SO_+$ since it maps a Wilson line to an 't~Hooft line.

The case of $N_f = N = 4$ is doubly special. First, the gauge algebras are self-dual since $N_f - N - 4 = N = 4$. Second, for $N_f = N$ we should consider the baryon operator $B$. Its Seiberg dual is the difference of the gauge kinetic terms for the two $\su(2)$ factors \cite{Intriligator:1995id}:
\be
B \quad\leftrightarrow\quad \Tr \bigl( \cW_{(1)}^2 - \cW_{(2)}^2 \bigr) \ec
\ee
and thus the dual of the KW superpotential is a shift of the relative coupling $\eta = \tau_1 - \tau_2$ and vice versa.%
\footnote{In this section we denote by $\tau_1$, $\tau_2$ the $\cN=1$ gauge couplings, which should not be confused with the exactly marginal gauge couplings parametrizing the two-dimensional conformal manifold that we discussed in Section~\ref{holosec}. We hope this does not cause confusion to the reader.}
It is convenient to parametrize a point on the $SU(4)$-preserving conformal manifold by $(\eta, \lambda)$, so that the total gauge coupling $\tau = \tau_1 + \tau_2$ is a specific function $\tau(\eta, \lambda)$ whose precise form is not known. We denote a KW theory by the data $\bigl( G + N_f,\, \eta, \,W \bigr)$ where $G$ is the global form of the gauge group. We then use the results of \cite{Aharony:2013hda} to build a self-duality. Starting from $\bigl( Spin(4) +4, \, \eta, \, \lambda B)$ we perform a first Seiberg duality to $\bigl( SO(4)_- + 4,\, \lambda' ,\, \eta' B' + M v' v')$, where the primed variables $\eta', \lambda'$ are some functions of the unprimed ones, $M$ is the meson field, and $v'$ are the dual quark fields. Then we perform a shift of $\theta_1$ by $2\pi$, which also maps the $SO(4)_-$ theory to $SO(4)_+$. The relative gauge coupling $\eta$ is shifted by 1, and so is $\tau$. This induces a nontrivial map $\lambda \to \tilde{\lambda}$ on the baryon coupling, such that
\be
\tau\bigl( \eta + 1,\, \tilde{\lambda} \bigr) = \tau(\eta, \lambda) + 1 \ed
\ee
Finally, we perform a second Seiberg duality from $SO(4)_+$ to $SO(4)_+$ and then integrate out the new meson field $M'$ through its equations of motion. In this way we obtain a dual description in terms of $\bigl( SO(4)_+ \,+\, 4,\; \eta'',\; \lambda'' \bigr)$. We interpret this duality as the $\cN=1$ version in the conifold of the $\cN=2$ S-duality element $\Sigma$, and we denote it as $\Sigma_{\cN=1}$. Importantly, one can derive that $\Sigma_{\cN=1}$ maps the global form $Spin(4)$ to $SO(4)_+$.

In order to define a self-duality defect we must assume that there exists a fixed point $\lambda = \lambda''$, $\eta=\eta''$, and then compose the map with a discrete gauging of the $\bZ\one_2$ symmetry. We propose that this gives rise to a non-invertible self-duality symmetry defect.%
\footnote{This duality is absent in the theory without superpotential, since Seiberg duality generates a Klebanov-Witten-type term \cite{Aharony:2013hda}.}
We summarize the construction in the following graph:
\bea
\label{seibergselfdual}
\begin{tikzcd}[row sep = scriptsize]
    Spin(4) + 4 ,\; \eta ,\; \lambda B \qquad  \arrow[rr, leftrightarrow, "\text{Seiberg duality}"] & & \qquad SO(4)_- + 4 ,\; \lambda' , \;  \eta' B' + M v' v'  \arrow[dd, leftrightarrow, "\text{shift } \theta_1"] \\
     & & \\ 
     SO(4)_+ + 4 ,\; \eta'' ,\; \lambda'' B'' \qquad \arrow[uu, leftrightarrow, "\Sigma_{\cN=1}"] & & \qquad SO(4)_+ + 4 ,\; \lambda' + 1 ,\; \tilde{\eta'} B' + M  v' v'  \arrow[ll, leftrightarrow, "\text{ Seiberg duality }"']
\end{tikzcd}
\eea

\paragraph{Acknowledgements.}
We thank Ofer Aharony, Michele Del Zotto, Ho Tat Lam, Kantaro Ohmori, and Yifan Wang for helpful discussions. J.A.D.~and R.A.~are respectively a Postdoctoral Researcher and a Research Director of the F.R.S.-FNRS (Belgium). The research  of J.A.D., R.A.~and L.T.~is supported by IISN-Belgium (convention 4.4503.15) and through an ARC advanced project. F.B. and C.C. are supported by the ERC-COG grant NP-QFT No.~864583 ``Non-perturbative dynamics of quantum fields: from new deconfined phases of matter to quantum black holes", by the MUR-FARE grant EmGrav No.~R20E8NR3HX ``The Emergence of Quantum Gravity from Strong Coupling Dynamics'', and by the INFN ``Iniziativa Specifica ST\&FI". S.B. is partially supported by the INFN ``Iniziativa Specifica GAST". L.T.~has been partially supported by funds from the Solvay Family.

\appendix

\section{TQFTs for \tps{\matht{\cN=1}}{N=1} SYM}
\label{app: N1TQFTs}

In this appendix we briefly review the TQFT description of the vacua of pure $\cN=1$ $\mathfrak{su}(N)$ SYM. Let us start with $G=SU(N)$ with its $N$ equivalent confining vacua. In the $k$-th vacuum the dyon $D_{(k,1)}$ condenses and the theory is described by%
\footnote{Here, for simplicity, we turn off the gauge field for the $\bZ_{2N}$ R-symmetry. That symmetry has a cubic remnant of the chiral anomaly, which is saturated in the IR and it becomes an 't~Hooft anomaly for the domain wall operators.}
\be
\text{SPT}_{-k}(B) = \exp\left( -\frac{2 \pi i k}{2 N} \int \fP(B)\right) \;.
\ee
These vacua form $N$ ``universes'', which can be described by the TQFT of a $\bZ_N$-valued scalar $\phi$ with the following partition function:
\be
\label{eq: actionN1}
Z(B) = \sum_{\phi \,\in H^0(X, \bZ_N)} \exp\biggl( \frac{2 \pi i}{2 N} \int \phi \cup \fP(B) \biggr) \;.
\ee
This TQFT reproduces the mixed anomaly
\be
I = \frac{2 \pi i (N-1)}{2 N} \int A \cup \fP(B)
\ee
between the $\bZ_{2N}$ R-symmetry and the 1-form symmetry. Denoting as $V_n = e^{i n \phi}$ the point-like operators, we can construct the projector $P_k$ on the $k$-th vacuum as
\be
P_k = \frac{1}{N} \sum_{n=0}^{N-1} e^{\frac{2 \pi i}{N} n k} \; V_n \;.
\ee
It is simple to show that the $\bZ_N$ 0-form symmetry defects $U_n$ satisfy
\be
U_n \;  P_k = P_{k + n} \; U_n \ed
\ee
These are the topological avatars of $\cN=1$ domain walls between confining vacua. In order to be consistent with the action \eqref{eq: actionN1}, the domain wall must also be stacked with an $\text{SPT}_{1}(B)$ theory when changing vacuum. This works as an inflow theory for a minimal $\cA^{N,1}$ TQFT, thus
\be
\text{DW} = U \times \cA^{N,1}(B)\ec
\ee
which is indeed a subsector of the Acharya-Vafa theory \cite{Acharya:2001dz}. 

In the $PSU(N)_0$ case, we can find the TQFT of vacua by gauging the 1-form symmetry:
\be
Z(C) = \frac{1}{\sqrt{ \lvert H^2(X, \bZ_N) \rvert}} \sum_{\substack{\phi \, \in H^0(X,\bZ_N) \\ b \, \in H^2(X, \bZ_N)}} \exp\biggl( \frac{2 \pi i}{2 N} \int \phi \cup \fP(b) + \frac{2 \pi i}{N} \int b \cup C \biggr) \ed
\ee
The vacua are now generically inequivalent and host a $\bZ_{\gcd(k,N)}$ TQFT. Following \cite{Kaidi:2021xfk}, the IR domain wall becomes a non-invertible defect $\fD = U \times \cA^{N,1}(b)$.

\section{Local order parameters for \tps{\matht{\cN=1^*}}{N=1*}}
\label{app: orderpar}

In this appendix we recall the transformation properties of the order parameter introduced in Section~\ref{ssec: su2}, following the discussion in  \cite{Aharony:2000nt}. Neglecting some prefactors, we can rewrite the superpotential \eqref{eq:onestarsuperpot} in terms of the following functions:
\begin{align}
h^{(N)}(\tau) &= E_2(\tau)-N E_2(N\tau) \ec \\
g^{(k)}_\ell(\tau) &= E_2(\tau) - \frac{k'}{k} E_2\biggl( \frac{k'\tau+\ell}{k} \biggr) \ec \quad N=kk'\ec \quad \ell=0,\ldots , k-1\ec
\nonumber
\end{align}
where $h^{(N)}(\tau)$ corresponds to the (unique) fully Higgsed vacuum. 

The Eisenstein series $E_2(\tau)$, that we can evaluate as
\be
E_2(\tau) = \frac{12}{\pi i} \, \frac{\eta'(\tau)}{\eta(\tau)}
\ee
in terms of the Dedekind eta function $\eta(\tau)$, is a quasi-modular form of degree 2, therefore
\be
\label{eq:quasimod}
E_2\left(\frac{a\tau + b}{c\tau +d}\right)= (c\tau+d)^2E_2(\tau)+\frac{6c(c\tau+d)}{\pi i} \ed
\ee 
An immediate consequence is that $h^{(N)}(\tau+1)=h^{(N)}(\tau)$ and $g^{(k)}_\ell(\tau+1)=g^{(k)}_{\ell+k'}(\tau)$, consistently with the expected properties of these vacua under the modular $T$ transformation. In addtion, from \eqref{eq:quasimod} it is straightforward to obtain
\be
\label{eq:Stransf1}
h^{(N)}\biggl( -\frac{1}{\tau} \biggr) = \tau^2 \, g_0^{(N)}(\tau) \;,\qquad\qquad g^{(N)}_0 \biggl( -\frac{1}{\tau} \biggr) = \tau^2 \, h^{(N)}(\tau) \ed
\ee 
Now let us consider an integer $\ell>0$, with $\gcd(\ell,N)=p$. Define $\tilde p = - \bigl( \ell/p \bigr)^{-1}_{\text{mod } N/p}$ in other words there exists an integer $r$ such that $-\tilde p \, \ell/p - r N/p = 1$. The the following holds:
\be
\frac{-\frac{1}{\tau}+\ell}{N}= M\cdot \tau'  \qquad\text{with}\qquad  M=\mat{ \ell/p & r \\ N/p & - \tilde p} \;,\qquad \tau'=\frac{p\,\tau+\tilde p}{N/p} \;,
\ee
where $M\in SL(2,\mathbb{Z})$ acts by fractional linear transformations.
Combining this with the transformation \eqref{eq:quasimod} yields
\be
\label{eq:Stransf2}
g^{(N)}_\ell \biggl( -\frac{1}{\tau} \biggr) = \tau^2 \, g^{(N/p)}_{\tilde p}(\tau) \ed
\ee
When $N$ is prime, the transformations \eqref{eq:Stransf1} and \eqref{eq:Stransf2} (with $p=1$) account for the mapping of the order parameter under the modular $S$ transformation for all the vacua.

For $N=kk'$ the procedure is similar. Denoting $\gcd(\ell,k)=p$ and $\tilde p = - \bigl( \ell/p \bigr)^{-1}_{\text{mod } k/p}$ so that there exists $r$ such that $- \tilde p \, \ell /p - r k /p = 1$, we note that
\be
\frac{-\frac{k'}{\tau}+\ell}{k} = \hat M \cdot \hat\tau \qquad\text{with}\qquad \hat M = \mat{ \ell / p & r \\ k/p & - \tilde p } \;,\qquad \hat \tau = \frac{\frac{p\tau}{k'} + \tilde p}{k/p} \;.
\ee  
This relation, together with \eqref{eq:quasimod}, leads to
\be
\label{eq:Stransf3}
g^{(k)}_\ell \biggl( -\frac{1}{\tau} \biggr) = \tau^2 \, g^{(N/p)}_{k' \tilde p}(\tau) \ed
\ee
The expression above perfectly matches the action of $S$-duality on the Lagrangian sublattices \eqref{TSactionL}, according to the correspondence between gapped vacua in $SU(N)$ $\cN=1^*$ SYM and global forms for the algebra $\mathfrak{su}(N)$.

When $N$ is a perfect square, \ie, when $k'=k$, there is a vacuum for which the order parameter vanishes identically for any value of $\tau$, namely
\be
g^{(k)}_0(\tau) = 0 \qquad \forall \tau \ed
\ee
This corresponds to the presence of a vacuum state which is invariant under the full $SL(2,\mathbb{Z})$, as we illustrate in the example of $N=4$ in the main text.

\section{TQFTs and topological manipulations}
\label{app: topmanip}

Here we provide some details about the topological field theories introduced in the main text as well as the non-trivial operations performed on them. 
Most of the calculations described below rely on the ability to decompose a given cocycle in terms of elements belonging to different cohomologies. Moreover, it is crucial for the consistency of the computations that the resulting expressions are themselves unrestricted sums over such cohomologies. However, this is not generally the case if one allows for an arbitrary spacetime topology. We then proceed to describe under which conditions these manipulations can be performed. 

Consider a dynamical 2-form field $b\in H^2(X,\bZ_N)$, where $X$ denotes the spacetime manifold. Take $N=k k'$ and perform the decomposition $b=k \, c + b'$. Naively, one would say that $c$ and $b'$ are arbitrary elements in $H^2(X,\bZ_{k'})$ and $H^2(X,\bZ_k)$, respectively. Instead, these two bundles are correlated by the fact that $d b$ must be trivial. To properly define this procedure consider the group extension
\be
1 \longrightarrow \bZ_{k'} \longrightarrow \bZ_N \longrightarrow \bZ_{k} \longrightarrow 1 \;.
\ee
This short exact sequence splits if and only if $\gcd(k,k')=1$. If $\gcd(k,k')\neq 1$, then consistency with the short exact sequence implies the following correlation between the bundles
\be
\label{eq:nonsplit}
dc =- \beta(b') \;,\qquad\qquad\text{where}\qquad  \beta: H^2(X,\bZ_{k}) \to H^3(X, \bZ_{k'})
\ee
is the Bockstein map. We could think of $\beta(b') \equiv b^{\prime\,*} e$ as the pull-back of an extension class $e \in H^3(B^2 \bZ_{k}, \, \bZ_{k'})$ \cite{Tachikawa:2017gyf} characterizing the bundle. 
Therefore, $b'$ is not an arbitrary element of $H^2(X, \bZ_k)$, rather it has to be an element such that $\beta(b')$ is trivial in cohomology. In other words, $b'$ is restricted to the cocycles that trivialize the extension class $e$ under the pull-back $b^{\prime\,*} e$. This would imply that $b'$ is not freely summed over, hence making it difficult to integrate it out. 

In order to highlight how this obstacle may be avoided, let us uplift $b'$ to a $\bZ$-valued cochain, such that the Bockstein map reads
\be
\beta(b') = \frac1k \, db' \;.
\ee
In order to understand if $\beta(b')$ can be non-trivial in cohomology, we integrate it on a 3-cycle $\gamma_3$. If $\partial \gamma_3 = 0$, by Stokes theorem we immediately find that the integral vanishes. However, if $H_2(X)$ has torsion, we could have $\partial \gamma_3 = n \gamma_2$ so that $\int_{\gamma_3} \beta(b') = \frac nk \int_{\gamma_2} b' \neq 0$. Thus, if $H_2(X;\bZ)$ is freely generated, this cannot happen and $\beta(b')$ is necessarily trivial in cohomology, meaning that (\ref{eq:nonsplit}) does not impose any constraint on $b'$. We will restrict to simply-connected four-manifolds, so as to guarantee that $H^2(X;\bZ)$ is freely generated.%
\footnote{A simply-connected manifold has $H_1(X; \bZ) = 0$. The universal coefficient theorem for cohomology states that there exists a short exact sequence $0 \to \operatorname{Ext}^1_\bZ\bigl( H_{i-1}(X; \bZ), \bZ \bigr) \to H^i(X; \bZ) \to \operatorname{Hom}_\bZ\bigl( H_i(X; \bZ), \bZ \bigr) \to 0$. Taking $i=2$, on a simply-connected manifold $H^2(X; \bZ) \cong \operatorname{Hom}_\bZ\bigl( H_2(X;\bZ) , \bZ \bigr)$ are isomorphic and the latter is freely generated.} 

Finally, notice that the solutions to \eqref{eq:nonsplit} for $c$ form a torsor over $H^2(X; \bZ_{k'})$. We can thus choose an arbitrary particular solution $c^*$ such that $dc^* = - \beta(b')$ and set $c = c^* + c'$ with $c' \in H^2(X; \bZ_{k'})$. In the following, with a slight abuse of notation, we will simply indicate a sum over $c$.

Now we provide a more detailed account of the manipulations presented in the main text. We start with the theories $\bZ_{N|k}^\ell$ describing vacua with partial breaking/confinement of the $\bZ_N$ 1-form symmetry. Their partition function is in (\ref{part func of ZNkl}). As mentioned in the main text, the SPT phase is evaluated in the preserved subgroup $\bZ_k$. This stems from the fact that the $\bZ_{k'}$ gauge theory has support on trivial $\bZ_{k'}$ backgrounds, \ie, its partition function is a delta function imposing $B = k' B'$ and in that case the partition function reduces to 
\be
Z[\bZ_{N | k}^\ell](k' B') = \sqrt{ \lvert H^2(X, \bZ_{k'}) \rvert} \; \exp\biggl( \frac{2 \pi i \ell}{2k} \int \fP(B') \biggr) \ec
\ee
which is a well defined $\bZ_k$ SPT phase. The extreme cases are a pure $\bZ_N$ gauge theory with partition function $Z[\bZ_N] = Z[\bZ_{N|1}^0]$, and $\text{SPT}_\ell$ with partition function $Z[\text{SPT}_\ell] = Z[\bZ_{N|N}^\ell]$.

Whenever ${\rm gcd}(\ell,k)=1$, there is an alternative expression:
\be
\label{ZNkellapp2}
Z[\bZ_{N | k}^{\ell}](B) = \frac{1}{\sqrt{ \lvert H^2(X, \bZ_N) \rvert} } \sum_{b \, \in H^2(X, \bZ_N)} \exp\biggl( -\frac{2 \pi i \ell\inv}{2 k} \int \fP(b) + \frac{2 \pi i}{N} \int b \cup B \biggr)  \;,
\ee
where $\ell\inv$ is the inverse of $\ell$ mod $k$. In order to show that the above expression equals (\ref{part func of ZNkl}), one decomposes  $b = k b' + b''$ with $b' \in H^2(X, \bZ_{k'})$ (as previously explained, we assume a simply connected spacetime). Summing over $b'$ yields a delta function $\delta_{H^2(X, \bZ_{k'})}(B)$, while completing the square and summing over $b''$ gives the same SPT phase as in (\ref{part func of ZNkl}).

Let us now describe some of the computations from Section~\ref{subsec: gappedTQFT}. In particular, the first two entries in equation \eqref{eq:gaugingTQFTs} follow almost by definition. In order to derive the third one, we need to make use of the following result:
\be
\label{Gauss}
\frac{1}{\sqrt{\lvert H^2(X, \bZ_k) \rvert}} \; \sum_{c \, \in H^2(X, \bZ_k)} \exp\biggl( \frac{2 \pi i p}{2 k} \int \fP(c) \biggr) = 1 
\ee
when $X$ is a simply-connected spin manifold \cite{Apte:2022xtu}. We therefore have, assuming $\gcd(\ell,N) = 1$:
\begin{align}
[\sigma \, \text{SPT}_\ell](B) &= \frac{1}{\sqrt{ \lvert H^2(X, \bZ_N) \rvert}} \; \sum_{b \, \in H^2(X, \bZ_N)} e^{\frac{2 \pi i \ell}{2 N} \!\int \fP(b) \, + \, \frac{2\pi  i}{N} \!\int\! b \, \cup B} \\
&= \frac{1}{\sqrt{ \lvert H^2(X, \bZ_N) \rvert }} \; \sum_{b \, \in H^2(X, \bZ_N)} e^{\frac{2 \pi i \ell}{2 N} \!\int \fP\left( b + \ell\inv B \right) \,-\, \frac{2\pi i \ell\inv}{2N} \!\int \fP(B)} = \text{SPT}_{- \ell\inv}(B) \;.\nonumber
\end{align}
To establish the last equality one uses \eqref{Gauss}. Here $\ell\inv$ is the inverse modulo $N$, which exists due to the assumption $\gcd(\ell,N)=1$. Let us introduce the generic notation $(\ell \,)^{-1}_x$ to denote the multiplicative inverse mod $x$.
When this condition fails, we define $p = \gcd(\ell,N)$ and split the dynamical field as $b=(N/p) \, c + b'$. As explained at the beginning of this section, the following manipulations are perfectly justified as long as the spacetime is restricted to a simply-connected four-manifold.
Given that, the sum over connections reads:
\begin{align}
&[\sigma \, \text{SPT}_\ell](B) = \frac{1}{\sqrt{ \lvert H^2(X, \bZ_N) \rvert}} \sum_{\substack{c \, \in H^2(X, \bZ_p) \\  b' \in H^2(X, \bZ_{N/p})}} \!\!\! \exp \biggl( \frac{2 \pi i \ell}{2 N} \!\int\! \fP(b') + \frac{2\pi i}{N} \!\int b'\cup B + \frac{2\pi i}{p} \!\int\! c \cup B \biggr) \nn \\
&= \frac{|H^2(X, \bZ_p)|}{\sqrt{|H^2(X, \bZ_N)|}} \sum_{b' \in H^2(X, \, \bZ_{N/p})} \exp \biggl( \frac{2 \pi i \ell/p}{2 N/p} \!\int\! \fP(b') + \frac{2\pi i}{N/p}\int b'\cup B/p \biggr) \, \delta_{H^2(X, \bZ_{p})}(B) \nn \\
&= \exp \biggl( -\frac{2\pi i (\ell/p)\inv_{N/p}}{2Np} \int\! \fP(B) \biggr) \, Z[ \bZ_p ](B) = Z \biggl[ \bZ_{N\vert N/p}^{(\ell/p)_{N/p}\inv} \biggr](B) \;,
\end{align}
thus leading to the last entry in \eqref{eq:gaugingTQFTs}. In the first line we used the fact that $\fP$ is a quadratic form, $\fP(A + B) = \fP(A) + \fP(B) + 2 A \cup B$, to drop the $c$ dependence. In the second line we performed the sum over $c$ and dropped terms which are trivial on the support of the delta function. Finally setting $B= p B'$ to solve the constraint, we performed the final sum over $b'$ using \eqref{Gauss} and reinstated the old variable $B$.

Let us now consider the general case and prove \eqref{sigmaZNkell}. Setting $N=kk'$, $p={\rm gcd}(\ell,k)$, we want to compute
\begin{align}
\sigma \, Z\Bigl[ \bZ^{(\ell)}_{N\vert k} \Bigr](B) &= \sqrt{ \frac{ |H^2(X, \bZ_{k'})| }{ |H^2(X, \bZ_N)| }} \sum_{ c \, \in \, H^2(X , \bZ_N) } \!\! \exp \biggl( \frac{2 \pi i \ell}{2 Nk'} \int \fP(c) + \frac{2\pi i}{N} \!\int\! c \cup B \biggr) \, \delta_{H^2(X, \bZ_{k'})}(c) \nn \\
&= \sqrt{ \frac{ |H^2(X, \bZ_{k'})| }{ |H^2(X, \bZ_N)| }} \sum_{ \tilde c \, \in \, H^2(X , \bZ_{k}) } \!\! \exp \biggl( \frac{2 \pi i \ell}{k} \!\int\! \fP(\tilde c) + \frac{2 \pi i}{k} \!\int\! \tilde c \cup B \biggr) \;.
\label{eqn C13}
\end{align}
To go to the second line we solved the delta function constraint by $c= k' \tilde c$. We are now in a familiar situation and we further decompose $\tilde c=(k/p)c'+c''$ so that $c'$ appears only linearly and gives rise to a $\bZ_p$ delta function enforcing $B=pB'$:
\begin{align}
(\ref{eqn C13}) &= \frac{ |H^2(X, \bZ_p)| }{ \sqrt{|H^2(X, \bZ_k)|}} \sum_{c'' \in H^2(X, \, \bZ_{k/p})} \exp \biggl( \frac{2 \pi i \ell/p}{k/p} \int \fP(c'') + \frac{ 2 \pi i}{k/p} \int c'' \cup B' \biggr) \nn \\
&= \sqrt{ |H^2(X, \bZ_p)| } \exp \biggl( - \frac{2 \pi i (\ell/p)^{-1}_{k/p}}{k/p} \int \fP(B') \biggr) \;.
\end{align}
Here we used that $H^2(X)$ is torsion-free and the sum \eqref{Gauss} to rearrange the prefactors. We can now reinstate the variable $B = p B'$ and the $\bZ_p$ gauge-theory factor explicitly to find: 
\be
\sigma \, Z \Bigl[ \bZ^{(\ell)}_{N\vert k} \Bigr] (B)  = \exp \biggl( -\frac{2\pi i k'(\ell/p)_{k/p}\inv}{2Np} \int \fP(B) \biggr) \, Z[\bZ_p ](B) = Z\Bigl[ \bZ_{N\vert N/p}^{\tilde \ell} \Bigr](B)
\ee
with $\tilde\ell \equiv -k'(\ell/p)\inv_{k/p}$, thus proving equation \eqref{sigmaZNkell}.

\bibliographystyle{ytphys}
\baselineskip=0.85\baselineskip
\bibliography{DualityFlows}

\providecommand{\href}[2]{#2}\begingroup\raggedright\begin{thebibliography}{10}

\bibitem{Gaiotto:2014kfa}
D.~Gaiotto, A.~Kapustin, N.~Seiberg, and B.~Willett, ``{Generalized Global
  Symmetries},'' \href{http://dx.doi.org/10.1007/JHEP02(2015)172}{{\em JHEP}
  {\bfseries 02} (2015) 172}, \href{http://arxiv.org/abs/1412.5148}{{\ttfamily
  arXiv:1412.5148 [hep-th]}}.

\bibitem{Verlinde:1988sn}
E.~P. Verlinde, ``{Fusion Rules and Modular Transformations in 2D Conformal
  Field Theory},'' \href{http://dx.doi.org/10.1016/0550-3213(88)90603-7}{{\em
  Nucl. Phys. B} {\bfseries 300} (1988) 360--376}.

\bibitem{Petkova:2000ip}
V.~B. Petkova and J.~B. Zuber, ``{Generalized twisted partition functions},''
  \href{http://dx.doi.org/10.1016/S0370-2693(01)00276-3}{{\em Phys. Lett. B}
  {\bfseries 504} (2001) 157--164},
  \href{http://arxiv.org/abs/hep-th/0011021}{{\ttfamily arXiv:hep-th/0011021}}.

\bibitem{Witten:1988hf}
E.~Witten, ``{Quantum Field Theory and the Jones Polynomial},''
  \href{http://dx.doi.org/10.1007/BF01217730}{{\em Commun. Math. Phys.}
  {\bfseries 121} (1989) 351--399}.

\bibitem{Barkeshli:2014cna}
M.~Barkeshli, P.~Bonderson, M.~Cheng, and Z.~Wang, ``{Symmetry
  Fractionalization, Defects, and Gauging of Topological Phases},''
  \href{http://dx.doi.org/10.1103/PhysRevB.100.115147}{{\em Phys. Rev. B}
  {\bfseries 100} (2019) 115147},
  \href{http://arxiv.org/abs/1410.4540}{{\ttfamily arXiv:1410.4540
  [cond-mat.str-el]}}.

\bibitem{Chang:2018iay}
C.-M. Chang, Y.-H. Lin, S.-H. Shao, Y.~Wang, and X.~Yin, ``{Topological Defect
  Lines and Renormalization Group Flows in Two Dimensions},''
  \href{http://dx.doi.org/10.1007/JHEP01(2019)026}{{\em JHEP} {\bfseries 01}
  (2019) 026}, \href{http://arxiv.org/abs/1802.04445}{{\ttfamily
  arXiv:1802.04445 [hep-th]}}.

\bibitem{Thorngren:2019iar}
R.~Thorngren and Y.~Wang, ``{Fusion Category Symmetry I: Anomaly In-Flow and
  Gapped Phases},'' \href{http://arxiv.org/abs/1912.02817}{{\ttfamily
  arXiv:1912.02817 [hep-th]}}.

\bibitem{Gaiotto:2020iye}
D.~Gaiotto and J.~Kulp, ``{Orbifold groupoids},''
  \href{http://dx.doi.org/10.1007/JHEP02(2021)132}{{\em JHEP} {\bfseries 02}
  (2021) 132}, \href{http://arxiv.org/abs/2008.05960}{{\ttfamily
  arXiv:2008.05960 [hep-th]}}.

\bibitem{Komargodski:2020mxz}
Z.~Komargodski, K.~Ohmori, K.~Roumpedakis, and S.~Seifnashri, ``{Symmetries and
  strings of adjoint QCD$_{2}$},''
  \href{http://dx.doi.org/10.1007/JHEP03(2021)103}{{\em JHEP} {\bfseries 03}
  (2021) 103}, \href{http://arxiv.org/abs/2008.07567}{{\ttfamily
  arXiv:2008.07567 [hep-th]}}.

\bibitem{Thorngren:2021yso}
R.~Thorngren and Y.~Wang, ``{Fusion Category Symmetry II: Categoriosities at
  $c=1$ and Beyond},'' \href{http://arxiv.org/abs/2106.12577}{{\ttfamily
  arXiv:2106.12577 [hep-th]}}.

\bibitem{Huang:2021zvu}
T.-C. Huang, Y.-H. Lin, and S.~Seifnashri, ``{Construction of two-dimensional
  topological field theories with non-invertible symmetries},''
  \href{http://dx.doi.org/10.1007/JHEP12(2021)028}{{\em JHEP} {\bfseries 12}
  (2021) 028}, \href{http://arxiv.org/abs/2110.02958}{{\ttfamily
  arXiv:2110.02958 [hep-th]}}.

\bibitem{Burbano:2021loy}
I.~M. Burbano, J.~Kulp, and J.~Neuser, ``{Duality defects in E$_{8}$},''
  \href{http://dx.doi.org/10.1007/JHEP10(2022)187}{{\em JHEP} {\bfseries 10}
  (2022) 186}, \href{http://arxiv.org/abs/2112.14323}{{\ttfamily
  arXiv:2112.14323 [hep-th]}}.

\bibitem{Lin:2023uvm}
Y.-H. Lin and S.-H. Shao, ``{Bootstrapping noninvertible symmetries},''
  \href{http://dx.doi.org/10.1103/PhysRevD.107.125025}{{\em Phys. Rev. D}
  {\bfseries 107} (2023) 125025},
  \href{http://arxiv.org/abs/2302.13900}{{\ttfamily arXiv:2302.13900
  [hep-th]}}.

\bibitem{Choi:2021kmx}
Y.~Choi, C.~Cordova, P.-S. Hsin, H.~T. Lam, and S.-H. Shao, ``{Noninvertible
  duality defects in 3+1 dimensions},''
  \href{http://dx.doi.org/10.1103/PhysRevD.105.125016}{{\em Phys. Rev. D}
  {\bfseries 105} (2022) 125016},
  \href{http://arxiv.org/abs/2111.01139}{{\ttfamily arXiv:2111.01139
  [hep-th]}}.

\bibitem{Kaidi:2021xfk}
J.~Kaidi, K.~Ohmori, and Y.~Zheng, ``{Kramers-Wannier-like Duality Defects in
  (3+1)D Gauge Theories},''
  \href{http://dx.doi.org/10.1103/PhysRevLett.128.111601}{{\em Phys. Rev.
  Lett.} {\bfseries 128} (2022) 111601},
  \href{http://arxiv.org/abs/2111.01141}{{\ttfamily arXiv:2111.01141
  [hep-th]}}.

\bibitem{Apruzzi:2021nmk}
F.~Apruzzi, F.~Bonetti, I.~Garc\'\i{}a~Etxebarria, S.~S. Hosseini, and
  S.~Schafer-Nameki, ``{Symmetry TFTs from String Theory},''
  \href{http://dx.doi.org/10.1007/s00220-023-04737-2}{{\em Commun. Math. Phys.}
  {\bfseries 402} (2023) 895--949},
  \href{http://arxiv.org/abs/2112.02092}{{\ttfamily arXiv:2112.02092
  [hep-th]}}.

\bibitem{Bhardwaj:2022yxj}
L.~Bhardwaj, L.~E. Bottini, S.~Schafer-Nameki, and A.~Tiwari, ``{Non-invertible
  higher-categorical symmetries},''
  \href{http://dx.doi.org/10.21468/SciPostPhys.14.1.007}{{\em SciPost Phys.}
  {\bfseries 14} (2023) 007}, \href{http://arxiv.org/abs/2204.06564}{{\ttfamily
  arXiv:2204.06564 [hep-th]}}.

\bibitem{Hayashi:2022fkw}
Y.~Hayashi and Y.~Tanizaki, ``{Non-invertible self-duality defects of
  Cardy-Rabinovici model and mixed gravitational anomaly},''
  \href{http://dx.doi.org/10.1007/JHEP08(2022)036}{{\em JHEP} {\bfseries 08}
  (2022) 036}, \href{http://arxiv.org/abs/2204.07440}{{\ttfamily
  arXiv:2204.07440 [hep-th]}}.

\bibitem{Choi:2022zal}
Y.~Choi, C.~Cordova, P.-S. Hsin, H.~T. Lam, and S.-H. Shao, ``{Non-invertible
  Condensation, Duality, and Triality Defects in 3+1 Dimensions},''
  \href{http://dx.doi.org/10.1007/s00220-023-04727-4}{{\em Commun. Math. Phys.}
  {\bfseries 402} (2023) 489--542},
  \href{http://arxiv.org/abs/2204.09025}{{\ttfamily arXiv:2204.09025
  [hep-th]}}.

\bibitem{Kaidi:2022uux}
J.~Kaidi, G.~Zafrir, and Y.~Zheng, ``{Non-invertible symmetries of
  $\mathcal{N}{=}4$ SYM and twisted compactification},''
  \href{http://dx.doi.org/10.1007/JHEP08(2022)053}{{\em JHEP} {\bfseries 08}
  (2022) 053}, \href{http://arxiv.org/abs/2205.01104}{{\ttfamily
  arXiv:2205.01104 [hep-th]}}.

\bibitem{Choi:2022jqy}
Y.~Choi, H.~T. Lam, and S.-H. Shao, ``{Noninvertible Global Symmetries in the
  Standard Model},''
  \href{http://dx.doi.org/10.1103/PhysRevLett.129.161601}{{\em Phys. Rev.
  Lett.} {\bfseries 129} (2022) 161601},
  \href{http://arxiv.org/abs/2205.05086}{{\ttfamily arXiv:2205.05086
  [hep-th]}}.

\bibitem{Cordova:2022ieu}
C.~C\'{o}rdova and K.~Ohmori, ``{Noninvertible Chiral Symmetry and Exponential
  Hierarchies},'' \href{http://dx.doi.org/10.1103/PhysRevX.13.011034}{{\em
  Phys. Rev. X} {\bfseries 13} (2023) 011034},
  \href{http://arxiv.org/abs/2205.06243}{{\ttfamily arXiv:2205.06243
  [hep-th]}}.

\bibitem{Antinucci:2022eat}
A.~Antinucci, G.~Galati, and G.~Rizi, ``{On continuous 2-category symmetries
  and Yang-Mills theory},''
  \href{http://dx.doi.org/10.1007/JHEP12(2022)061}{{\em JHEP} {\bfseries 12}
  (2022) 061}, \href{http://arxiv.org/abs/2206.05646}{{\ttfamily
  arXiv:2206.05646 [hep-th]}}.

\bibitem{Damia:2022rxw}
J.~A. Damia, R.~Argurio, and L.~Tizzano, ``{Continuous Generalized Symmetries
  in Three Dimensions},'' \href{http://dx.doi.org/10.1007/JHEP05(2023)164}{{\em
  JHEP} {\bfseries 23} (2023) 164},
  \href{http://arxiv.org/abs/2206.14093}{{\ttfamily arXiv:2206.14093
  [hep-th]}}.

\bibitem{Damia:2022bcd}
J.~A. Damia, R.~Argurio, and E.~Garc\'{i}a-Valdecasas, ``{Non-invertible
  defects in 5d, boundaries and holography},''
  \href{http://dx.doi.org/10.21468/SciPostPhys.14.4.067}{{\em SciPost Phys.}
  {\bfseries 14} (2023) 067}, \href{http://arxiv.org/abs/2207.02831}{{\ttfamily
  arXiv:2207.02831 [hep-th]}}.

\bibitem{Bhardwaj:2022lsg}
L.~Bhardwaj, S.~Schafer-Nameki, and J.~Wu, ``{Universal Non-Invertible
  Symmetries},'' \href{http://dx.doi.org/10.1002/prop.202200143}{{\em Fortsch.
  Phys.} {\bfseries 70} (2022) 2200143},
  \href{http://arxiv.org/abs/2208.05973}{{\ttfamily arXiv:2208.05973
  [hep-th]}}.

\bibitem{Bartsch:2022mpm}
T.~Bartsch, M.~Bullimore, A.~E.~V. Ferrari, and J.~Pearson, ``{Non-invertible
  Symmetries and Higher Representation Theory I},''
  \href{http://arxiv.org/abs/2208.05993}{{\ttfamily arXiv:2208.05993
  [hep-th]}}.

\bibitem{Apruzzi:2022rei}
F.~Apruzzi, I.~Bah, F.~Bonetti, and S.~Schafer-Nameki, ``{Noninvertible
  Symmetries from Holography and Branes},''
  \href{http://dx.doi.org/10.1103/PhysRevLett.130.121601}{{\em Phys. Rev.
  Lett.} {\bfseries 130} (2023) 121601},
  \href{http://arxiv.org/abs/2208.07373}{{\ttfamily arXiv:2208.07373
  [hep-th]}}.

\bibitem{Mekareeya:2022spm}
N.~Mekareeya and M.~Sacchi, ``{Mixed anomalies, two-groups, non-invertible
  symmetries, and 3d superconformal indices},''
  \href{http://dx.doi.org/10.1007/JHEP01(2023)115}{{\em JHEP} {\bfseries 01}
  (2023) 115}, \href{http://arxiv.org/abs/2210.02466}{{\ttfamily
  arXiv:2210.02466 [hep-th]}}.

\bibitem{Chen:2022cyw}
S.~Chen and Y.~Tanizaki, ``{Solitonic symmetry beyond homotopy: invertibility
  from bordism and noninvertibility from topological quantum field theory},''
  \href{http://dx.doi.org/10.1103/PhysRevLett.131.011602}{{\em Phys. Rev.
  Lett.} {\bfseries 131} (2023) 011602},
  \href{http://arxiv.org/abs/2210.13780}{{\ttfamily arXiv:2210.13780
  [hep-th]}}.

\bibitem{Choi:2022fgx}
Y.~Choi, H.~T. Lam, and S.-H. Shao, ``{Non-invertible Gauss law and axions},''
  \href{http://dx.doi.org/10.1007/JHEP09(2023)067}{{\em JHEP} {\bfseries 09}
  (2023) 067}, \href{http://arxiv.org/abs/2212.04499}{{\ttfamily
  arXiv:2212.04499 [hep-th]}}.

\bibitem{Yokokura:2022alv}
R.~Yokokura, ``{Non-invertible symmetries in axion electrodynamics},''
  \href{http://arxiv.org/abs/2212.05001}{{\ttfamily arXiv:2212.05001
  [hep-th]}}.

\bibitem{Bhardwaj:2022kot}
L.~Bhardwaj, S.~Schafer-Nameki, and A.~Tiwari, ``{Unifying Constructions of
  Non-Invertible Symmetries},''
  \href{http://arxiv.org/abs/2212.06159}{{\ttfamily arXiv:2212.06159
  [hep-th]}}.

\bibitem{Bhardwaj:2022maz}
L.~Bhardwaj, L.~E. Bottini, S.~Schafer-Nameki, and A.~Tiwari, ``{Non-Invertible
  Symmetry Webs},'' \href{http://arxiv.org/abs/2212.06842}{{\ttfamily
  arXiv:2212.06842 [hep-th]}}.

\bibitem{Bartsch:2022ytj}
T.~Bartsch, M.~Bullimore, A.~E.~V. Ferrari, and J.~Pearson, ``{Non-invertible
  Symmetries and Higher Representation Theory II},''
  \href{http://arxiv.org/abs/2212.07393}{{\ttfamily arXiv:2212.07393
  [hep-th]}}.

\bibitem{Heckman:2022xgu}
J.~J. Heckman, M.~Hubner, E.~Torres, X.~Yu, and H.~Y. Zhang, ``{Top down
  approach to topological duality defects},''
  \href{http://dx.doi.org/10.1103/PhysRevD.108.046015}{{\em Phys. Rev. D}
  {\bfseries 108} (2023) 046015},
  \href{http://arxiv.org/abs/2212.09743}{{\ttfamily arXiv:2212.09743
  [hep-th]}}.

\bibitem{Montonen:1977sn}
C.~Montonen and D.~I. Olive, ``{Magnetic Monopoles as Gauge Particles?},''
  \href{http://dx.doi.org/10.1016/0370-2693(77)90076-4}{{\em Phys. Lett. B}
  {\bfseries 72} (1977) 117--120}.

\bibitem{Aharony:2013hda}
O.~Aharony, N.~Seiberg, and Y.~Tachikawa, ``{Reading between the lines of
  four-dimensional gauge theories},''
  \href{http://dx.doi.org/10.1007/JHEP08(2013)115}{{\em JHEP} {\bfseries 08}
  (2013) 115}, \href{http://arxiv.org/abs/1305.0318}{{\ttfamily arXiv:1305.0318
  [hep-th]}}.

\bibitem{Roumpedakis:2022aik}
K.~Roumpedakis, S.~Seifnashri, and S.-H. Shao, ``{Higher Gauging and
  Non-invertible Condensation Defects},''
  \href{http://dx.doi.org/10.1007/s00220-023-04706-9}{{\em Commun. Math. Phys.}
  {\bfseries 401} (2023) 3043--3107},
  \href{http://arxiv.org/abs/2204.02407}{{\ttfamily arXiv:2204.02407
  [hep-th]}}.

\bibitem{Intriligator:1998ig}
K.~A. Intriligator, ``{Bonus symmetries of $\mathcal{N}{=}4$ superYang-Mills
  correlation functions via AdS duality},''
  \href{http://dx.doi.org/10.1016/S0550-3213(99)00242-4}{{\em Nucl. Phys. B}
  {\bfseries 551} (1999) 575--600},
  \href{http://arxiv.org/abs/hep-th/9811047}{{\ttfamily arXiv:hep-th/9811047}}.

\bibitem{Kapustin:2006pk}
A.~Kapustin and E.~Witten, ``{Electric-Magnetic Duality And The Geometric
  Langlands Program},''
  \href{http://dx.doi.org/10.4310/CNTP.2007.v1.n1.a1}{{\em Commun. Num. Theor.
  Phys.} {\bfseries 1} (2007) 1--236},
  \href{http://arxiv.org/abs/hep-th/0604151}{{\ttfamily arXiv:hep-th/0604151}}.

\bibitem{Argyres:1999xu}
P.~C. Argyres, K.~A. Intriligator, R.~G. Leigh, and M.~J. Strassler, ``{On
  inherited duality in $\mathcal{N}{=}1$ $d{=}4$ supersymmetric gauge
  theories},'' \href{http://dx.doi.org/10.1088/1126-6708/2000/04/029}{{\em
  JHEP} {\bfseries 04} (2000) 029},
  \href{http://arxiv.org/abs/hep-th/9910250}{{\ttfamily arXiv:hep-th/9910250}}.

\bibitem{Garcia-Etxebarria:2015wns}
I.~Garc\'{i}a~Etxebarria and D.~Regalado, ``{$\mathcal{N}{=}3 $
  four-dimensional field theories},''
  \href{http://dx.doi.org/10.1007/JHEP03(2016)083}{{\em JHEP} {\bfseries 03}
  (2016) 083}, \href{http://arxiv.org/abs/1512.06434}{{\ttfamily
  arXiv:1512.06434 [hep-th]}}.

\bibitem{Aharony:2016kai}
O.~Aharony, Y.~Tachikawa, and K.~Gomi, ``{S-folds and 4d $\mathcal{N}{=}3$
  superconformal field theories},''
  \href{http://dx.doi.org/10.1007/JHEP06(2016)044}{{\em JHEP} {\bfseries 06}
  (2016) 044}, \href{http://arxiv.org/abs/1602.08638}{{\ttfamily
  arXiv:1602.08638 [hep-th]}}.

\bibitem{Argyres:2016yzz}
P.~C. Argyres and M.~Martone, ``{4d $\mathcal{N}{=}2$ theories with
  disconnected gauge groups},''
  \href{http://dx.doi.org/10.1007/JHEP03(2017)145}{{\em JHEP} {\bfseries 03}
  (2017) 145}, \href{http://arxiv.org/abs/1611.08602}{{\ttfamily
  arXiv:1611.08602 [hep-th]}}.

\bibitem{Donagi:1995cf}
R.~Donagi and E.~Witten, ``{Supersymmetric Yang-Mills theory and integrable
  systems},'' \href{http://dx.doi.org/10.1016/0550-3213(95)00609-5}{{\em Nucl.
  Phys. B} {\bfseries 460} (1996) 299--334},
  \href{http://arxiv.org/abs/hep-th/9510101}{{\ttfamily arXiv:hep-th/9510101}}.

\bibitem{Dorey:1999sj}
N.~Dorey, ``{An elliptic superpotential for softly broken $\mathcal{N}{=}4$
  supersymmetric Yang-Mills theory},''
  \href{http://dx.doi.org/10.1088/1126-6708/1999/07/021}{{\em JHEP} {\bfseries
  07} (1999) 021}, \href{http://arxiv.org/abs/hep-th/9906011}{{\ttfamily
  arXiv:hep-th/9906011}}.

\bibitem{Dorey:2000fc}
N.~Dorey and S.~P. Kumar, ``{Softly broken $\mathcal{N}{=}4$ supersymmetry in
  the large $N$ limit},''
  \href{http://dx.doi.org/10.1088/1126-6708/2000/02/006}{{\em JHEP} {\bfseries
  02} (2000) 006}, \href{http://arxiv.org/abs/hep-th/0001103}{{\ttfamily
  arXiv:hep-th/0001103}}.

\bibitem{Dorey:2001qj}
N.~Dorey, T.~J. Hollowood, and S.~P. Kumar, ``{An exact elliptic superpotential
  for $\mathcal{N}{=}1^*$ deformations of finite $\mathcal{N}{=}2$ gauge
  theories},'' \href{http://dx.doi.org/10.1016/S0550-3213(01)00647-2}{{\em
  Nucl. Phys. B} {\bfseries 624} (2002) 95--145},
  \href{http://arxiv.org/abs/hep-th/0108221}{{\ttfamily arXiv:hep-th/0108221}}.

\bibitem{Hsieh:2019iba}
C.-T. Hsieh, Y.~Tachikawa, and K.~Yonekura, ``{Anomaly of the Electromagnetic
  Duality of Maxwell Theory},''
  \href{http://dx.doi.org/10.1103/PhysRevLett.123.161601}{{\em Phys. Rev.
  Lett.} {\bfseries 123} (2019) 161601},
  \href{http://arxiv.org/abs/1905.08943}{{\ttfamily arXiv:1905.08943
  [hep-th]}}.

\bibitem{Hsieh:2020jpj}
C.-T. Hsieh, Y.~Tachikawa, and K.~Yonekura, ``{Anomaly Inflow and $p$-Form
  Gauge Theories},'' \href{http://dx.doi.org/10.1007/s00220-022-04333-w}{{\em
  Commun. Math. Phys.} {\bfseries 391} (2022) 495--608},
  \href{http://arxiv.org/abs/2003.11550}{{\ttfamily arXiv:2003.11550
  [hep-th]}}.

\bibitem{Debray:2023yrs}
A.~Debray, M.~Dierigl, J.~J. Heckman, and M.~Montero, ``{The Chronicles of
  IIBordia: Dualities, Bordisms, and the Swampland},''
  \href{http://arxiv.org/abs/2302.00007}{{\ttfamily arXiv:2302.00007
  [hep-th]}}.

\bibitem{Bashmakov:2022jtl}
V.~Bashmakov, M.~Del~Zotto, and A.~Hasan, ``{On the 6d origin of non-invertible
  symmetries in 4d},'' \href{http://dx.doi.org/10.1007/JHEP09(2023)161}{{\em
  JHEP} {\bfseries 09} (2023) 161},
  \href{http://arxiv.org/abs/2206.07073}{{\ttfamily arXiv:2206.07073
  [hep-th]}}.

\bibitem{Bashmakov:2022uek}
V.~Bashmakov, M.~Del~Zotto, A.~Hasan, and J.~Kaidi, ``{Non-invertible
  symmetries of class $\mathcal{S}$ theories},''
  \href{http://dx.doi.org/10.1007/JHEP05(2023)225}{{\em JHEP} {\bfseries 05}
  (2023) 225}, \href{http://arxiv.org/abs/2211.05138}{{\ttfamily
  arXiv:2211.05138 [hep-th]}}.

\bibitem{Antinucci:2022cdi}
A.~Antinucci, C.~Copetti, G.~Galati, and G.~Rizi, ``{``Zoology'' of
  non-invertible duality defects: the view from class $\mathcal{S}$},''
  \href{http://arxiv.org/abs/2212.09549}{{\ttfamily arXiv:2212.09549
  [hep-th]}}.

\bibitem{Carta:2023bqn}
F.~Carta, S.~Giacomelli, N.~Mekareeya, and A.~Mininno, ``{Comments on
  Non-invertible Symmetries in Argyres-Douglas Theories},''
  \href{http://dx.doi.org/10.1007/JHEP07(2023)135}{{\em JHEP} {\bfseries 07}
  (2023) 135}, \href{http://arxiv.org/abs/2303.16216}{{\ttfamily
  arXiv:2303.16216 [hep-th]}}.

\bibitem{Klebanov:1998hh}
I.~R. Klebanov and E.~Witten, ``{Superconformal field theory on three-branes at
  a Calabi-Yau singularity},''
  \href{http://dx.doi.org/10.1016/S0550-3213(98)00654-3}{{\em Nucl. Phys. B}
  {\bfseries 536} (1998) 199--218},
  \href{http://arxiv.org/abs/hep-th/9807080}{{\ttfamily arXiv:hep-th/9807080}}.

\bibitem{DelZotto:2015isa}
M.~Del~Zotto, J.~J. Heckman, D.~S. Park, and T.~Rudelius, ``{On the Defect
  Group of a 6D SCFT},''
  \href{http://dx.doi.org/10.1007/s11005-016-0839-5}{{\em Lett. Math. Phys.}
  {\bfseries 106} (2016) 765--786},
  \href{http://arxiv.org/abs/1503.04806}{{\ttfamily arXiv:1503.04806
  [hep-th]}}.

\bibitem{Bhardwaj:2021pfz}
L.~Bhardwaj, M.~Hubner, and S.~Schafer-Nameki, ``{1-form Symmetries of 4d
  $\mathcal{N}{=}2$ Class $\mathcal{S}$ Theories},''
  \href{http://dx.doi.org/10.21468/SciPostPhys.11.5.096}{{\em SciPost Phys.}
  {\bfseries 11} (2021) 096}, \href{http://arxiv.org/abs/2102.01693}{{\ttfamily
  arXiv:2102.01693 [hep-th]}}.

\bibitem{Bhardwaj:2021mzl}
L.~Bhardwaj, S.~Giacomelli, M.~H\"ubner, and S.~Sch\"afer-Nameki, ``{Relative
  defects in relative theories: Trapped higher-form symmetries and irregular
  punctures in class $\mathcal{S}$},''
  \href{http://dx.doi.org/10.21468/SciPostPhys.13.4.101}{{\em SciPost Phys.}
  {\bfseries 13} (2022) 101}, \href{http://arxiv.org/abs/2201.00018}{{\ttfamily
  arXiv:2201.00018 [hep-th]}}.

\bibitem{Witten:1979ey}
E.~Witten, ``{Dyons of Charge $e \theta/2 \pi$},''
  \href{http://dx.doi.org/10.1016/0370-2693(79)90838-4}{{\em Phys. Lett. B}
  {\bfseries 86} (1979) 283--287}.

\bibitem{Bergman:2022otk}
O.~Bergman and S.~Hirano, ``{The holography of duality in $\mathcal{N}{=}4$
  Super-Yang-Mills theory},''
  \href{http://dx.doi.org/10.1007/JHEP11(2022)069}{{\em JHEP} {\bfseries 11}
  (2022) 069}, \href{http://arxiv.org/abs/2208.09396}{{\ttfamily
  arXiv:2208.09396 [hep-th]}}.

\bibitem{Cordova:2019uob}
C.~C\'ordova, D.~S. Freed, H.~T. Lam, and N.~Seiberg, ``{Anomalies in the space
  of coupling constants and their dynamical applications II},''
  \href{http://dx.doi.org/10.21468/SciPostPhys.8.1.002}{{\em SciPost Phys.}
  {\bfseries 8} (2020) 002}, \href{http://arxiv.org/abs/1905.13361}{{\ttfamily
  arXiv:1905.13361 [hep-th]}}.

\bibitem{Ang:2019txy}
J.~P. Ang, K.~Roumpedakis, and S.~Seifnashri, ``{Line Operators of Gauge
  Theories on Non-Spin Manifolds},''
  \href{http://dx.doi.org/10.1007/JHEP04(2020)087}{{\em JHEP} {\bfseries 04}
  (2020) 087}, \href{http://arxiv.org/abs/1911.00589}{{\ttfamily
  arXiv:1911.00589 [hep-th]}}.

\bibitem{Bhardwaj:2020ymp}
L.~Bhardwaj, Y.~Lee, and Y.~Tachikawa, ``{$SL(2,\mathbb{Z})$ action on QFTs
  with $\mathbb{Z}_2$ symmetry and the Brown-Kervaire invariants},''
  \href{http://dx.doi.org/10.1007/JHEP11(2020)141}{{\em JHEP} {\bfseries 11}
  (2020) 141}, \href{http://arxiv.org/abs/2009.10099}{{\ttfamily
  arXiv:2009.10099 [hep-th]}}.

\bibitem{Leigh:1995ep}
R.~G. Leigh and M.~J. Strassler, ``{Exactly marginal operators and duality in
  four-dimensional $\mathcal{N}{=}1$ supersymmetric gauge theory},''
  \href{http://dx.doi.org/10.1016/0550-3213(95)00261-P}{{\em Nucl. Phys. B}
  {\bfseries 447} (1995) 95--136},
  \href{http://arxiv.org/abs/hep-th/9503121}{{\ttfamily arXiv:hep-th/9503121}}.

\bibitem{Beem:2013hha}
C.~Beem, L.~Rastelli, A.~Sen, and B.~C. van Rees, ``{Resummation and S-duality
  in $\mathcal{N}{=}4$ SYM},''
  \href{http://dx.doi.org/10.1007/JHEP04(2014)122}{{\em JHEP} {\bfseries 04}
  (2014) 122}, \href{http://arxiv.org/abs/1306.3228}{{\ttfamily arXiv:1306.3228
  [hep-th]}}.

\bibitem{Polchinski:2000uf}
J.~Polchinski and M.~J. Strassler, ``{The String dual of a confining
  four-dimensional gauge theory},''
  \href{http://arxiv.org/abs/hep-th/0003136}{{\ttfamily arXiv:hep-th/0003136}}.

\bibitem{Kaidi:2022cpf}
J.~Kaidi, K.~Ohmori, and Y.~Zheng, ``{Symmetry TFTs for Non-Invertible
  Defects},'' \href{http://arxiv.org/abs/2209.11062}{{\ttfamily
  arXiv:2209.11062 [hep-th]}}.

\bibitem{Antinucci:2022vyk}
A.~Antinucci, F.~Benini, C.~Copetti, G.~Galati, and G.~Rizi, ``{The holography
  of non-invertible self-duality symmetries},''
  \href{http://arxiv.org/abs/2210.09146}{{\ttfamily arXiv:2210.09146
  [hep-th]}}.

\bibitem{Apte:2022xtu}
A.~Apte, C.~Cordova, and H.~T. Lam, ``{Obstructions to gapped phases from
  noninvertible symmetries},''
  \href{http://dx.doi.org/10.1103/PhysRevB.108.045134}{{\em Phys. Rev. B}
  {\bfseries 108} (2023) 045134},
  \href{http://arxiv.org/abs/2212.14605}{{\ttfamily arXiv:2212.14605
  [hep-th]}}.

\bibitem{Antinucci:2023ezl}
A.~Antinucci, F.~Benini, C.~Copetti, G.~Galati, and G.~Rizi, ``{Anomalies of
  non-invertible self-duality symmetries: fractionalization and gauging},''
  \href{http://arxiv.org/abs/2308.11707}{{\ttfamily arXiv:2308.11707
  [hep-th]}}.

\bibitem{Benini:2018reh}
F.~Benini, C.~C\'ordova, and P.-S. Hsin, ``{On 2-Group Global Symmetries and
  their Anomalies},'' \href{http://dx.doi.org/10.1007/JHEP03(2019)118}{{\em
  JHEP} {\bfseries 03} (2019) 118},
  \href{http://arxiv.org/abs/1803.09336}{{\ttfamily arXiv:1803.09336
  [hep-th]}}.

\bibitem{Tachikawa:2017gyf}
Y.~Tachikawa, ``{On gauging finite subgroups},''
  \href{http://dx.doi.org/10.21468/SciPostPhys.8.1.015}{{\em SciPost Phys.}
  {\bfseries 8} (2020) 015}, \href{http://arxiv.org/abs/1712.09542}{{\ttfamily
  arXiv:1712.09542 [hep-th]}}.

\bibitem{Niro:2022ctq}
P.~Niro, K.~Roumpedakis, and O.~Sela, ``{Exploring non-invertible symmetries in
  free theories},'' \href{http://dx.doi.org/10.1007/JHEP03(2023)005}{{\em JHEP}
  {\bfseries 03} (2023) 005}, \href{http://arxiv.org/abs/2209.11166}{{\ttfamily
  arXiv:2209.11166 [hep-th]}}.

\bibitem{Kaidi:2023maf}
J.~Kaidi, E.~Nardoni, G.~Zafrir, and Y.~Zheng, ``{Symmetry TFTs and Anomalies
  of Non-Invertible Symmetries},''
  \href{http://arxiv.org/abs/2301.07112}{{\ttfamily arXiv:2301.07112
  [hep-th]}}.

\bibitem{Witten:1997sc}
E.~Witten, ``{Solutions of four-dimensional field theories via M-theory},''
  \href{http://dx.doi.org/10.1016/S0550-3213(97)00416-1}{{\em Nucl. Phys. B}
  {\bfseries 500} (1997) 3--42},
  \href{http://arxiv.org/abs/hep-th/9703166}{{\ttfamily arXiv:hep-th/9703166}}.

\bibitem{Halmagyi:2004ju}
N.~Halmagyi, C.~Romelsberger, and N.~P. Warner, ``{Inherited duality and quiver
  gauge theory},'' \href{http://dx.doi.org/10.4310/ATMP.2006.v10.n2.a1}{{\em
  Adv. Theor. Math. Phys.} {\bfseries 10} (2006) 159--179},
  \href{http://arxiv.org/abs/hep-th/0406143}{{\ttfamily arXiv:hep-th/0406143}}.

\bibitem{Diaconescu:1997br}
D.-E. Diaconescu, M.~R. Douglas, and J.~Gomis, ``{Fractional branes and wrapped
  branes},'' \href{http://dx.doi.org/10.1088/1126-6708/1998/02/013}{{\em JHEP}
  {\bfseries 02} (1998) 013},
  \href{http://arxiv.org/abs/hep-th/9712230}{{\ttfamily arXiv:hep-th/9712230}}.

\bibitem{Bertolini:2000dk}
M.~Bertolini, P.~Di~Vecchia, M.~Frau, A.~Lerda, R.~Marotta, and I.~Pesando,
  ``{Fractional D-branes and their gauge duals},''
  \href{http://dx.doi.org/10.1088/1126-6708/2001/02/014}{{\em JHEP} {\bfseries
  02} (2001) 014}, \href{http://arxiv.org/abs/hep-th/0011077}{{\ttfamily
  arXiv:hep-th/0011077}}.

\bibitem{Polchinski:2000mx}
J.~Polchinski, ``{$\mathcal{N}{=}2$ gauge/gravity duals},''
  \href{http://dx.doi.org/10.1142/S0217751X01003834}{{\em Int. J. Mod. Phys. A}
  {\bfseries 16} (2001) 707--718},
  \href{http://arxiv.org/abs/hep-th/0011193}{{\ttfamily arXiv:hep-th/0011193}}.

\bibitem{Klebanov:1999rd}
I.~R. Klebanov and N.~A. Nekrasov, ``{Gravity duals of fractional branes and
  logarithmic RG flow},''
  \href{http://dx.doi.org/10.1016/S0550-3213(00)00016-X}{{\em Nucl. Phys. B}
  {\bfseries 574} (2000) 263--274},
  \href{http://arxiv.org/abs/hep-th/9911096}{{\ttfamily arXiv:hep-th/9911096}}.

\bibitem{Klebanov:2000nc}
I.~R. Klebanov and A.~A. Tseytlin, ``{Gravity duals of supersymmetric $SU(N)
  \times SU(N+M)$ gauge theories},''
  \href{http://dx.doi.org/10.1016/S0550-3213(00)00206-6}{{\em Nucl. Phys. B}
  {\bfseries 578} (2000) 123--138},
  \href{http://arxiv.org/abs/hep-th/0002159}{{\ttfamily arXiv:hep-th/0002159}}.

\bibitem{Klebanov:2000hb}
I.~R. Klebanov and M.~J. Strassler, ``{Supergravity and a confining gauge
  theory: Duality cascades and $\chi$SB resolution of naked singularities},''
  \href{http://dx.doi.org/10.1088/1126-6708/2000/08/052}{{\em JHEP} {\bfseries
  08} (2000) 052}, \href{http://arxiv.org/abs/hep-th/0007191}{{\ttfamily
  arXiv:hep-th/0007191}}.

\bibitem{Strassler:2005qs}
M.~J. Strassler, \href{http://dx.doi.org/10.1142/9789812775108_0005}{``{The
  Duality cascade},''} in {\em {Theoretical Advanced Study Institute in
  Elementary Particle Physics (TASI 2003): Recent Trends in String Theory}},
  pp.~419--510.
\newblock 2005.
\newblock \href{http://arxiv.org/abs/hep-th/0505153}{{\ttfamily
  arXiv:hep-th/0505153}}.

\bibitem{Intriligator:1995id}
K.~A. Intriligator and N.~Seiberg, ``{Duality, monopoles, dyons, confinement
  and oblique confinement in supersymmetric $SO(N_c)$ gauge theories},''
  \href{http://dx.doi.org/10.1016/0550-3213(95)00159-P}{{\em Nucl. Phys. B}
  {\bfseries 444} (1995) 125--160},
  \href{http://arxiv.org/abs/hep-th/9503179}{{\ttfamily arXiv:hep-th/9503179}}.

\bibitem{Acharya:2001dz}
B.~S. Acharya and C.~Vafa, ``{On domain walls of $\mathcal{N}{=}1$
  supersymmetric Yang-Mills in four-dimensions},''
  \href{http://arxiv.org/abs/hep-th/0103011}{{\ttfamily arXiv:hep-th/0103011}}.

\bibitem{Aharony:2000nt}
O.~Aharony, N.~Dorey, and S.~P. Kumar, ``{New modular invariance in the
  $\mathcal{N}{=}1^*$ theory, operator mixings and supergravity
  singularities},'' \href{http://dx.doi.org/10.1088/1126-6708/2000/06/026}{{\em
  JHEP} {\bfseries 06} (2000) 026},
  \href{http://arxiv.org/abs/hep-th/0006008}{{\ttfamily arXiv:hep-th/0006008}}.

\end{thebibliography}\endgroup

\end{document}